%% file: ANA-TRIG-2024-01-PAPER.tex
\documentclass[PAPER, cernpreprint, UKenglish, texmf, orcidlogo]{atlasdoc}
\usepackage[full]{atlaspackage}
\usepackage{cleveref}
\usepackage{multirow}
\usepackage{import}
\usepackage{makecell}
\usepackage{subcaption}
\usepackage{float}
\usepackage{atlasbiblatex}

\usepackage[jetetmiss=true, process=true]{atlasphysics}
\graphicspath{{logos/}{figures/}}

\usepackage{ANA-TRIG-2024-01-PAPER-defs}

\AtlasTitle{Performance of the ATLAS tau trigger in Run 3 and future prospects}

\AtlasAbstract{%
The algorithms and selections used at the trigger level in the ATLAS experiment to select events with hadronically decaying tau leptons are described. The performance of the hadronic tau trigger is presented using proton--proton collision data recorded by the ATLAS experiment in Run 3 at $\sqrt{s} = 13.6$ TeV at the Large Hadron Collider. In the plateau region of the efficiency curve for a given single tau trigger, efficiency varies between $90$--$98\%$. Correction factors for the simulated tau trigger efficiency in Monte Carlo samples were derived using 2022 and 2023 data enriched with hadronically decaying tau lepton final states, and these range from $0.95$--$1.05$ with an  uncertainty of $2$--$3\%$ over a wide range of the transverse momentum of the hadronic tau candidates.
In addition, potential improvements and their performance for the identification of hadronic tau candidates at trigger level are discussed using data recorded in 2026, having been implemented in selected tau trigger test chains.
}

\AtlasRefCode{TRIG-2024-01}

\PreprintIdNumber{CERN-EP-2026-290}

\AtlasJournal{JINST}
\AtlasCoverEgroupAnalysisTeam{atlas-TRIG-2024-01-analysis-team@cern.ch}

\hypersetup{pdftitle={ATLAS document},pdfauthor={The ATLAS Collaboration}}

\begin{document}

\maketitle

\tableofcontents

\let\clearpagebak\clearpage
\let\clearpage\relax

\section{Introduction}%

Tau leptons are a key signature for many measurements of Standard Model (SM) processes and for searches for new physics at the Large Hadron Collider (LHC)~\cite{Evans:2008zzb}. The most significant signal for measurements of the SM Higgs boson coupling to fermions~\cite{HIGG-2013-32, HIGG-2022-07} is provided by tau lepton pairs. Final states containing tau leptons are also favoured by heavy Higgs bosons or other new resonances in many scenarios beyond the SM~\cite{HDBS-2018-39, SUSY-2018-04, HDBS-2018-46}. Most searches for these new particles rely on triggering exclusively on hadronically decaying tau (\tauhad) leptons. Achieving high acceptance with optimal reconstruction quality for such hadronic signatures within the bandwidth and computing constraints of the ATLAS trigger system~\cite{TRIG-2022-01} is very challenging.

With a mass of $1776.86 \pm 0.12$ \MeV and a proper decay length of $87.03 \pm 0.15$ \si{\micro\meter}, tau leptons decay either leptonically ($\tau \rightarrow \ell + \nu_\ell + \nu_\tau$, $\ell = e, \mu$) or hadronically ($\tau \rightarrow \mathrm{~hadrons~} + \nu_\tau$) and do so typically before reaching active regions of the ATLAS detector~\cite{PERF-2007-01}. The hadronic decays represent $65\%$ of all possible tau lepton decay modes, and their decay products are typically one (``one-prong'' decay) or three (``three-prong'' decay) charged pions and possible additional neutral pions~\cite{ParticleDataGroup:2026zbx}. The neutral and charged hadrons arising from a hadronic tau lepton decay constitute the visible decay products of the tau lepton. These visible decay products of the tau lepton are denoted by \tauhadvis. Leptonic tau lepton decays, on the other hand, are denoted via $\tau_e$ or $\tau_\mu$, depending on the flavour of the light lepton in the final state.

In ATLAS, dedicated tau trigger algorithms are designed and implemented based on the main features of hadronic tau decays: narrow calorimeter energy deposits and a small number of associated tracks. Jets have similar features to hadronic tau decays and are abundantly produced at LHC. %
Keeping the \RunThr (2022--2026) tau trigger rate under control is even more challenging than in Run 1 (2009--2012) or Run 2 (2015--2018) of the LHC, because of the higher production rate of jets from the increase in collision energy and the higher number of collisions per bunch crossing (\pileup).

This paper describes the tau trigger used by ATLAS to select events with hadronically decaying tau leptons in proton--proton (\pp) collisions at a centre-of-mass energy of $\sqrt{s} = \SI{13.6}{\TeV}$ at the LHC\@. The paper is organised as follows: After a brief overview of the ATLAS detector and LHC operating conditions for \RunThr in \cref{sec:lhc-atlas} and the ATLAS trigger system in \cref{sec:trigger}, the details of the data and simulated samples along with the object reconstruction are described in \cref{sec:datasets-reco}. The ATLAS hardware and software tau trigger algorithms and selection criteria are described in Sections~\ref{sec:L1} and~\ref{sec:HLT}, respectively, while \cref{sec:menu} details the tau triggers used during \RunThr along with the topological selection requirement and observed online rates. \cref{sec:perf} shows the performance and efficiency of the hadronic tau triggers as measured in a dataset of \SI{56.2}{\ifb} collected in 2022 and 2023. Finally, \cref{sec:future} discusses some possible future improvements to tau trigger algorithms that were investigated and tested using data collected in 2026.


\section{The LHC operations and the ATLAS detector in \RunThr}\label{sec:lhc-atlas}

\newcommand{\AtlasCoordFootnote}{%
ATLAS uses a right-handed coordinate system with its origin at the nominal interaction point (IP)
in the centre of the detector and the \(z\)-axis along the beam pipe.
The \(x\)-axis points from the IP to the centre of the LHC ring,
and the \(y\)-axis points upwards.
Polar coordinates \((r,\phi)\) are used in the transverse plane,
\(\phi\) being the azimuthal angle around the \(z\)-axis.
The pseudorapidity is defined in terms of the polar angle \(\theta\) as \(\eta = -\ln \tan(\theta/2)\) and is equal to the rapidity
$ y = \frac{1}{2} \ln \left( \frac{E + p_z}{E - p_z} \right) $ in the relativistic limit.
Angular distance is measured in units of \(\Delta R \equiv \sqrt{(\Delta \eta)^{2} + (\Delta\phi)^{2}}\). Transverse momentum and energy are defined by their  projection on the transverse plane, $\pt = |\vec{p}|\sin\theta$ and $\ET = E \sin\theta$.}

After three years of long shutdown in 2019--2021 for maintenance and upgrades of the LHC machine~\cite{LHCupgradeRun3} and its experiments, the LHC at CERN produced \pp collisions at higher collision energies from 2022 to 2026 for \RunThr. The centre-of-mass energy of these collisions was increased from \SI{13}{\TeV} in \RunTwo to \SI{13.6}{\TeV} in \RunThr. The bunch intensity in \RunThr increased from \num{1.5e11} protons per bunch in 2022 to \num{1.6e11} protons per bunch. The average number of interactions per bunch crossing, $\langle\mu\rangle$, and the integrated luminosity delivered by the LHC during \RunThr are summarised in \cref{tab:lhc-atlas:lumi}.

\begin{table}[ht]
\caption{The average number of interactions per bunch crossing, $\langle\mu\rangle$, and luminosity delivered by the LHC during \RunThr.}
\label{tab:lhc-atlas:lumi}
\centering
\begin{tabular}{c|c|c}
\toprule
Year    & $\left\langle \mu \right\rangle$  & Luminosity [\ifb] \\
\midrule
2022    & 42                                & 38.5              \\
2023    & 51                                & 31.7              \\
2024    & 57                                & 123               \\
2025    & 60                                & 125               \\
2026    & 59                                & 32.9              \\
\bottomrule
\end{tabular}
\end{table}

The ATLAS detector at the LHC covers nearly the entire solid angle around the collision point\footnote{\AtlasCoordFootnote}. It consists of an inner tracking detector surrounded by a thin superconducting solenoid, electromagnetic (EM) and hadronic (HAD) calorimeters, and a muon spectrometer incorporating three large superconducting air-core toroidal magnets.

The inner-detector system (ID) is immersed in a \qty{2}{\tesla} axial magnetic field and provides charged-particle tracking in the range \(|\eta| < 2.5\). The high-granularity silicon pixel detector covers the vertex region and typically provides four measurements per track, the first hit generally being in the insertable B-layer (IBL). It is followed by the semiconductor tracker (SCT), which usually provides eight measurements per track. These silicon detectors are complemented by the transition radiation tracker (TRT), which enables radially extended track reconstruction up to \(|\eta| = 2.0\). The TRT also provides electron identification information based on the fraction of hits (typically 30 in total) above a high energy-deposit threshold corresponding to transition radiation.

The calorimeter system covers the pseudorapidity range \(|\eta| < 4.9\). Within the region \(|\eta|< 3.2\), EM calorimetry is provided by barrel and endcap high-granularity lead/liquid-argon (LAr) calorimeters, with an additional thin LAr presampler (PS) covering \(|\eta| < 1.8\) to correct for energy loss in material upstream of the calorimeters. HAD calorimetry is provided by the steel/scintillator-tile calorimeter, segmented into three barrel structures within \(|\eta| < 1.7\), and two copper/LAr HAD endcap calorimeters. The solid angle coverage is completed with forward copper/LAr and tungsten/LAr calorimeter modules optimised for EM and HAD energy measurements, respectively.

The muon spectrometer (MS) comprises separate trigger and high-precision tracking chambers measuring the deflection of muons in a magnetic field generated by the superconducting air-core toroidal magnets. The field integral of the toroids ranges between \num{2.0} and \qty{6.0}{\tesla\metre} across most of the detector. Three layers of precision chambers, each consisting of layers of monitored drift tubes, cover the region \(|\eta| < 2.7\). These were complemented in the innermost layer of the endcap region by cathode-strip chambers in \RunTwo, which were replaced in \RunThr by layers of small-strip thin-gap chambers and micromegas chambers, both providing redundant precision tracking in the region \(1.3 < |\eta| < 2.7\) and a large reduction of fake muon triggers. The muon trigger system covers the range \(|\eta| < 2.4\) with resistive-plate chambers in the barrel region (\(|\eta| < 1.05\)), thin-gap chambers in the endcap regions (\(|\eta| > 1.05\)), and, in \RunThr, the aforementioned small-strip thin-gap chambers and micromegas chambers in the innermost layer of the endcap~\cite{GENR-2019-02}.

The luminosity is measured mainly by the LUCID--2 detector that records Cherenkov light produced in the quartz windows of photomultipliers located close to the beampipe~\cite{ATL-DAPR-PUB-2025-001}.


\section{The ATLAS trigger system in \RunThr}\label{sec:trigger}

The \RunThr ATLAS trigger system~\cite{TRIG-2022-01} uses a two-level trigger system to select events of interest to reduce the nominal $40$ MHz bunch crossing rate to about $3$ kHz of recorded physics events. The trigger system is composed of a custom, hardware-based first-level trigger (L1) followed by software-based algorithms implemented in the High Level Trigger (HLT).

The L1 trigger is a synchronous, pipelined system consisting of a Central Trigger Processor (CTP) fed by signals coming primarily from dedicated trigger hardware in the calorimeter (L1Calo) and muon (L1Muon) detector systems. The L1 topological processor (L1Topo) system uses kinematic information from objects reconstructed in the L1Calo and L1Muon systems and applies topological selections.  The initial rate of $40$ MHz is reduced to a maximum of $100$ kHz.

Events accepted at L1 are then processed in the HLT system, which is deployed in a farm of computers using full-granularity data. Algorithms similar to those employed offline are used to reconstruct tau candidates and employed to select events and reduce the output event rate to about 3~kHz of recorded events.

\subsection{L1 trigger}

During the Phase-I upgrade of the ATLAS detector between \RunTwo and \RunThr, the L1Calo trigger was upgraded to perform on-detector digitisation of transverse energies from the EM calorimeters (ECal)~\cite{ATLAS-TDR-22}. With this change in \RunThr, L1Calo receives finer-granularity input data from the ECal. While in \RunOne and \RunTwo, these inputs consisted of trigger towers (TT) spanning $0.1 \times 0.1$ in $\eta \times \phi$, the ECal information in \RunThr is provided in the form of supercells (SCs) containing sums of four or eight calorimeter cells. Each TT is segmented in four layers along its longitudinal axis, with the ECal presampler and back layer SCs retaining the $0.1 \times 0.1$ granularity of the TTs, while the middle two layers have a finer SC size of $0.025 \times 0.1$. The new digital readout system performs a per-cell energy calibration, and transparently handles the different longitudinal segmentation of the calorimeters, as well as the geometrical irregularities in the transition region between the central barrel and the endcap sectors of the ECal ($1.37 < |\eta| < 1.52$).
The Tile calorimeter data are still received from the detector electronics in analogue format, digitised by pre-processors, and transmitted by the new Tile Rear EXtension (TREX) modules to both the feature extractor (FEX) processor and the \RunTwo L1Calo system (hereafter referred to as the legacy system).

The upgraded L1Calo system includes new EM, jet, and global FEXs labelled as eFEX, jFEX and gFEX, respectively. The full ECal SCs granularity is available at the eFEX to reconstruct EM objects and hadronically decaying tau leptons, as well as shower-shape variables used for their identification.  The L1Topo modules were also replaced with more capable processors and receive information from L1Muon and the new L1Calo FEX processors.

In the legacy L1 system, analogue detector signals are digitised and calibrated in the multi-chip modules (nMCM) in the pre-processor system and sent in parallel to the cluster processors (CP) and the jet/energy-sum processors (JEP). The CP system identifies electron, photon, and tau lepton candidates above a programmable energy threshold, and the JEP system identifies jet candidates and produces global sums of total and missing transverse energy. Trigger chains using this legacy system are referred to as legacy chains, while trigger chains using the new upgraded L1Calo system are referred to as Phase-I trigger chains, since their development was part of the ATLAS Phase-I upgrade. During data-taking in $2022$ and $2023$, the legacy trigger chains operated in parallel to their Phase-I counterparts.

\subsection{High level trigger}

The HLT is implemented in a CPU farm where computers run AthenaMT, a multi-threading version of the ATLAS reconstruction software suite, which allows inter-event and intra-event parallel processing~\cite{SOFT-2022-02}. Algorithms similar to those used for offline reconstruction are implemented. The HLT reconstruction proceeds in steps, reducing the rate in each step.  Calorimeter clustering is executed first, while CPU-intensive algorithms, such as tracking, are executed in later steps and only for candidate objects fulfilling the previous selection criteria. Only parts of the HLT relevant to \tauhadvis reconstruction are described in the following.

Calorimetric object reconstruction in the HLT has access to the full detector granularity and is harmonised with the offline reconstruction but not fully calibrated. The topo-clustering algorithm~\cite{TRIG-2022-01} is executed on all events accepted by the L1 tau triggers. Bunch-by-bunch cell corrections are implemented to mitigate out-of-time pile-up effects and improve the HLT energy resolution with respect to offline clusters.

Track reconstruction in the HLT~\cite{TRIG-2019-03} is run in two steps. In first step, the fast track finder (FTF) algorithm is used to run trigger-specific pattern recognition algorithms and a fast Kalman filter track fit. Following this, a precision tracking stage runs the offline track reconstruction algorithms to refit the FTF tracks found in the fast tracking stage, and then extrapolates them into the TRT to include TRT information in the track fit. Details of the HLT selection for \tauhadvis candidates are described in \cref{sec:HLT}.


\section{Data samples and offline object reconstruction}\label{sec:datasets-reco}

The results presented in this paper make use of both simulated Monte Carlo (MC) samples and collision data. The selection criteria for the trigger, to be run online, are optimised using simulated MC samples to estimate how efficiently the signal events are selected. For optimisation purposes, the combined rate of signal and background events, on the other hand, is estimated by using enhanced bias data overweighted in high \pt events ~\cite{ATL-DAQ-PUB-2016-002}.

\subsection{Data and MC samples}\label{sec:datamc}

Various data samples of \(pp\) collision data in \RunThr are employed in this paper. A data sample recorded in between May and August 2024 is used to compare the efficiency of legacy and Phase-I tau triggers in Section~\ref{sec:L1}. Typical trigger accept rates are shown in Section~\ref{sec:menu} for data samples recorded in May 2023 (legacy tau triggers) and October 2025 (Phase-I tau triggers).
Data samples corresponding to $29$ and $27.2$ \ifb, recorded in 2022 and 2023, respectively, are used for measurements of the tau trigger efficiencies presented in Section~\ref{sec:perf}. Data recorded in May 2026 are used to demonstrate the performance of tau trigger test chains discussed in Section~\ref{sec:future}.

A summary of the simulated MC samples, used for various studies including optimisation and performance, is given in \cref{tab:samples}.
In particular, the samples consisting of \gtautau events are generated with \PYTHIA[8.212]~\cite{Sjostrand:2014zea}.
This is used for the matrix-element calculation, as well as the simulation of parton shower, hadronisation, underlying event, and $\tau$ lepton decay. The A14~\cite{ATL-PHYS-PUB-2014-021} set of parameter tunes and \NNPDF[2.3lo] or \NNPDF[2.3nnlo]~\cite{Ball:2012cx} as the PDF are employed, depending on the availability of samples for different studies over time.
The \gtautau events use a biased phase space sampling to populate the generated spectrum towards high invariant masses ($m_{\gamma^*}$) to provide sufficient numbers of high-\pt{} \tauhadvis{} candidates up to 1~\TeV. In addition, the absence of amplitudes mediated by a $Z$ boson provides a more uniform sample of tau leptons in terms of helicity as a function of  \tauhad \pT.
All $\tau$ leptons in the simulated \gtautau events are forced to decay hadronically.

The effect of \pileup was modelled by overlaying~\cite{SIMU-2020-01} the simulated hard-scattering event with inelastic \pp events following the \pileup profile of the corresponding data sample and generated from a mix of \EPOS[2.0.1.4]~\cite{Werner:2005jf} and \PYTHIA[8.308]~\cite{Bierlich:2022pfr}. The \EPOS events were generated with the \EPOS LHC tune~\cite{Pierog:2013ria} and the \PYTHIA events with the A3 tune~\cite{ATL-PHYS-PUB-2016-017} and the \NNPDF[2.3lo]~\cite{Ball:2012cx} set of parton distribution functions (PDF). \PYTHIA \pileup events include either a high \pt jet, a prompt photon, or a lepton from a $b$-hadron decay, while \EPOS was filtered to simulate all remaining \pileup events in the overlay sample. The individual simulations were first reweighted to ensure a smooth connection across jet \pt then the combination was reweighted to match the distribution of $\left<\mu\right>$ measured in data.

\begin{table}[ht]
\caption{Summary of the simulated MC samples used for optimisation and performance of the hadronic tau trigger selections.}
\label{tab:samples}
\centering
\resizebox{.7\textwidth}{!}{
\centering
\begin{tabular}{c|l|c|c}
\toprule
Process                 & \multicolumn{1}{c|}{ Event Generator }    & PDFs                              & Relevant Section \\
\midrule
\Ztautau                & \POWHEG~\cite{Nason:2004rx, Frixione:2007vw, Alioli:2010xd} + \PYTHIA[8]                      & \NNPDF[2.3nnlo]                   & \ref{sec:L1} \\
\ttbar                  & \POWHEG + \PYTHIA[8]                      & \NNPDF[3.0nlo]                    & \ref{sec:future} \\
\midrule
\gtautau                & \PYTHIA[8]                                & \NNPDF[2.3nnlo]                   & \ref{sec:L1}, \ref{sec:HLT} \\
\gtautau                & \PYTHIA[8]                                & \NNPDF[2.3lo]                     & \ref{sec:future} \\
\Ztautau                & \SHERPA[2.2]~\cite{Hoeche:2009rj}         & \NNPDF[3.0nnlo]                   & \ref{sec:HLT} \\
\ttbar                  & \POWHEG + \PYTHIA[8]                      & \NNPDF[3.0nnlo]                   & \ref{sec:HLT}, \ref{sec:perf} \\
Dijet                   & \PYTHIA[8]                                & \NNPDF[2.3nnlo]                   & \ref{sec:HLT} \\
\midrule
\Zjets                  & \POWHEG + \PYTHIA[8]                      & \CT[10]~\cite{Lai:2010vv}         & \ref{sec:perf} \\
\Wjets                  & \POWHEG + \PYTHIA[8]                      & \CT[10]                           & \ref{sec:perf} \\
Diboson                 & \SHERPA[2.2]                              & \NNPDF[3.0nnlo]                   & \ref{sec:perf} \\
Single top              & \POWHEG + \PYTHIA[8]                      & \NNPDF[3.0nnlo]                   & \ref{sec:perf} \\
\bottomrule
\end{tabular}
}
\end{table}

\subsection{Offline object reconstruction}\label{sec:Objectreco}
The optimisation and performance of trigger algorithms are evaluated with the 2022 and 2023 data sample mentioned in Section~\ref{sec:datamc}, enriched in $Z\rightarrow \tau_{e/\mu}\tau_{\text{had}}$ events where one tau lepton decays into an electron or muon and the other decays hadronically, with associated neutrinos. The chosen tag-and-probe approach consists of selecting events triggered by the presence of an electron or muon (tag) that also contain a \tauhad candidate (probe) in the final state. Details of the tag-and-probe measurement are given in~\cref{sec:perf}. This section describes the criteria used to reconstruct objects such as electrons, muons, \tauhad leptons, jets, \btagged jets and missing transverse energy.

Electrons are reconstructed from topological clusters of energy deposits in the ECal that are matched to a track reconstructed in the ID\@.
The electrons used in the analysis are required to have transverse momentum ($\pT$) above $27$~\GeV, $|\eta|<2.47$, excluding the LAr barrel-endcap transition region $1.37 < |\eta| < 1.52$, and satisfy the \emph{Tight} identification and isolation criteria, as described in Ref.~\cite{EGAM-2021-01}.

Muons are reconstructed from signals in the MS matched with tracks in the ID and are required to have $\pT>\SI{25}{\GeV}$, $|\eta|<2.5$ and satisfy the \emph{Medium} identification and \emph{Loose} isolation criteria based on track information, as described in Ref.~\cite{MUON-2018-03}.

The constituents for jet reconstruction are identified by combining measurements from both the ID and the calorimeter using a particle flow (PFlow) algorithm~\cite{PERF-2015-09}. Jet candidates are reconstructed from these PFlow objects using the anti-\(k_{t}\) algorithm~\cite{Cacciari:2008gp, Fastjet} with a radius parameter of $R = 0.4$. They are calibrated using simulation with corrections obtained by using in situ techniques in data~\cite{JETM-2018-05}.

A jet vertex tagger (JVT)~\cite{PERF-2014-03} is used to remove jets with $\pT<\SI{60}{\GeV}$ and $\abseta<2.5$ that are identified as not being associated with the production vertex of the event. Similarly, \pileup jets in the forward region are suppressed with a \enquote{forward JVT}~\cite{PERF-2016-06} algorithm that exploits jet shapes and topological jet correlations in \pileup interactions and is applied to all jets with $\pT<\SI{60}{\GeV}$ and $|\eta|>2.5$.

Jets containing $b$-hadrons are identified (\btagged) via the DL1d algorithm~\cite{FTAG-2019-07} that uses a deep-learning neural network based on the distinctive features of $b$-hadron decays, primarily the impact parameters of tracks and the displaced vertices reconstructed in the ID\@. Additional input to this network is provided by discriminating variables constructed by a recurrent neural network (RNN), which exploits the spatial and kinematic correlations between tracks originating from the same $b$-hadron. A multivariate $b$-tagging discriminant value is calculated for each jet.

The reconstruction and identification procedure for \tauhadvis candidates is described in Refs.~\cite{PERF-2013-06, ATLAS-CONF-2017-029, ATL-PHYS-PUB-2019-033, ATL-PHYS-PUB-2022-044}. The \tauhadvis candidates are seeded by jets formed using the anti-\(k_{t}\) algorithm with a radius parameter of $R=0.4$. Three-dimensional clusters of calorimeter cells called topoclusters~\cite{PERF-2014-07}, calibrated using a local hadronic calibration (LC)~\cite{PERF-2014-07}, serve as inputs to the jet algorithm. Jets seeding \tauhadvis candidates are required to have $\pt > 5~\GeV$ and $|\eta| < 2.5$.

Dedicated algorithms are run to identify the production vertex, associate and classify reconstructed ID tracks, and calibrate the energy of the \tauhadvis candidate. Only \tauhadvis candidates with one or three associated tracks are considered. After reconstruction, the \tauhadvis candidate is classified by the identification algorithm, based on a RNN classifier, to determine its likelihood of being the decay product of a $\tauhad$ lepton.

The missing transverse momentum is an estimate of the imbalance in \pt in the detector~\cite{PERF-2016-07}. It is calculated as the negative vector sum of the transverse momenta of all reconstructed final-state objects ($e/\mu/\tauhad$ and jets) and a soft term including tracks associated with the production vertex of the event but not to any reconstructed object.  Its magnitude is denoted by \met.

Since the objects used in the analysis are reconstructed from the same set of tracks and calorimetric energy clusters, some constituents can be associated with multiple objects. These overlaps between different objects are resolved with the procedure described in Ref.~\cite{HIGG-2019-09}.


\section{L1 tau triggers}\label{sec:L1}

The L1 hadronic tau triggers are part of the L1Calo trigger system, which uses reduced information from the EM and HAD calorimeters to determine regions of interest (RoIs) in the detector within $|\eta| < 2.5$. A set of selection criteria for the RoIs, based on the minimum multiplicity, quality, and potential topological distributions, are required to be fulfilled by the L1 trigger. The tau RoIs selected at L1 are later used as input to the HLT hadronic tau reconstruction sequence.

In the legacy trigger system, L1 tau RoIs were created using an algorithm with a square 4-TT sliding window (covering an area of $0.2 \times 0.2$ in the $\eta-\phi$ plane) to identify local energy maxima, and calculate the transverse energy (\ET) of the RoI based upon the four hadronic TTs in the window and the two most energetic contiguous EM TTs. Additional isolation criteria, based on the energy deposited in a surrounding ring of 12 EM TTs around the selected 4-TT window, were applied to RoIs with $\ET < \SI{40}{\GeV}$ and used in the primary unprescaled tau triggers.

The migration of the L1 hadronic tau triggers from legacy to Phase-I trigger system was performed in steps, replacing the non-isolated L1 tau triggers in $2023$, and the isolated lower-threshold and topological triggers in $2024$. A detailed description of the Phase-I tau triggers is described below.

\subsection{Non-isolated tau triggers at L1}

In the Phase-I system, the \ET of the L1 RoI is calculated from a subset of SCs in all layers. As shown in~\cref{fig:L1:ET_sums}, an asymmetric prescription in the $\eta-\phi$ plane was used during $2023$ data-taking, including only SCs in a reduced $0.3 \times 0.2$ window, determined by the $\eta$ coordinate of the RoI, and the most energetic of the two other SCs in the second layer of ECal with the same $\eta$. This algorithm was superseded from 2024 onwards by a simpler symmetric calculation, whose final configuration is also shown in \cref{fig:L1:ET_sums}.

The construction of the non-isolated tau RoIs in L1 trigger hardware is done by the L1Calo eFEX subsystem, comprised of 24 modules with four FPGAs each, covering overlapping sectors of the detector. Each eFEX tau (eTAU) RoI covers an area of $0.3 \times 0.3$, including $9$ TT\@. The RoIs are built using a seed finder algorithm, which selects the central TT as the one whose total energy (added from all layers of the calorimeters) is a local maxima amongst the eight closest surrounding TTs. The $\eta$ coordinate of the RoI is further refined by selecting the $\eta$ of the highest energy SC in the second layer of the ECal within the central TT, and is used in topological selections and the HLT reconstruction seeding.

The design of the \ET sum prescription was driven by its energy reconstruction accuracy, estimated by comparing the \ET of the eTAU with the visible transverse energy ($\ET^{\text{vis}}$) of a matched tau in $Z\to\tau\tau$ simulated MC events described in \cref{tab:samples}. The slope of the L1 trigger efficiency turn-on curve as a function of $\ET^{\text{vis}}$ of the tau candidates was taken into consideration in the design, choosing the algorithm resulting into the sharpest efficiency turn-on.

\begin{figure}[ht!]
\centering
\includegraphics[width=0.6\linewidth]{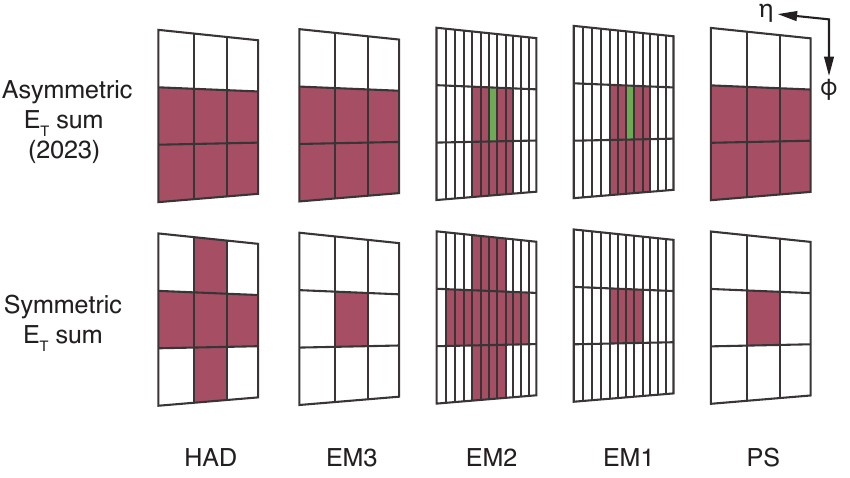}
\caption{Supercells included in the \ET sum for eTAU RoIs: an asymmetric energy sum was used in $2023$ for non-isolated triggers, and a symmetric sum starting in 2024. In the former, the highest \ET supercell that determines the centre of the RoI and its shape is highlighted.}
\label{fig:L1:ET_sums}
\end{figure}

\subsection{Isolated tau triggers at L1}

To suppress background events and have manageable trigger rates, L1 hadronic tau triggers with low thresholds ($\ET \lesssim \SI{40}{\GeV}$) use a combination of two additional quality criteria: a boosted-decision-tree (BDT) based algorithm and an energy-based isolation ring algorithm.

The BDT-based algorithm for eTAUs, implemented in the eFEX FPGAs, is designed to reject background events based on the shape of the energy deposits in the SCs that constitute eTAU RoIs. It uses $11$ input variables, $10$ of which are built from information of the high-granularity SCs in the first and second ECal layers, incorporating information about the shape of the energy deposition. The last parameter, with the greatest impact on the background rejection, is the energy sum over all layers of the central TT, providing the BDT with the \ET scale of the RoI.
Based on the minimum \ET requirement of each isolated L1 tau trigger, a selection on the BDT score is applied only on RoIs with $\SI{12}{\GeV} < \ET < \SI{50}{\GeV}$. For RoIs with $\ET > \SI{50}{\GeV}$, the L1 trigger rates are low enough that the BDT isolation isolation criteria is not applied.

The second criteria, based on an isolation requirement of energy deposits, exploits the narrow radius of \tauhadvis jets in the calorimeters, compared with jets from light quarks or gluon-induced jets. A requirement of absence of energy deposits in the surroundings of the tau RoI can significantly reduce the background contamination~\cite{PERF-2013-06}. However, in the Phase-I L1Calo system, the small eFEX RoIs do not provide a large enough surrounding area for an effective estimate of their isolation energy.

This isolation algorithm for \tauhadvis candidates at L1 makes use of an alternative RoI construction in the jFEX\@. Similar to the eFEX, it receives information from the calorimeters in $0.1 \times 0.1$ TTs, but trades the higher granularity of the ECal SCs with the capability of processing larger $0.7 \times 0.7$ areas. A dedicated algorithm is implemented to generate RoIs for \tauhadvis candidates, named jTAUs, for which the energy in the central $0.3 \times 0.3$ area, with $9$ TTs, is summed and referred to $\ET^{\text{jTAU}}$. A $2$ TT wide isolation ring is formed surrounding the central jTAU area, and its respective total energy is summed and referred to as $\ET^{\text{jTAU, Iso}}$. A geometrical matching is performed between the eTAU and the jTAU central area, as shown in \cref{fig:L1:cTAU}, to build a new collection of combined L1 tau RoIs (cTAUs). If no jTAU is matched to a seeding eTAU, the cTAU is considered to be isolated, otherwise the condition on the isolation fraction is applied in the cTAU algorithm. To reduce the sensitivity on the energy leakage of \tauhadvis candidates into the central isolation area at different energies, an additional regulating parameter $\alpha$ is included. The isolation fraction, $r$, is defined after taking into account contributions from both the external isolation ring and the central isolation area, formed of calorimeter cells included in the $0.3 \times 0.3$ central area but not in $\ET^{\text{eTAU}}$. The isolation criteria is then applied:
\[ \frac{\ET^{\text{jTAU, Iso}} + \alpha \left(\ET^{\text{jTAU}} - \ET^{\text{eTAU}}\right)}{\ET^{\text{eTAU}}} < r. \]

\begin{figure}[ht!]
\centering
\includegraphics[width=0.6\linewidth]{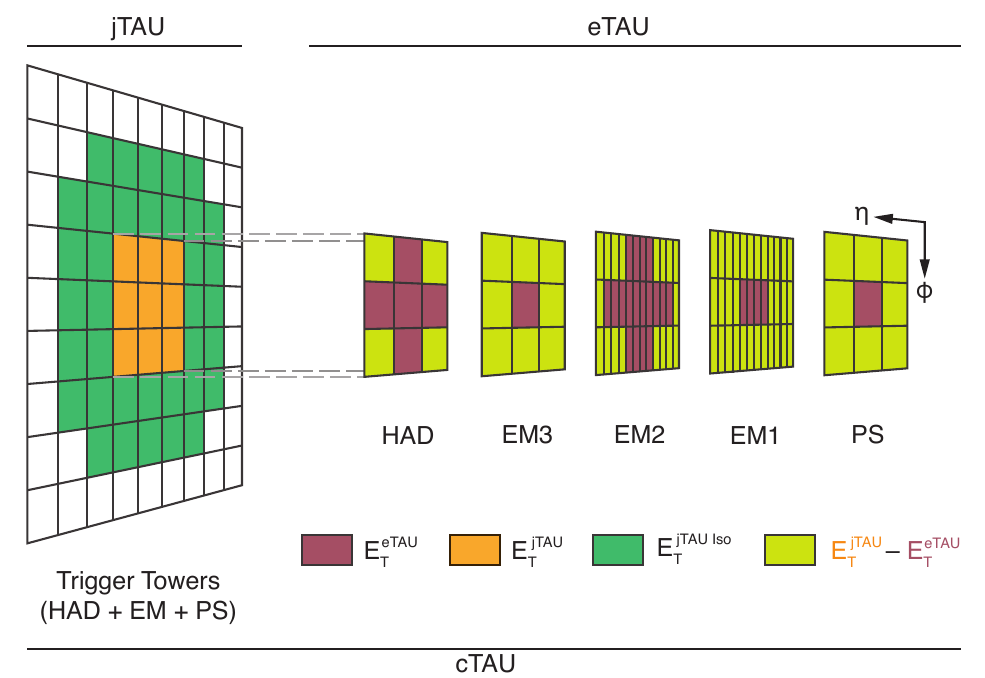}
\caption{Representation of the eTAU and jTAU RoIs. The energy sums calculated in both RoIs are indicated in colours. In a cTAU, the seeding eTAU RoI is attempted to be geometrically matched to a jTAU’s central region.}
\label{fig:L1:cTAU}
\end{figure}

A multi-variable optimisation procedure per $\ET$ threshold at L1 was used to find the best values for requirements on the isolation parameters ($r$ and $\alpha$) and the BDT output. The expected rates of the single and di-tau triggers were required to match with that of the legacy triggers, while maximising the selection efficiency of events with \tauhadvis signatures. As described below, the Phase-I system shows an improved efficiency relative to the Legacy system in the plateau region for the same rate.

The efficiencies of the isolated and non-isolated tau triggers at L1 are evaluated using \gtautau simulated MC events. The denominator, in the efficiency fraction, includes all particle-level \tauhadvis candidates with $\pT^{\text{vis}} > \SI{20}{\GeV}$ and $|\eta| < 2.5$ classified according to their visible decay products into one-prong and three-prong \tauhadvis candidates. The numerator additionally requires the \tauhadvis candidates to be matched, within $\Delta R < 0.15$, to a cTAU RoI with a given \ET threshold, and the isolation and BDT configurations stated in the figure. \cref{fig:L1:truth_efficiencies} shows the efficiency of tau trigger at L1, with $\ET > 10 $ \GeV, as a function of the generator-level $\pT^{\text{vis}}$ of the matched \tauhadvis candidate. The slope of the curve was taken into consideration in the design of the \ET sum prescription, where a sharper turn-on performance is preferred. The non-isolated tau trigger was found to be $95\%$ efficient for 1-prong (3-prong) \tauhad leptons with $\pT^{\text{vis}} > \SI{35} (\SI{40}){\GeV}$. Almost no efficiency is lost with the BDT selection, while the isolation requirement introduces a $5$-$10$\% efficiency loss primarily in the turn-on region.

\begin{figure}[ht!]
\centering
\subfloat[]{\label{fig:L1:truth_efficiencies:1p}\includegraphics[width=0.45\textwidth]{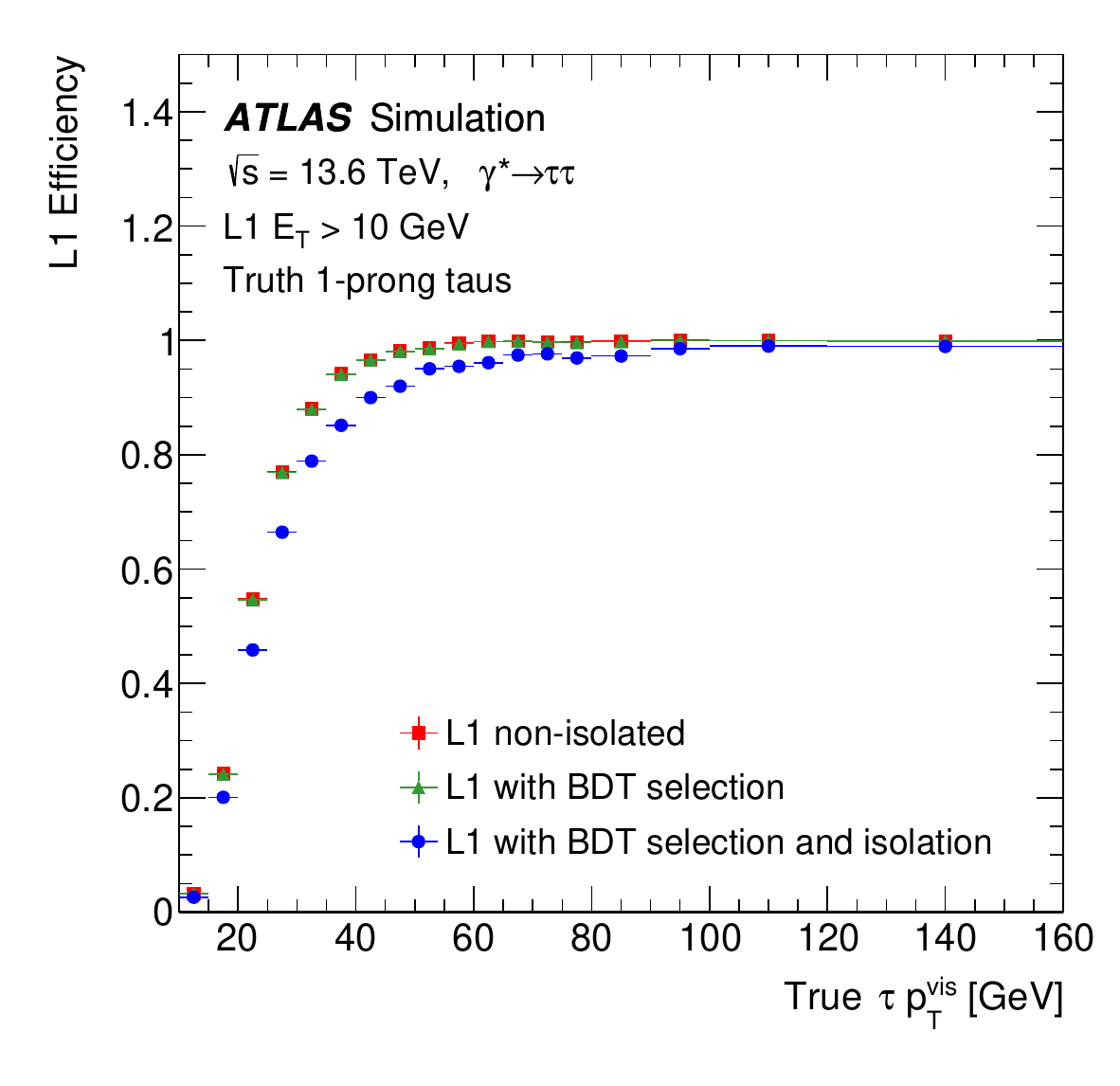}}
\subfloat[]{\label{fig:L1:truth_efficiencies:3p}\includegraphics[width=0.45\textwidth]{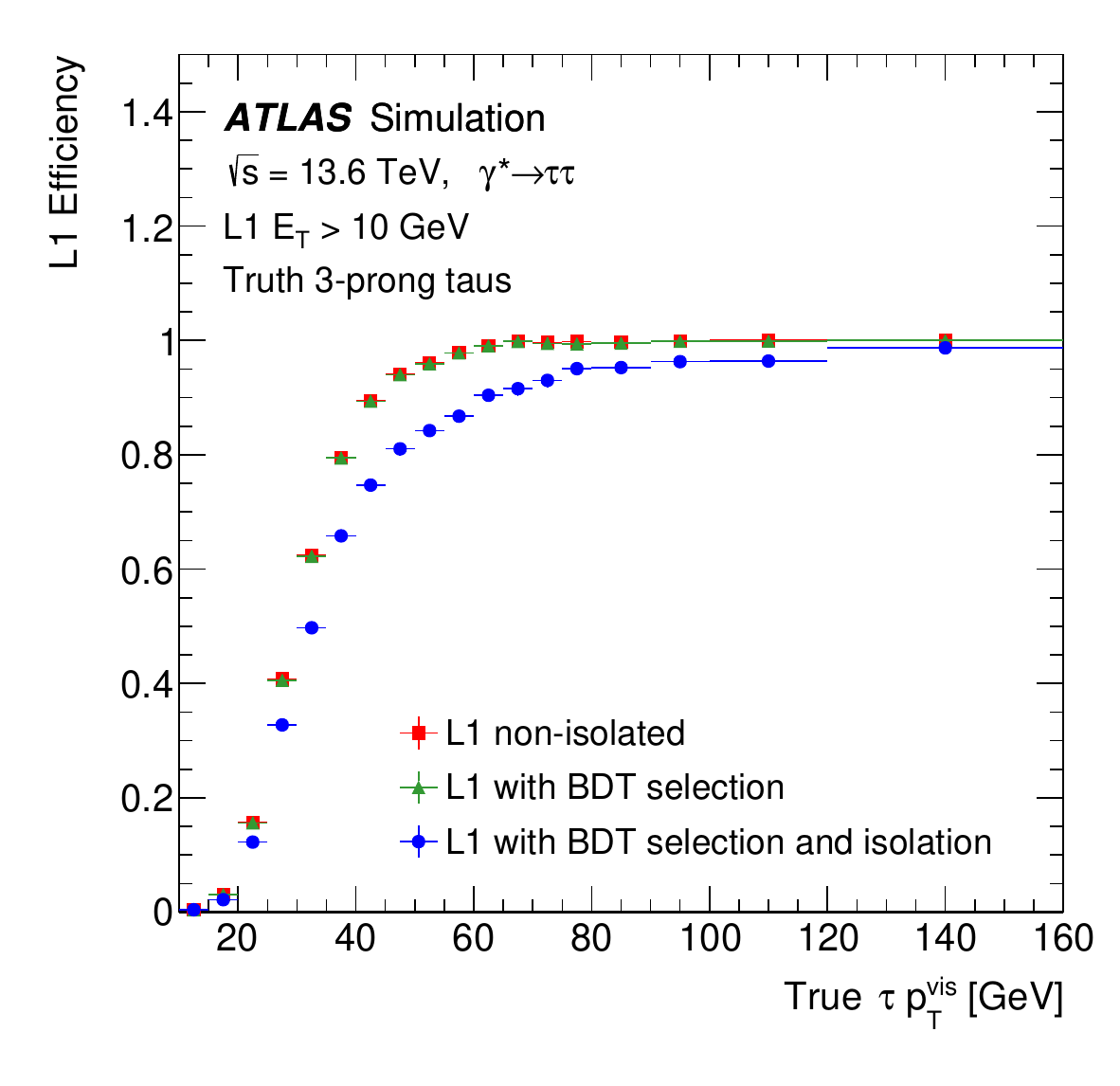}}
\caption{Expected performance of the L1 tau trigger, in \gtautau simulated MC events, as a function of the generator-level $\pt^{\text{vis}}$ for one-prong \protect\subref{fig:L1:truth_efficiencies:1p} and three-prong \protect\subref{fig:L1:truth_efficiencies:3p} \tauhad candidates. The L1 efficiency for non-isolated triggers along with the efficiency for requirements on the BDT score and isolation fraction for isolated triggers are shown for $\ET > \SI{10}{\GeV}$ candidates.}
\label{fig:L1:truth_efficiencies}
\end{figure}

A comparison of the tau trigger performance as observed in data, between legacy and Phase-I triggers is shown in \cref{fig:L1:2024_efficiencies} using data taken from May to August 2024. The details of the selections applied to select events with true \tauhad leptons from \Ztautau and \ttbar processes are described in \cref{sec:perf}. These efficiencies are measured without subtracting any backgrounds. The performance of the Phase-I triggers is at least as good as the legacy triggers, with a sharper efficiency turn-on curve, and an efficiency improvement of approximately 10\% in the plateau region for three-prong \tauhad candidates.

\begin{figure}[ht]
\centering
\subfloat[]{\label{fig:L1:data_efficiencies:1p}\includegraphics[width=0.45\textwidth]{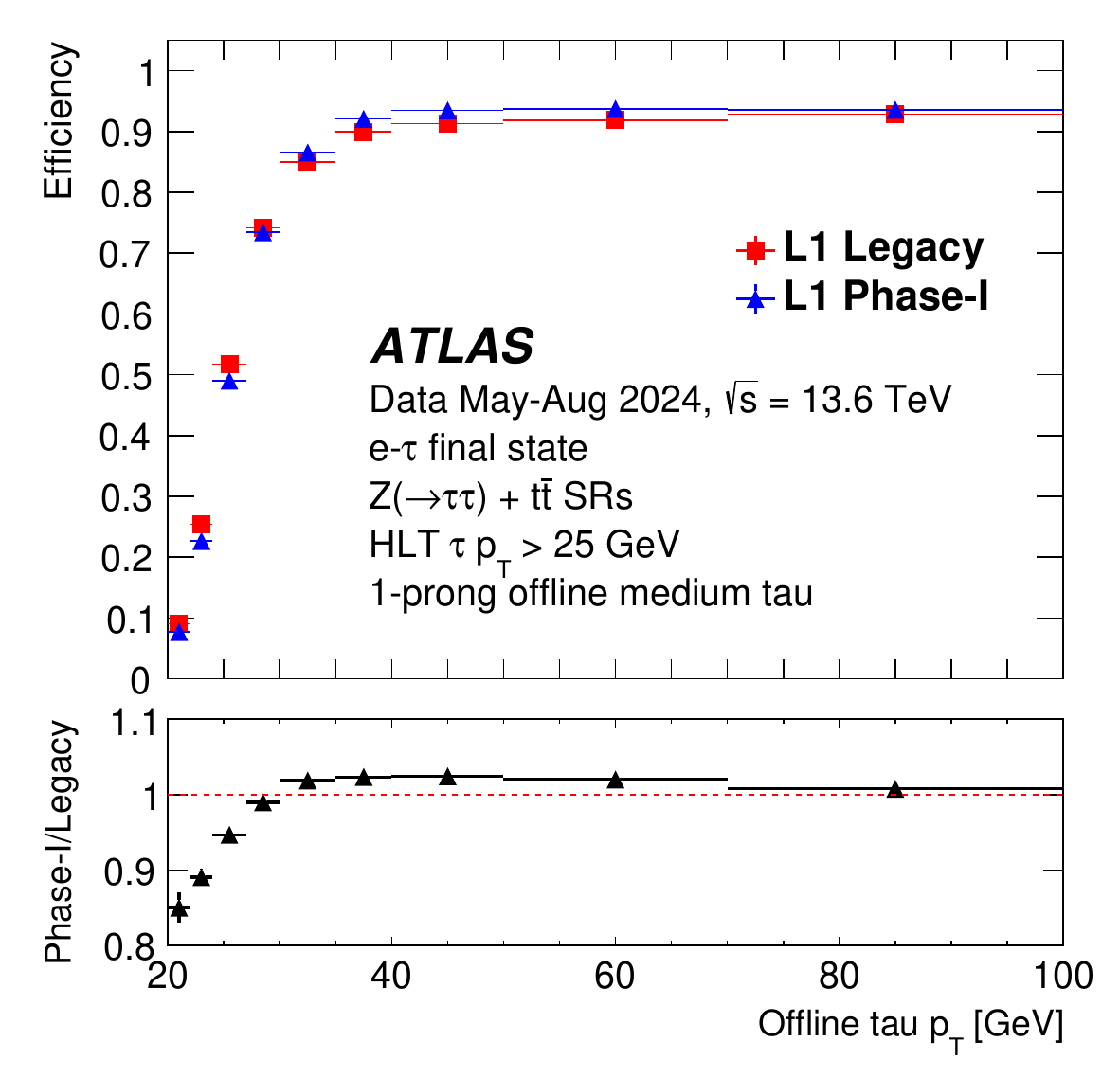}}
\subfloat[]{\label{fig:L1:data_efficiencies:3p}\includegraphics[width=0.45\textwidth]{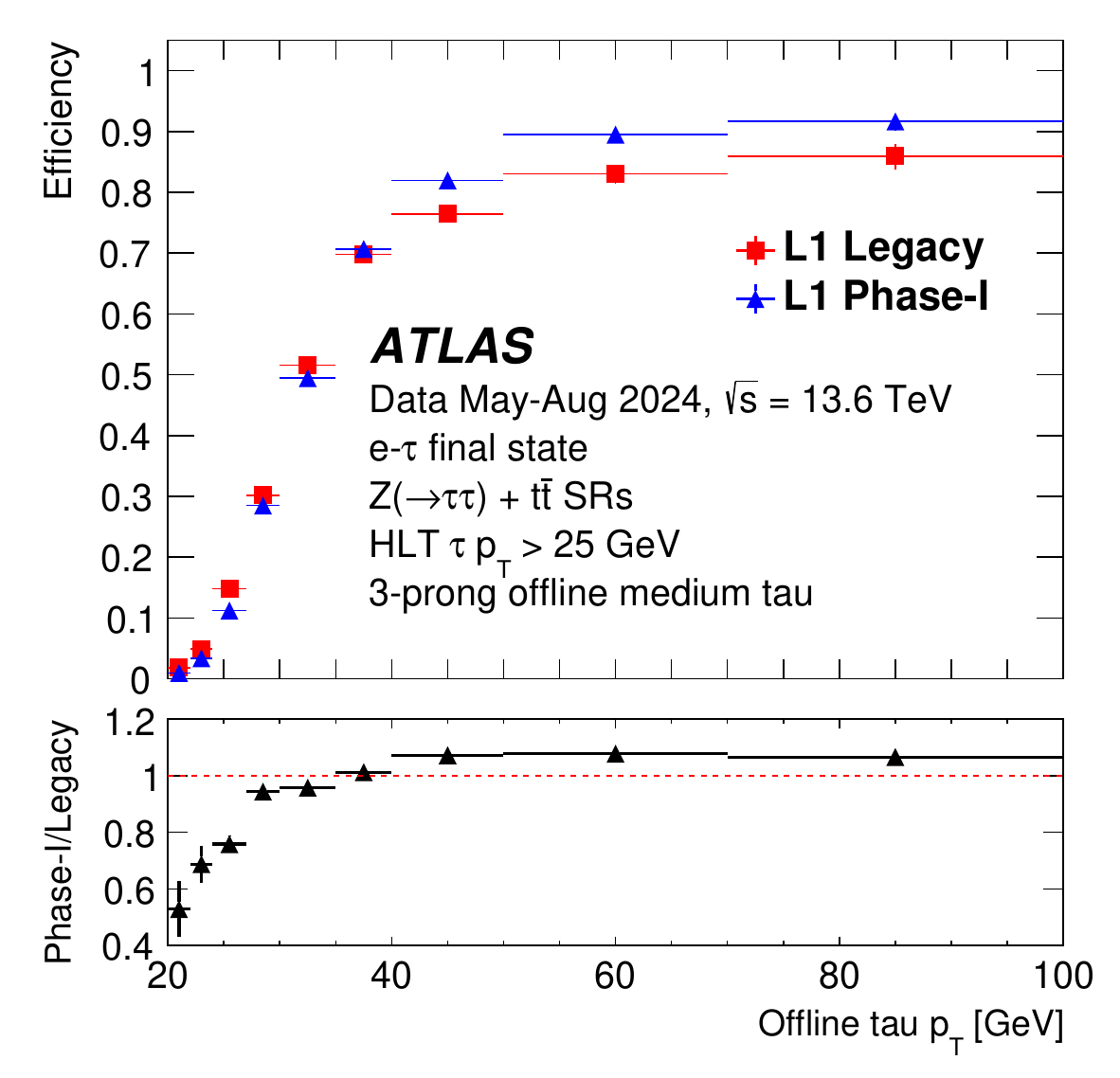}}
\caption{Comparison of the observed performance of the L1 tau trigger between legacy and Phase-I triggers in data collected in 2024. The trigger efficiency is shown as a function of the offline tau \pT for one-prong \protect\subref{fig:L1:data_efficiencies:1p} and three-prong \protect\subref{fig:L1:data_efficiencies:3p} \tauhad candidates. The \tauhad candidates are classified as one-prong or three-prong according to the number of offline reconstructed tracks.}
\label{fig:L1:2024_efficiencies}
\end{figure}


\section{HLT tau trigger}\label{sec:HLT}

The HLT tau object reconstruction and selection is divided in three steps: calorimeter-based selection, track-based selection and tau identification.
In each step, different reconstruction algorithms are used and selection requirements applied to reduce the CPU usage and trigger rate.

\subsection{Calorimeter-based selection}

At the start of the HLT reconstruction, the \tauhadvis candidate is built using only calorimeter-based information. A topo-clustering algorithm using calorimeter cells falling within the L1 tau RoIs is executed. Topo-clusters are calibrated using a local hadronic calibration and their 4-momentum sum defines the \tauhadvis direction~\cite{TRIG-2022-01}. Topo-clusters within a cone of radius $R=0.2$ around that direction are used in the calculation of the transverse momentum. A dedicated tau energy calibration is applied on top of the LC calibration to get a precise estimate of \pt. The calibration is based on Boosted Regression Trees (BRT) following the procedure described in Ref.~\cite{ATL-PHYS-PUB-2022-044}. The BRT is trained and evaluated using a \gtautau simulated sample. \cref{fig:HLT:Calo:TES_resolution} shows the transverse energy resolution as a function of \pileup and the true \tauhadvis dressed transverse momentum ($\pT^{\mathrm{vis,dressed}}$), which includes final-state photons within a cone of radius $R=0.2$ around the true \tauhadvis direction. This resolution is defined as half of the 68\% central quantile of the distribution of the ratio of the BRT-calibrated \pt of the \tauhadvis candidate and the $\pT^{\mathrm{vis,dressed}}$. Because the \tauhadvis candidates used at this stage have fulfilled L1 tau selection criteria, the average \pt of three-prong candidates is higher than those of one-prong candidates. The \tauhadvis candidates satisfying a minimum requirement on the BRT-calibrated \pt are considered for the next step.

\begin{figure}[ht!]
\centering
\subfloat[]{\label{fig:HLT:Calo:TES_resolution:mu}\includegraphics[width=0.485\linewidth]{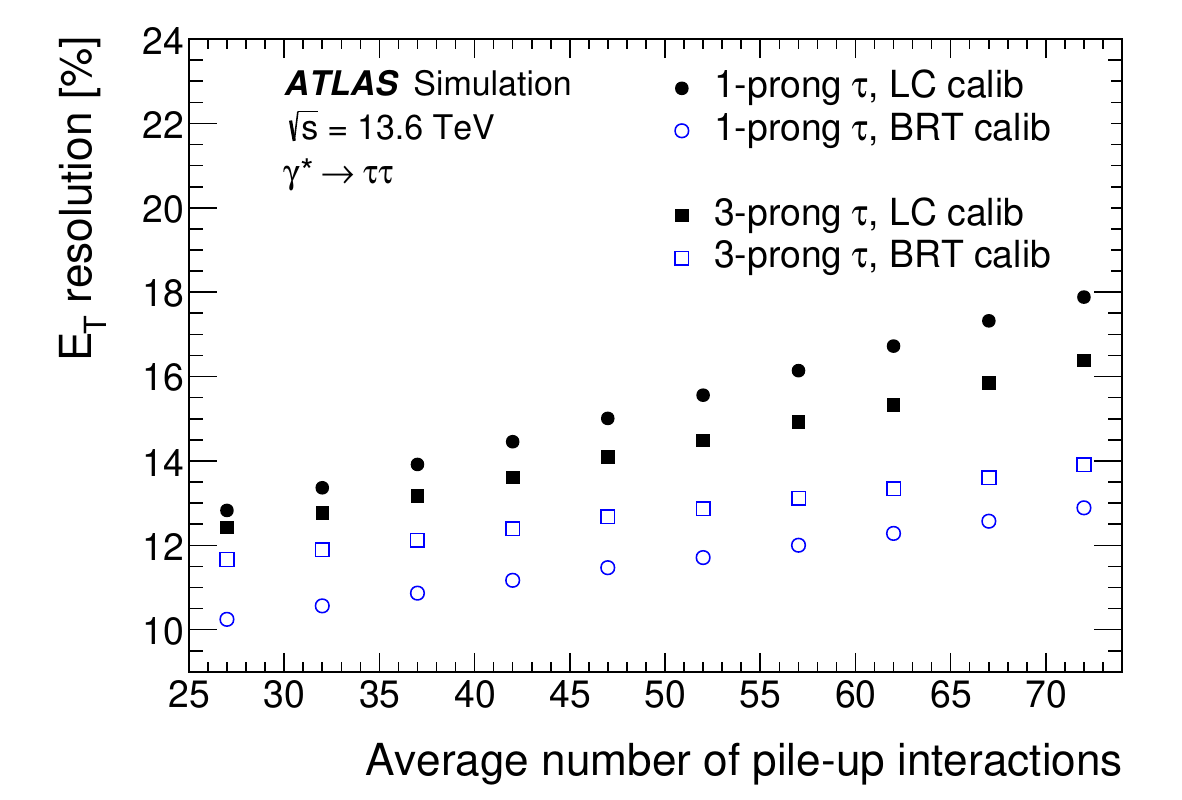}}
\subfloat[]{\label{fig:HLT:Calo:TES_resolution:pt}\includegraphics[width=0.485\linewidth]{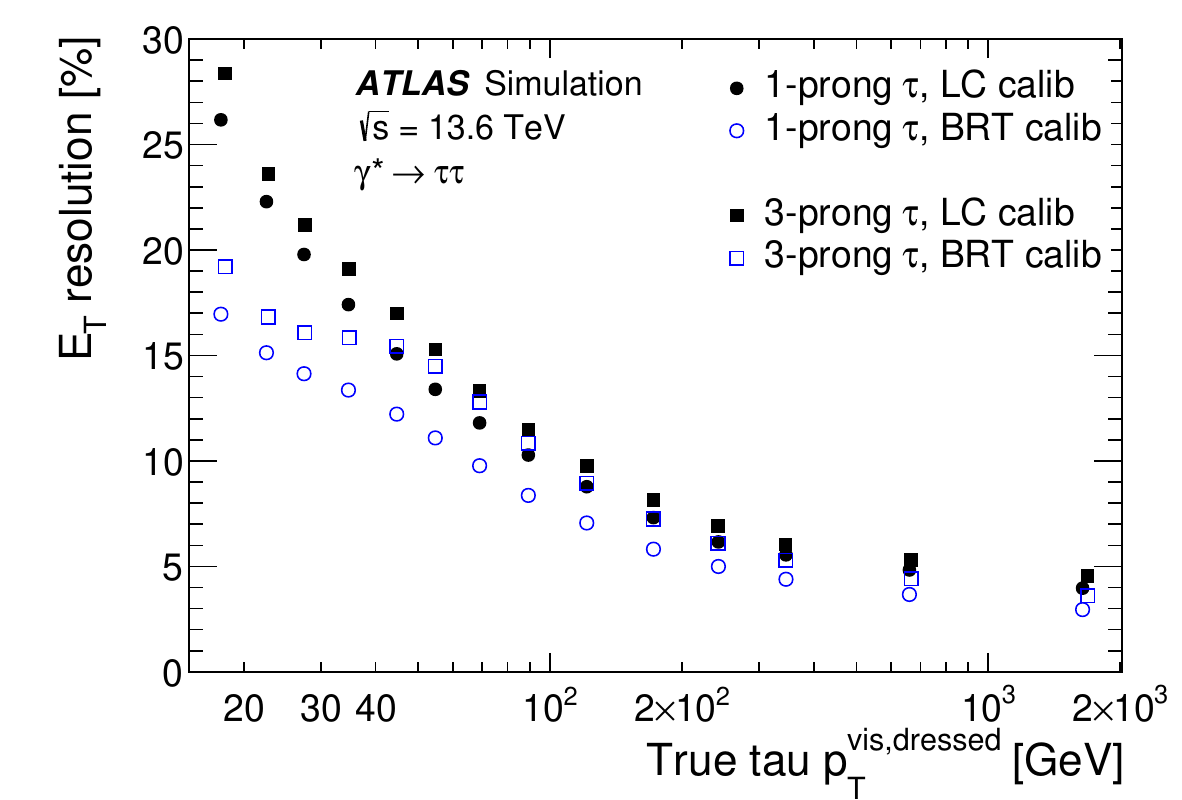}}
\caption{Transverse energy resolution as function of \protect\subref{fig:HLT:Calo:TES_resolution:mu} the average number of \pileup interactions and \protect\subref{fig:HLT:Calo:TES_resolution:pt} the true \tauhadvis dressed transverse momentum. The plots show the comparison between the local hadronic calibration (before tau energy calibration) and the tau energy calibration from Boosted Regression Trees. The comparison is reported separately for one-prong and three-prong \tauhadvis candidates classified according to the number of true charged pions. Given that the \tauhadvis candidates used at this stage have satisfied L1 tau selection criteria, the average \pt of three-prong candidates is higher than those of one-prong candidates. Statistical uncertainties are smaller than the marker size.}
\label{fig:HLT:Calo:TES_resolution}
\end{figure}

\subsection{Track-based selection}\label{sec:HLT:Tracks}

Following the calorimeter reconstruction, track information is added to the \tauhadvis reconstruction using a two-stage tracking approach. In the first stage, tracks are reconstructed using the fast tracking in a narrow wedge RoI with half-width $\Delta z = \SI{180}{mm}$ (reduced from \SI{225}{mm} at the beginning of 2023) that spans the entire region in $z$, and $\Delta\phi = 0.1$ (centred around the calorimetric \tauhadvis candidate direction). The extrema of the wedge at $\pm\SI{180}{mm}$ expand in the $\pm \eta$ direction with half widths $\Delta\eta = 0.1$ around the $\eta$ direction of the \tauhadvis candidate. These tracks include information only from the Pixel and SCT detectors. The highest-\pt track fulfilling track quality criteria to reduce the possibility of tracks from random combinations of hits is selected and serves as the starting point for the second stage. Tracks are then reconstructed in a larger $\eta \times \phi$ window but in a narrow $z$ window around the leading track, with $\Delta z = \SI{7}{mm}$ and $\Delta\eta = \Delta\phi = 0.4$ (expanding in the $\pm\eta$ direction at the ends of the wedge along the $z$-axis), as shown in \cref{fig:HLT:Tracks:RoIs}. In the rare (<2\%) cases where no tracks are found in the first stage, the $z$ width of the window is maintained at $\Delta z = \SI{180}{mm}$ during the second stage while increasing the $\eta \times \phi$ size. The resulting tracks are refitted including information from the TRT detector using precision tracking algorithms to reduce fit ambiguities, improve the track measurement, and discard fake tracks~\cite{TRIG-2022-01}. All precision tracks with $\pT > \SI{1}{\GeV}$ are retained and classified as \emph{core} ($\Delta R < 0.2$) or \emph{isolation} ($0.2 < \Delta R < 0.4$) tracks depending on the angular distance between the track and \tauhadvis direction. Based on the number of core tracks, the \tauhadvis candidates are classified as zero-prong, one-prong, or multi-prong (two or more tracks).%

\begin{figure}[ht!]
\centering
\includegraphics[width=0.35\linewidth]{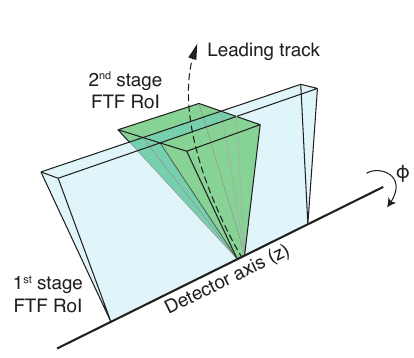}
\caption{A pictorial representation of the first and second stage RoIs used for the fast track finder (FTF) algorithms. The first stage RoI is centred at the detector origin in the z-axis. The second stage RoI is centred around the highest \pt track found at the first FTF stage.}
\label{fig:HLT:Tracks:RoIs}
\end{figure}

A requirement is applied on the number of core and isolation tracks to reduce background from quark and gluon jets. The \tauhadvis candidates with $\pT < \SI{200}{\GeV}$ must have at most three core tracks and at most one isolation track, while those with $\SI{200}{\GeV} < \pT < \SI{440}{\GeV}$ are required to have at most three core tracks without any requirements on the number of isolation tracks. The upper bound of \SI{440}{\GeV} was lowered to \SI{430}{\GeV} in 2023 in order to maximise trigger acceptance without adding significant rate. For \tauhadvis candidates with \pt above these thresholds, no track requirement is applied, and all such candidates are thus kept. To reduce inefficiency at high \pileup, only tracks above a \pt threshold are considered in this requirement. The track \pt threshold was raised from \SI{1}{\GeV} in 2022 to \SI{1.5}{\GeV} in 2023, which was particularly beneficial for three-prong \tauhadvis candidates. The efficiency of the track-based selection at HLT, shown in \cref{fig:HLT:Tracks:sel_eff}, is evaluated using \gtautau simulated MC samples where HLT \tauhadvis candidates are required to match to offline \tauhadvis candidates passing the \emph{Medium} RNN identification criteria and having one or three tracks. As it can be seen in \cref{fig:HLT:Tracks:sel_eff:mu}, the \pileup dependence can be reduced by increasing the track \pt threshold.

\begin{figure}[ht!]
\centering
\subfloat[]{\label{fig:HLT:Tracks:sel_eff:mu}\includegraphics[width=0.485\linewidth]{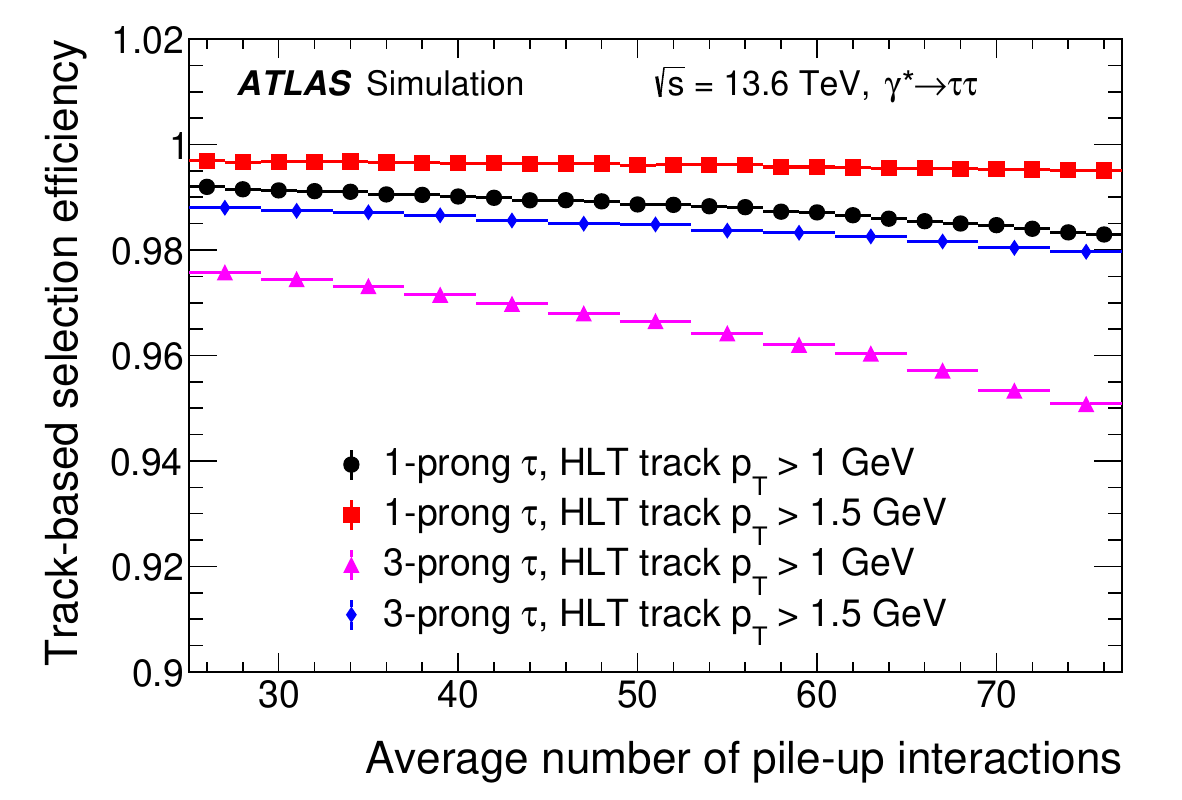}}
\subfloat[]{\label{fig:HLT:Tracks:sel_eff:pt}\includegraphics[width=0.485\linewidth]{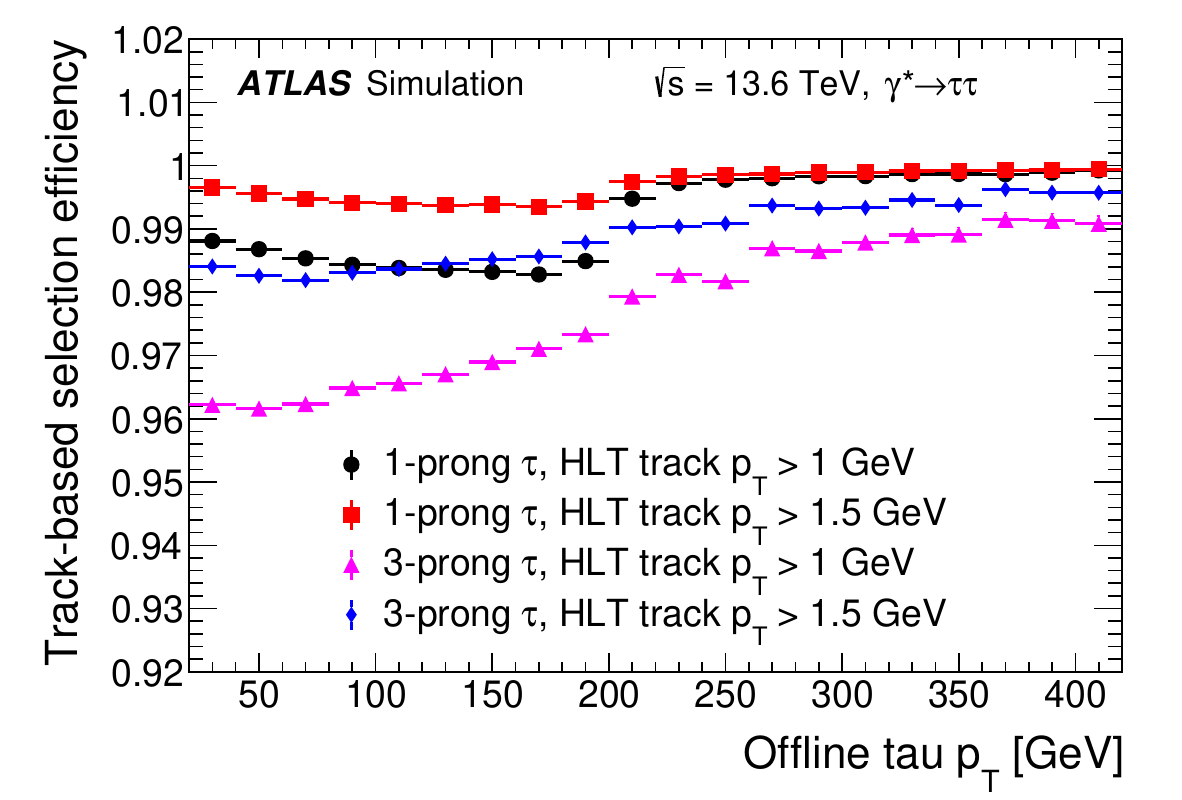}}
\caption{Efficiency of the requirement on the number of core and isolation tracks in the HLT as a function of \protect\subref{fig:HLT:Tracks:sel_eff:mu} the average number of \pileup interactions and \protect\subref{fig:HLT:Tracks:sel_eff:pt} the \pT of offline \tauhadvis. The \tauhadvis candidates are classified as one-prong or three-prong according to the number of offline reconstructed tracks. Only HLT tracks passing a \pt threshold of either \SI{1}{\GeV} or \SI{1.5}{\GeV} are considered. The efficiencies are evaluated in \gtautau simulated MC samples. Only statistical uncertainties are shown.}
\label{fig:HLT:Tracks:sel_eff}
\end{figure}

\subsection{Tau identification}%

Throughout \RunThr of the LHC, various machine learning techniques were employed for the identification of \tauhadvis candidates. An RNN was used in 2022, just as in \RunTwo, which was then replaced by a DeepSets network~\cite{zaheerdeepsets} in 2023, followed by a Graph Neural Network (GNN)~\cite{vaswani2023attentionneed, Zhou:2020GraphNeuralNet} in 2025. The DeepSets network and GNN were retrained with \textsc{Keras}~\cite{chollet:2015ker} and \textsc{PyTorch}~\cite{Paszke2019}, respectively, using samples simulating the Run~3 collisions. The training signal sample consists of \gtautau events, while background samples consist of multijet events. For the GNN, additional samples with $Z \to \tauhad\tauhad$ and $\ttbar \to b\bar{b}\tauhad\tauhad$ events were also used as signals to reduce bias across different kinematic phase spaces. The details of the samples can be found in \cref{tab:samples}. All dedicated algorithms are applied only to \tauhadvis candidates with $\pT < \SI{440}{\GeV}$, while candidates with higher \pt are accepted without any identification requirements.  The upper bound of \SI{440}{\GeV} was lowered to \SI{430}{\GeV} in 2023.

All three networks were trained independently for one-prong and multi-prong cases using a combination of track, cluster, and high-level variables as inputs to gain separation power between \tauhadvis and other hadronic jets. A linear or logarithmic transformation is applied to the input variables to have distributions with a mean of roughly zero and a standard deviation of roughly one, thus resulting with input variables of a similar scale. Independent zero-prong trainings were also performed, although such trigger selections focus on selecting events only for control region studies. Generally, high-level variables provide the strongest separation power between signal and background. \cref{fig:HLT:ID:vars} shows the distributions of signal and background processes for the following four high-level variables used in the training of the three networks.

\begin{description}
\item[Central energy fraction ($f_\mathrm{cent}$):] Fraction of transverse energy at EM scale deposited in the region $\Delta R < 0.1$ with respect to all energy deposited in $\Delta R < 0.2$ around the \tauhadvis candidate axis, calculated by summing the energy deposited in all cells with a barycentre in these regions. The calorimeter cells are required to be a part of a topocluster belonging to the \tauhadvis candidate.
\item[Ratio of EM energy and track momentum ($f_\mathrm{track}^\mathrm{EM}$):] Ratio of the sum of cluster energy deposited in the electromagnetic part of the topoclusters (pre-sampler, first and second layers of the LAr calorimeter) to the sum of momentum of core tracks. Clusters are calibrated at the LC energy scale.
\item[Maximum track $\Delta R$ ($\Delta R_\mathrm{max}$):] The maximum $\Delta R$ between a core track associated with the \tauhadvis candidate and the \tauhadvis direction.
\item[Fraction of track-plus-EM-system \pT ($\pT^\mathrm{EM+track} / \pT$):] Ratio of the \tauhadvis \pT, estimated using the vector sum of core track momenta and up to two most energetic EM clusters in the core region to the calibrated calorimeter-only measurement of \tauhadvis \pT.
\end{description}

\begin{figure}[hbt!]
\centering
\subfloat[]{\label{fig:HLT:ID:vars:centFrac}\includegraphics[width=0.48\linewidth]{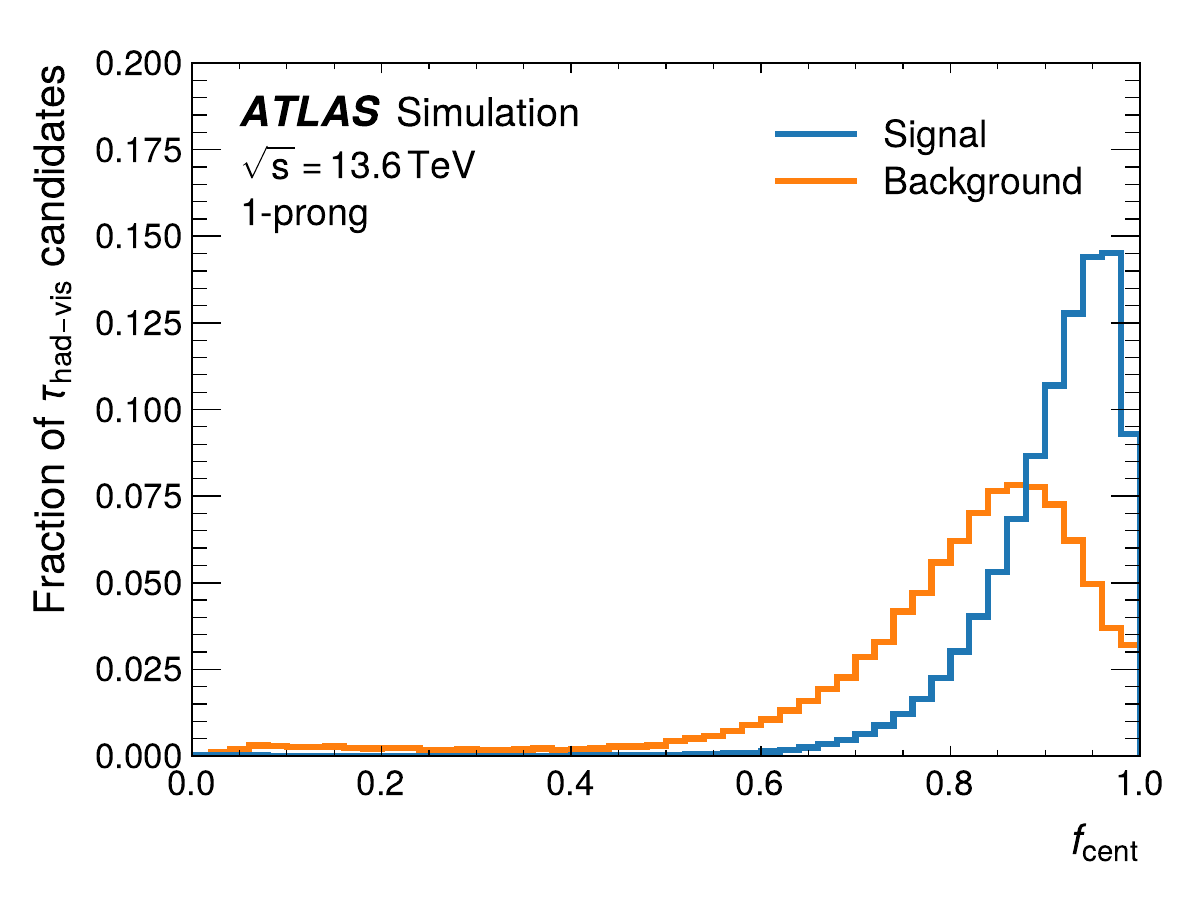}}
\subfloat[]{\label{fig:HLT:ID:vars:EMPOverTrkSysP}\includegraphics[width=0.48\linewidth]{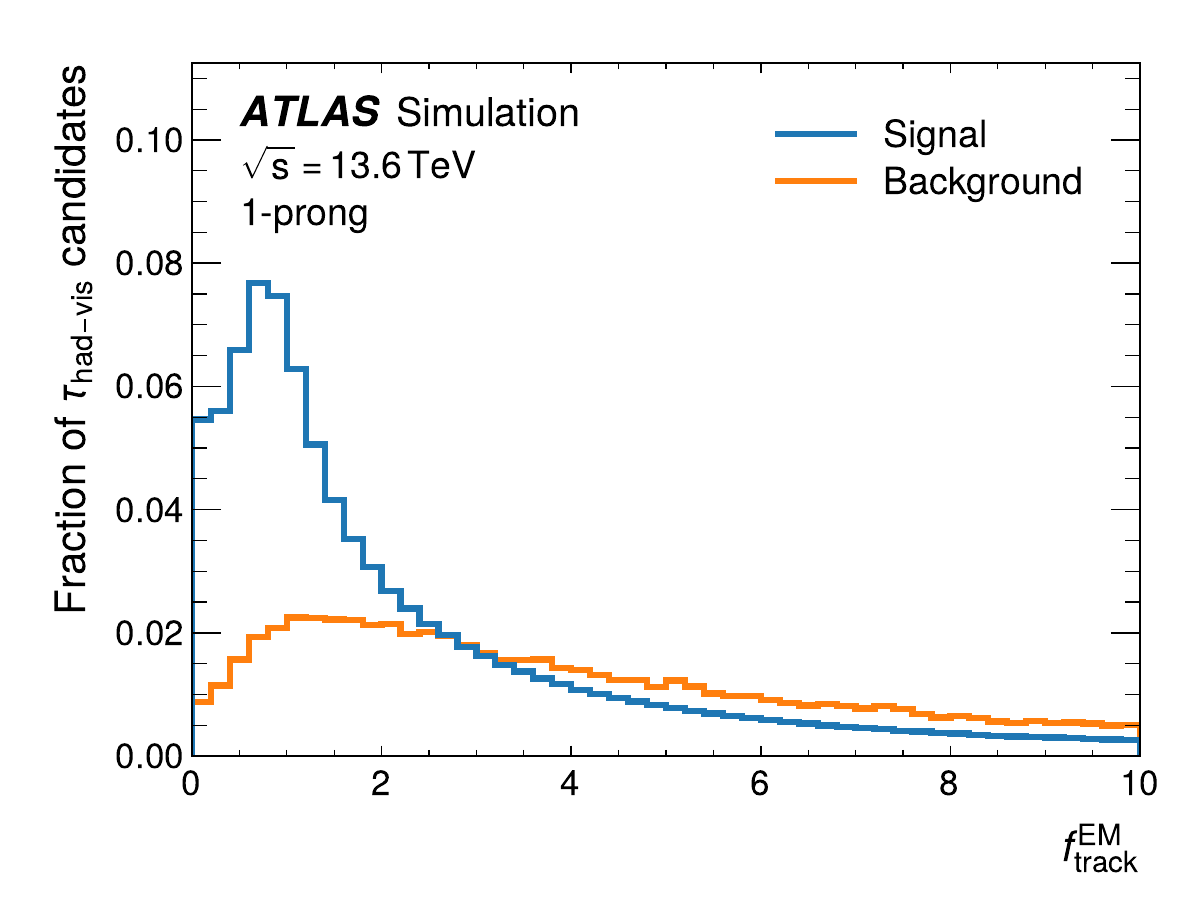}}

\subfloat[]{\label{fig:HLT:ID:vars:dRmax}\includegraphics[width=0.48\linewidth]{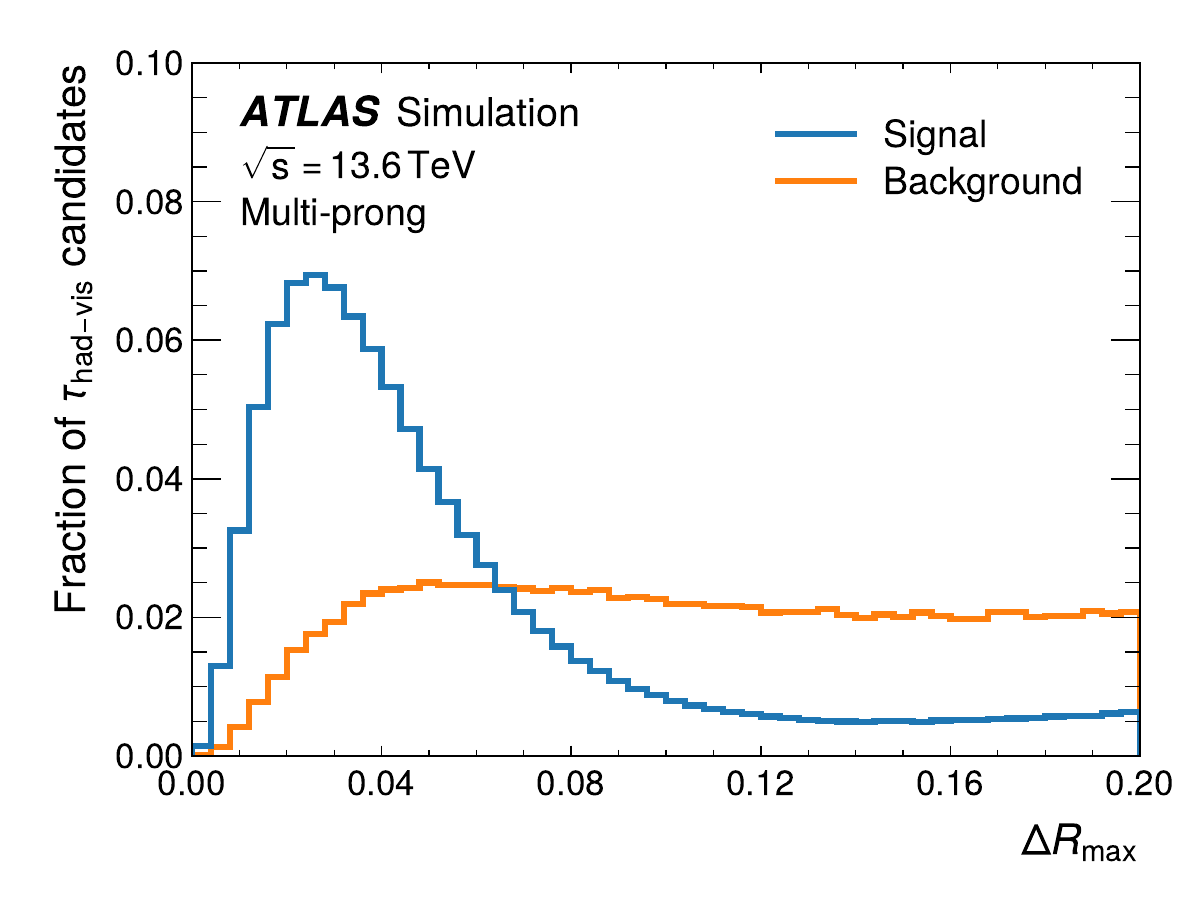}}
\subfloat[]{\label{fig:HLT:ID:vars:ptRatioEflowApprox}\includegraphics[width=0.48\linewidth]{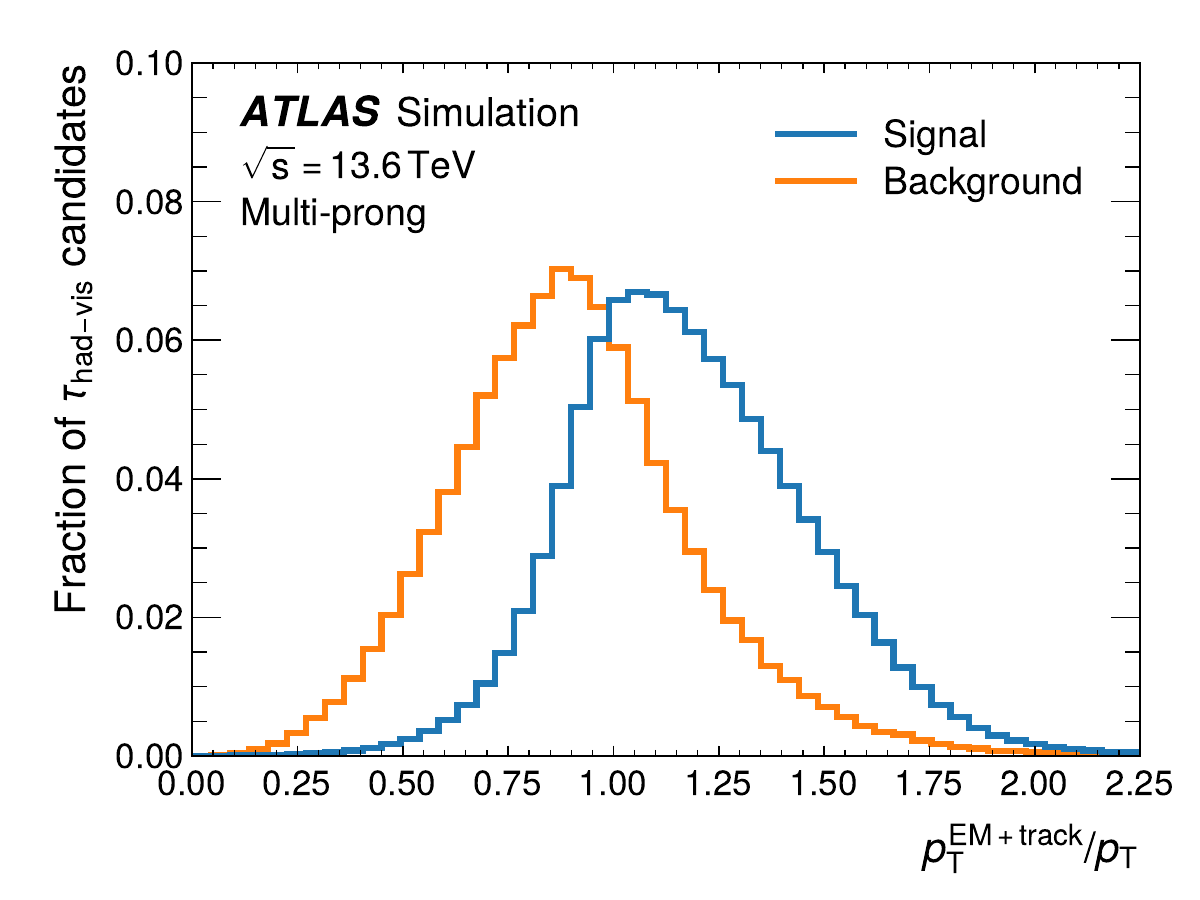}}
\caption{Distributions of high-level input variables for \gtautau signal and multijet background samples used in the training of various ML-based \tauhadvis identification algorithms, \protect\subref{fig:HLT:ID:vars:centFrac} $f_\mathrm{cent}$, \protect\subref{fig:HLT:ID:vars:EMPOverTrkSysP} $f_\mathrm{track}^\mathrm{EM}$, \protect\subref{fig:HLT:ID:vars:dRmax} $\Delta R_\mathrm{max}$, \protect\subref{fig:HLT:ID:vars:ptRatioEflowApprox} $\pT^\mathrm{EM+track} / \pT$. The $\tauhadvis$ candidates are classified as one-prong or multi-prong according to the number of offline reconstructed tracks.}
\label{fig:HLT:ID:vars}
\end{figure}

A complete list of input variables, along with the architecture and training procedure used by the RNN, is provided in Ref.~\cite{ATL-PHYS-PUB-2019-033}. The hyperparameters concerning the architecture of the DeepSets and GNN are optimised using a grid search. The improvements made in the architecture, and in the set of input variables for DeepSets and the GNN are described in the following sections.

\subsubsection{DeepSets}

DeepSets, used for the identification of \tauhad candidates at the HLT in 2023, is a type of Neural Network architecture characterised by the ability to process inputs in the form of sets with variable sizes. The key advantage of DeepSets over an RNN is its permutation invariance, i.e. the output of the DeepSets network does not depend on the ordering of the elements in the input sets. This is achieved by sharing the off-diagonal terms in the weight matrix and equating all its diagonal elements. This permutation invariance makes it especially suitable for \tauhadvis identification, where the sequence of tracks and clusters should be irrelevant to the outcome. The sharing of terms in the weight matrix significantly reduces the number of free parameters to approximately 33000, which represents a reduction of about 40\% with respect to the RNN. As a consequence, the DeepSets achieves 3-4 times faster training times compared to the RNN, depending on the \tauhadvis prongness.

In the context of the ATLAS tau trigger, the main distinction between the RNN and DeepSets architectures lies in the four long-short-term-memory layers of the tracks and clusters branches, which were replaced by DeepSets layers. These new layers consist of 32 nodes each in the tracks branch and 24 for the clusters branch. The dense and shared dense layers do not change in going from the RNN to DeepSets. Two new track variables are also introduced as inputs to the DeepSets network: the significance of the transverse ($d_0^\mathrm{sig.}$) and longitudinal ($z_0^\mathrm{sig.} \sin\theta$) impact parameters. These are calculated by dividing the signed impact parameters by their respective uncertainties.

\subsubsection{Graph Neural Network}\label{sec:HLT:ID:GNTau}

While the DeepSets based architectures offer clear advantages such as permutation invariance and reduced model complexity, they are inherently limited in capturing the relational structure between the input constituents. To overcome this issue, \texttt{GNTau} was introduced in 2025. This is a transformer-based GNN, designed specifically to exploit the complex correlations between tracks and clusters.

The architecture of \texttt{GNTau} is inspired from the GN2 algorithm~\cite{FTAG-2023-05}, used in ATLAS for flavour tagging. The model begins by concatenating the jet-level feature vector with the feature vectors of individual tracks and clusters. These concatenated features are processed through a series of dense layers to generate an initial latent representation for each constituent. A transformer layer is then used to capture contextual dependencies among the constituents, enabling the model to produce context aware embeddings by updating each representation based on its interactions with all others. In the final stage, these embeddings serve as nodes in a fully connected graph, upon which a graph-level classification is performed to yield the output probabilities.

\subsubsection{Performance}%

The achieved performances of the three Neural Network architectures used for the identification of \tauhadvis are compared in \cref{fig:HLT:ID:perf:ROC}, which shows the background rejection as a function of signal efficiency of the three algorithms. The performance comparison is evaluated on a test subset consisting of \gtautau events as signal and multijet events as background, with reconstructed \tauhadvis candidates in the transverse momentum range from \SI{20}{GeV} to \SI{440}{GeV}. Compared to the RNN used in 2022, DeepSets and \texttt{GNTau} deliver a significantly higher rejection against mis-identified \tauhadvis candidates. The improvement in background rejection allows for a loosening of the \emph{Medium} identification working point from \SI{96.5}{\%} (\SI{86.5}{\%}) with the RNN, to \SI{97}{\%} (\SI{89.5}{\%}) with DeepSets, and then to 97\% (92\%) with \texttt{GNTau} for one-prong (multi-prong) \tauhad candidates. However, this improvement in rejection is not only due to the change in network architecture, but also to the fact that the RNN used in 2022 was trained during the previous Run of the LHC for operations in 2018, whereas \texttt{GNTau} and DeepSets are trained with the updated samples simulating \RunThr conditions. In addition, \texttt{GNTau} benefits from a more powerful transformer-based architecture, and both \texttt{GNTau} and DeepSets benefit from additional variables used in their respective training.

\begin{figure}[ht]
\centering
\subfloat[]{\label{fig:HLT:ID:perf:ROC:1p}\includegraphics[width=0.45\linewidth]{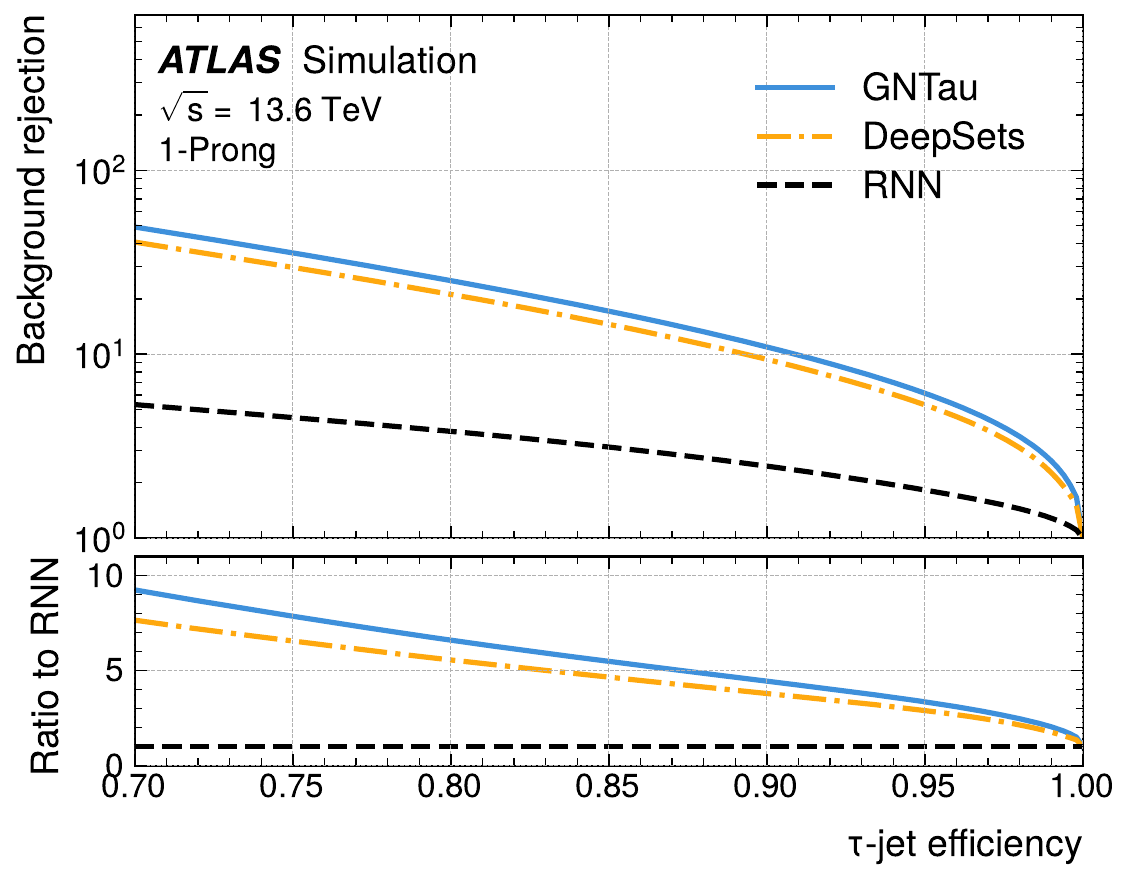}}
\subfloat[]{\label{fig:HLT:ID:perf:ROC:mp}\includegraphics[width=0.45\linewidth]{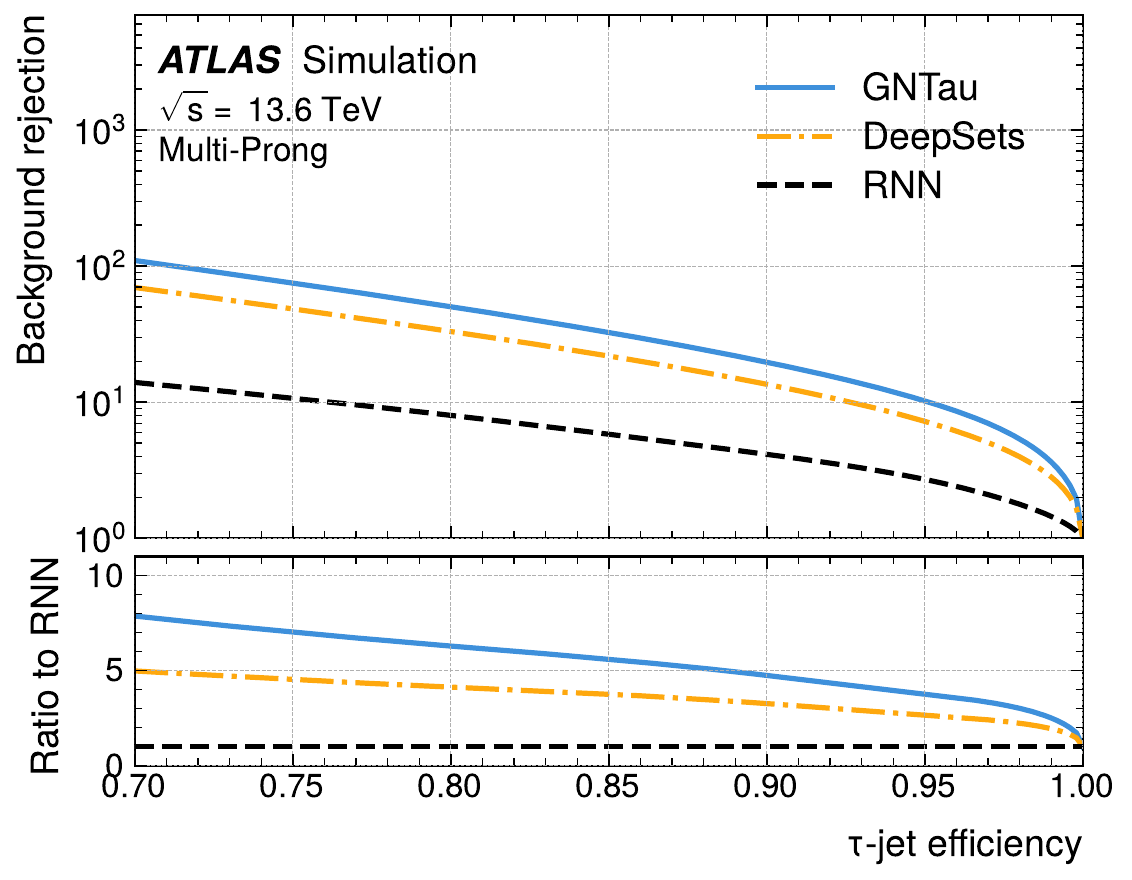}}
\caption{Background rejection as a function of signal efficiency for three discriminants -- \texttt{GNTau} (solid), DeepSets (dash-dotted), and RNN (dashed) -- evaluated separately for \protect\subref{fig:HLT:ID:perf:ROC:1p} one-prong and \protect\subref{fig:HLT:ID:perf:ROC:mp} multi-prong \tauhadvis candidates. Signal efficiency is defined as the fraction of true \tauhad candidates passing selection for a given working point; background rejection is the inverse of the fraction of background jets passing the same selection. The \tauhadvis candidates are classified as one-prong or multi-prong according to the number of offline reconstructed tracks. The lower panels show the rejection ratio relative to the RNN baseline, showing the relative improvement achieved by \texttt{GNTau} and DeepSets (and the expanded set of input information).}
\label{fig:HLT:ID:perf:ROC}
\end{figure}


\section{Trigger rates and menu}\label{sec:menu}

A diverse list of unprescaled tau triggers is used online to select a large spectrum of final states that involve \tauhad candidates. The list includes categories of triggers requiring one or two \tauhad objects and requiring one \tauhad object together with other objects such as $e/\mu$, \MET or a jet, as shown in~\cref{tab:menu}. The combined triggers help in keeping the event rate of the trigger within the allowed rate allocations, which are set based on the physics goals of the ATLAS experiment.

\begin{figure}[ht!]
\centering
\includegraphics[width=0.55\textwidth]{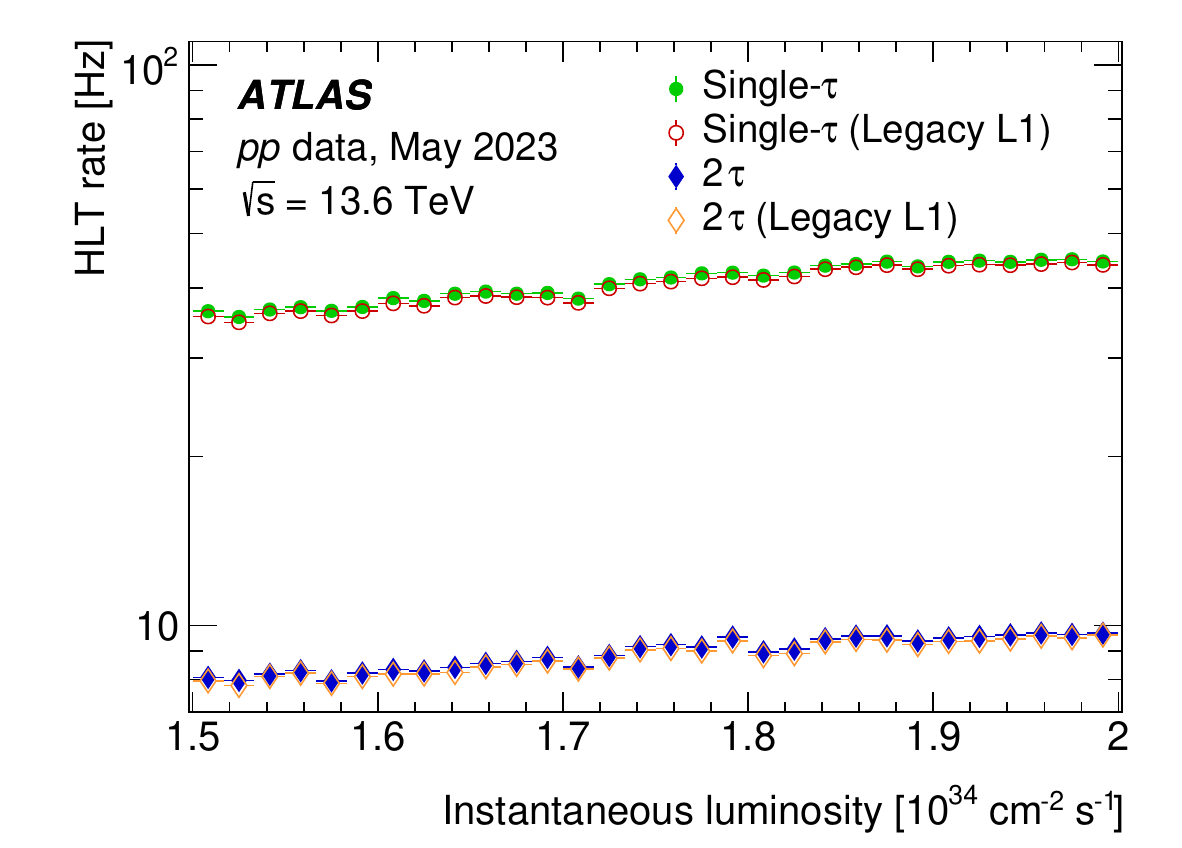}
\caption{HLT rates as a function of instantaneous luminosity for the unprescaled single-$\tau$ and di-$\tau$ triggers using non-isolated L1 \tauhadvis candidates from the legacy and Phase-I algorithms in \(pp\) collisions recorded at $\sqrt{s} = 13.6$ \tev in May 2023. Statistical uncertainties are smaller than the marker size. Details of these trigger chains can be found in~\cref{tab:menu}.}
\label{fig:menu:2023_hlt_leg_vs_ph1}
\end{figure}

In each category, the definition of the trigger is based on the \pT thresholds and the identification requirements. The triggers that require more than one object use topological cuts to select events in a specific phase-space. Some di-tau triggers require an additional number of jets in the L1 trigger to further reduce the L1 rate, targeting specific physics signatures such as $H\to\tau^+\tau^-$ or $HH \to b\bar{b} \tau^+\tau^-$. All unprescaled triggers containing either two \tauhad or one \tauhad combined with an electron, muon, or jet are required to have a minimum angular separation of $\Delta R(\tauhad, \tauhad/e/\mu) > 0.3$ between the HLT object candidates.

The transition from the legacy to Phase-I triggers during \RunThr was performed gradually to ensure smooth commissioning of the new L1 trigger system while maintaining stable data taking. The first Phase-I non-isolated \tauhad triggers were validated in 2023 while being run in parallel with the legacy triggers. A similar trigger rate is observed in the HLT with both variants of the L1 algorithm, as shown in \cref{fig:menu:2023_hlt_leg_vs_ph1}. The Phase-I tau triggers were progressively enabled as the commissioning advanced, until the legacy triggers were fully decommissioned in 2024. The \pT thresholds of the unprescaled tau triggers at HLT were kept constant during the \RunThr data taking period. Conversely, the \ET thresholds of the L1 Phase-I triggers were modified accordingly with each L1 algorithm configuration upgrade.  This was to maintain a minimum efficiency of 50\% at L1 corresponding to the \pT threshold of the \tauhad triggers at the HLT. In their final configurations, the combined L1 and HLT efficiency for \tauhad triggers using Phase-I L1 algorithms matched or surpassed the efficiency of their legacy counterparts while maintaining the total L1 rate within the allocated budget.

To increase the acceptance of events with two \btagged jets and two \tauhad candidates originating from the $HH \to b\bar{b} \tau^+\tau^-$ process, di-$\tau$ triggers with reduced \pt thresholds at HLT were introduced in 2023~\cite{HIGP-2024-37}. Events selected by these triggers are written to a delayed stream, which is reconstructed offline when resources allow.~\cref{tab:menu} shows the L1 and HLT requirements of the main unprescaled legacy and Phase-I \tauhad triggers as well as those in the delayed stream. The L1 and HLT rates are indicated for a luminosity of $\SI{1.95e34}{cm^{-2}s^{-1}}$.

\begin{table}[htbp]
\caption{Configuration and rates of the unprescaled triggers including a \tauhad lepton in 2023 and 2025. L1 tau candidates are required to be isolated, unless specified. All triggers including multiple $\tau$ leptons, electrons or muons are required to have an additional minimum angular separation of $\Delta R > 0.3$ between the HLT reconstructed particle candidates, ensuring no overlapping candidates between objects reconstructed by the HLT. Different types of objects at L1 can overlap, unless a minimum $\Delta R$ criterion prevents this, such as in the $\tau$ + $\mu$ trigger in 2025 or the two $\tau$ + one jet triggers. Delayed indicates that the selected events are sent to the delayed stream. All tau triggers used for data-taking used the legacy L1 tau algorithms and DeepSet tau identification in the HLT in 2023, and the Phase-I L1 algorithms and GNTau identification in the HLT in 2025. The total trigger rates for the different tau triggers are reported for an instantaneous luminosity of $\SI{1.95e34}{cm^{-2}s^{-1}}$.}
\label{tab:menu}
\centering
\resizebox{\textwidth}{!}{
\begin{tabular}{l|lll|cc|cc}
\toprule
\multirow[c]{3}{2.3cm}[-4pt]{\centering \textbf{Trigger}}                 & \multicolumn{3}{c|}{\textbf{Thresholds [GeV]}}                                                                                      & \multicolumn{4}{c}{\textbf{Rates at $\SI{1.95e34}{cm^{-2}s^{-1}}$}}                \\
\cmidrule{2-8}
& \multirow[c]{2}{4cm}{\centering \textbf{L1 Legacy (2023)}} & \multirow[c]{2}{4cm}{\centering \textbf{L1 Phase-I (2025)}} & \multirow[c]{2}{2.3cm}{\centering \textbf{HLT}}                                               & \multicolumn{2}{c|}{\textbf{Legacy (2023)}} & \multicolumn{2}{c}{\textbf{Phase-I (2025)}}    \\
& & &           & \textbf{L1 [kHz]}          & \textbf{HLT [Hz]}      & \textbf{L1 [kHz]}          & \textbf{HLT [Hz]}          \\
\midrule
Single-$\tau$           & $\ET^{\tau} > 100$ non-iso                & $\ET^{\tau} > 100$ non-iso                & $\pT^{\tau} > 160$                & 1.4               & 43            & 1.5               & 45                \\
\midrule
2 $\tau$                & $\ET^{\tau_{1(2)}} > 60 (40)$ non-iso     & $\ET^{\tau_{1(2)}} > 60 (40)$ non-iso     & $\pT^{\tau_{1(2)}} > 80 (60)$     & 3.3               & 9             & 3.0               & 9                 \\
\midrule
2 $\tau$ + 1 jet        & $\ET^{\tau_{1(2)}} > 20 (12)$             & $\ET^{\tau_{1(2)}} > 17.7 (10.1)$         & $\pT^{\tau_{1(2)}} > 35 (25)$     & 6.4               & 50            & 8.5               & 42                \\
& $\ET^{\text{jet}} > 25$                   & $\ET^{\text{jet}} > 41$                   & $\Delta R(\tau_1, \tau_2) < 3.0$  &                   &               &                   &                   \\
& $\Delta R(\tau_1, \tau_2) < 2.9$          & $\Delta R(\tau_1, \tau_2) < 3.0$          &                                   &                   &               &                   &                   \\
& $\Delta R(\tau_{1, 2}, \text{jet}) > 0.2$ & $\Delta R(\tau_{1, 2}, \text{jet}) > 0.2$ &                                   &                   &               &                   &                   \\
\midrule
2 $\tau$ + 1 jet        & $\ET^{\tau_{1(2)}} > 20 (12)$             & $\ET^{\tau_{1(2)}} > 17.7 (10.1)$         & $\pT^{\tau_{1(2)}} > 30 (20)$     & 6.4               & 90            & 8.5               & 73                \\
(Delayed)               & $\ET^{\text{jet}} > 25$                    & $\ET^{\text{jet}} > 41$                    & $\Delta R(\tau_1, \tau_2) < 3.0$  &                   &               &                   &                   \\
& $\Delta R(\tau_1, \tau_2) < 2.9$          & $\Delta R(\tau_1, \tau_2) < 3.0$          &                                   &                   &               &                   &                   \\
& $\Delta R(\tau_{1, 2}, \text{jet}) > 0.2$ & $\Delta R(\tau_{1, 2}, \text{jet}) > 0.2$ &                                   &                   &               &                   &                   \\
\midrule
2 $\tau$ + 3 jets       & $\ET^{\tau_{1(2)}} > 25 (20)$             & $\ET^{\tau_{1(2)}} > 23.2 (17.7)$         & $\pT^{\tau_{1(2)}} > 40 (35)$     & 2.8               & 13            & 4.9               & 17                \\
& $\ET^{\text{jet}_{1, 2}} > 25$             & $\ET^{\text{jet}_{1, 2}} > 41$             &                                   &                   &               &                   &                   \\
& $\ET^{\text{jet}_{3}} > 20$               & $\ET^{\text{jet}_{3}} > 31$               &                                   &                   &               &                   &                   \\
\midrule
2 $\tau$ + 4 jets       & $\ET^{\tau_{1(2)}} > 20 (12)$             & $\ET^{\tau_{1(2)}} > 17.7 (10.1)$         & $\pT^{\tau_{1(2)}} > 35 (25)$     & 2.7               & 24            & 3.3               & 28                \\
& $\ET^{\text{jet}_{1, 2, 3, 4}} > 12$, $|\eta^{\text{jet}}|<2.5$       & $\ET^{\text{jet}_{1, 2, 3, 4}} > 22$, $|\eta^{\text{jet}}|<2.5$       &                                   &                   &               &                   &                   \\
\midrule
2 $\tau$ + 4 jets       & $\ET^{\tau_{1(2)}} > 20 (12)$             & $\ET^{\tau_{1(2)}} > 17.7 (10.1)$         & $\pT^{\tau_{1(2)}} > 30 (20)$     & 2.7               & 41            & 3.3               & 46                \\
(Delayed)               & $\ET^{\text{jet}_{1, 2, 3, 4}} > 12$, $|\eta^{\text{jet}}|<2.5$       & $\ET^{\text{jet}_{1, 2, 3, 4}} > 22$, $|\eta^{\text{jet}}|<2.5$       &                                   &                   &               &                   &                   \\
\midrule
2 $\tau$ + \MET         & $\ET^{\tau_1} > 40$ non-iso               & $\ET^{\tau_1} > 40$ non-iso               & $\pT^{\tau_{1(2)}} > 60 (25)$     & 2.4               & 7             & 3.8               & 7                 \\
& $\ET^{\tau_2} > 12$                       & $\ET^{\tau_2} > 10.1$                     & $\MET > 50$                       &                   &               &                   &                   \\
& $\MET > 40$                               & $\MET > 34$                               &                                   &                   &               &                   &                   \\
\midrule
$\tau$ + $\mu$          & $\ET^{\tau} > 20$                         & $\ET^{\tau} > 17.7$                       & $\pT^{\tau} > 35$                 & 4.4               & 5             & 3.5               & 5                 \\
& $\pT^{\mu} > 8$                           & $\pT^{\mu} > 8$                           & $\pT^{\mu} > 14$                  &                   &               &                   &                   \\
&                                           & $0.3 < \Delta R(\tau, \mu) < 3.5$         &               & & & & \\
\midrule

$\tau$ + $\mu$ + 3 jets & $\ET^{\tau} > 12$                         & $\ET^{\tau} > 10.1$                       & $\pT^{\tau} > 25$                 & 1.3               & 3             & 0.6               & 2                 \\
& $\pT^{\mu} > 8$                           & $\pT^{\mu} > 8$                           & $\pT^{\mu} > 14$                  &                   &               &                   &                   \\
& $\ET^{\text{jet}_{1,2,3}} > 12$            & $\ET^{\text{jet}_{1,2,3}} > 26$      &                                   &                   &               &                   &                   \\
\midrule
2 $\tau$ + $e$ + 4 jets & $\ET^{\tau_{1,2}} > 12$                   & $\ET^{\tau_{1,2}} > 10.1$                 & $\pT^{\tau} > 25$                 & 1.6               & 2             & 0.9               & 2                 \\
& $\ET^{e} > 15$                            & $\ET^{e} > 16$                            & $\pT^{e} > 17$                    &                   &               &                   &                   \\
& $\ET^{\text{jet}_{1,2,3,4}} > 12$          & $\ET^{\text{jet}_{1,2,3,4}} > 26$       &                                   &                   &               &                   &                   \\
\midrule
$\tau$ + $b$-jet + 3 jets       & -- & $\ET^{\text{jet}_{1(2, 3)}} > 70 (25)$    & $\pT^{\tau} > 25$             & --               & --            & 8.1               & 94                \\
&    &                                          & $\pT^{\text{jet}_{1(2)}} > 65 (40)$, & & & & \\
&    &                                          & $\pT^{\text{jet}_{3(4)}} > 25(20)$   & & & & \\
&    &                                          & (with $b$+$\tau$ presel.                             & & & & \\
&    &                                          & and $b$ selection)                                        & & & & \\
\bottomrule
\end{tabular}

}
\end{table}

In 2024, an additional dedicated trigger for the $HH \to b\bar{b} \tau^+\tau^-$ search was included, requiring a single \tauhad candidate and three additional jets, one of which is expected to have originated in the remaining $\tau$ lepton in signal events, and another required to be \btagged.

To reduce the usage of HLT processing resources, an initial calorimeter-only jet reconstruction and four-jet requirement is applied, followed by fast $b$-tag~\cite{TRIG-2022-03} and hadronic tau-tag preselections using tracks reconstructed in the vicinity of these jets. Final event candidates must contain a \tauhad with a minimum momentum of $\SI{25}{\GeV}$, reconstructed and selected as described in the previous sections of this publication. This strategy increases the $HH \to b\bar{b} \tau^+\tau^-$ signal acceptance aproximately by 25\% compared to the original di-$\tau$ trigger.

In the Phase-I system, the energy values in the L1 items reflect the mid-point of the turn-on curve while in the legacy system the value represents the actual threshold applied. The reconstruction of \tauhadvis candidates in the HLT is similar in both legacy and Phase-I systems except the identification algorithm used, with 2023 triggers using the DeepSets identification, and 2025 triggers using the GNTau algorithm. Differences in rates of similar triggers in the two years are mainly due to the use of different \ET thresholds and identification for \tauhadvis candidates.

The rates as a function of the instantaneous luminosity for these triggers are presented in \cref{fig:menu:2023:tau:rates,fig:menu:2023:taupx:rates} for 2023 and in \cref{fig:menu:2025:tau:rates,fig:menu:2025:taupx:rates} for 2025. The corresponding trigger cross-sections, computed as the trigger rate divided by the instantaneous luminosity, as a function of $\left<\mu\right>$ are shown in \cref{fig:menu:2023:tau:xsection,fig:menu:2023:taupx:xsection} for 2023 and in \cref{fig:menu:2025:tau:xsection,fig:menu:2025:taupx:xsection} for 2025. The single and di-\tauhad L1 triggers without additional non-\tauhad L1 requirements have a stable trigger cross-section as a function of $\left<\mu\right>$, while L1 triggers combining particle candidates of different types usually exhibit a linear dependence, consistent with the combinatorics from the \pileup.

\begin{figure}[ht!]
\centering
\subfloat[]{\label{fig:menu:2023:tau:rates}\includegraphics[width=0.45\textwidth]{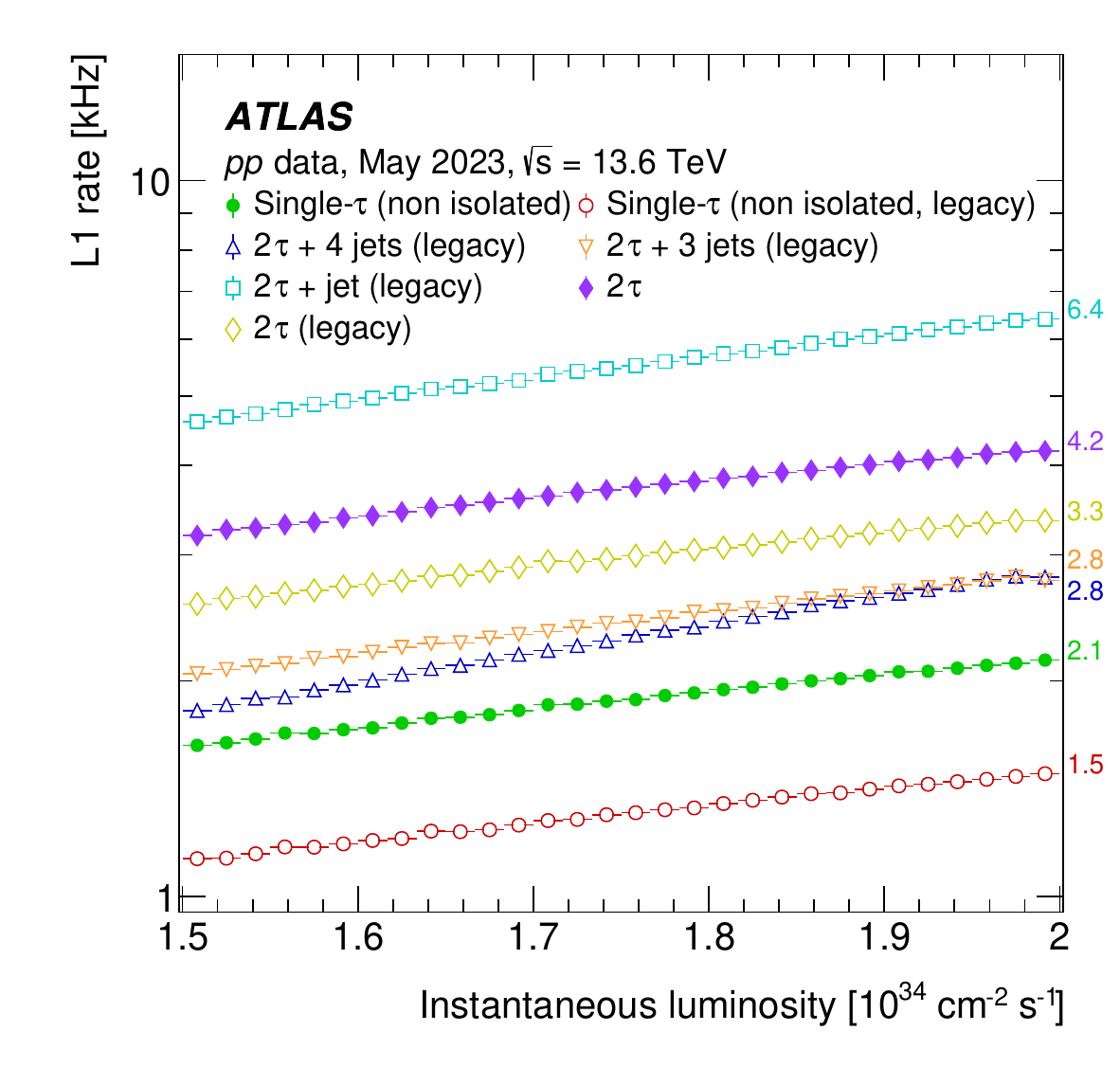}}
\subfloat[]{\label{fig:menu:2023:tau:xsection}\includegraphics[width=0.45\textwidth]{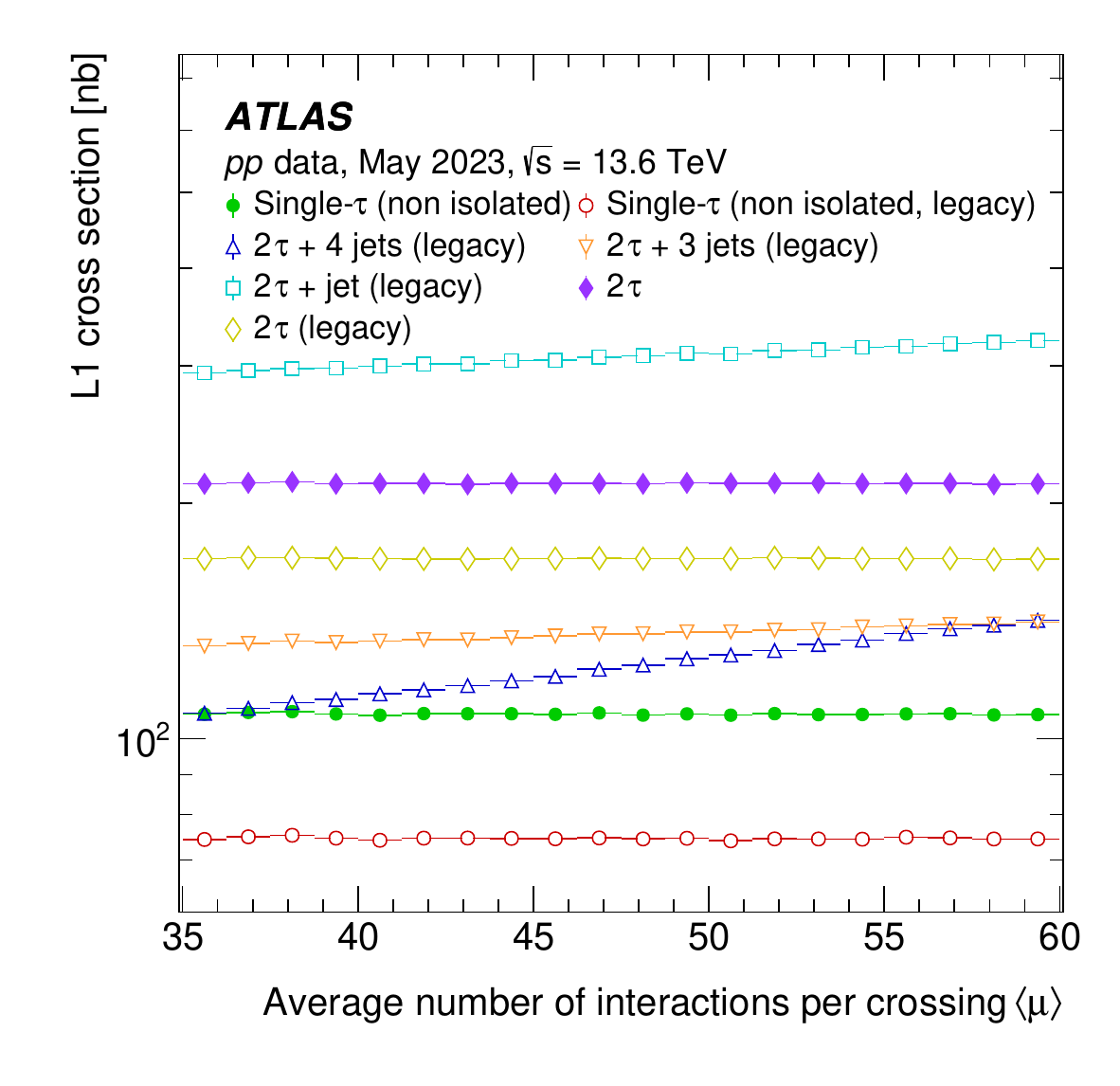}}

\subfloat[]{\label{fig:menu:2023:taupx:rates}\includegraphics[width=0.45\textwidth]{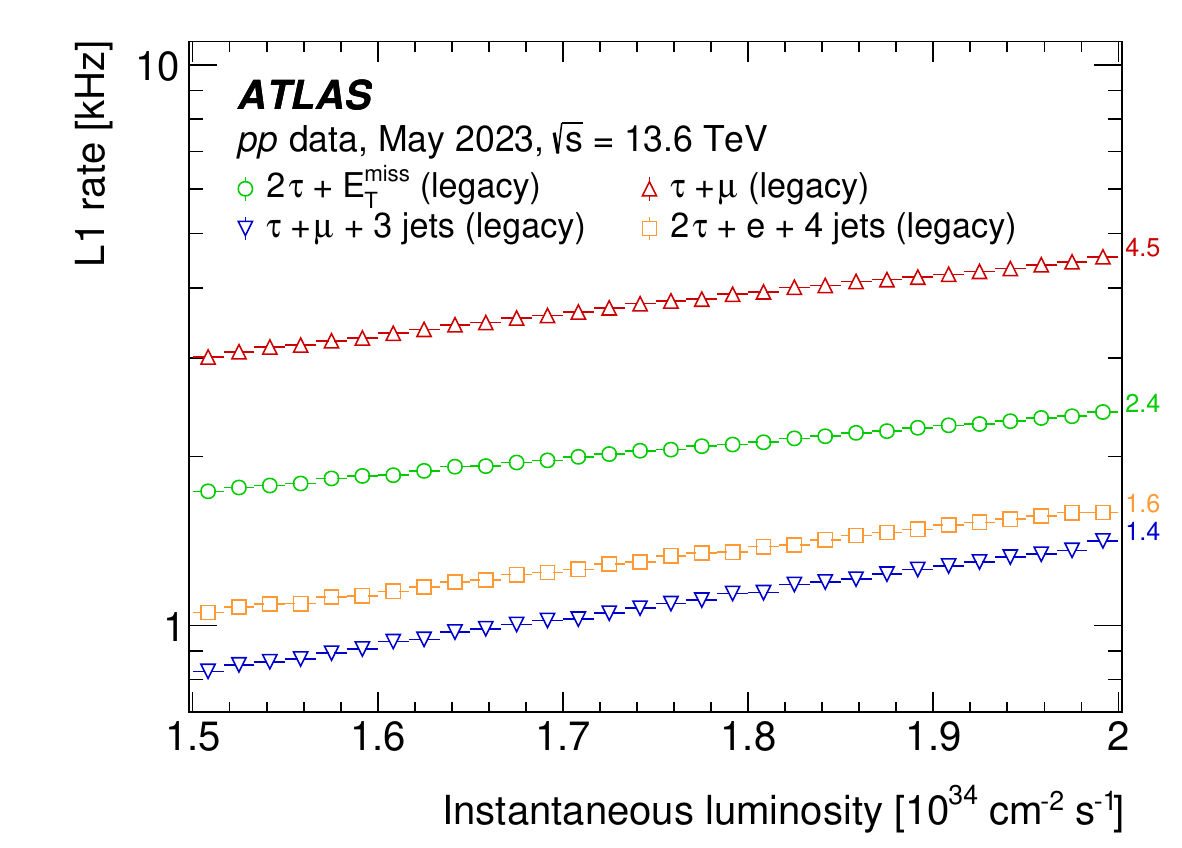}}
\subfloat[]{\label{fig:menu:2023:taupx:xsection}\includegraphics[width=0.45\textwidth]{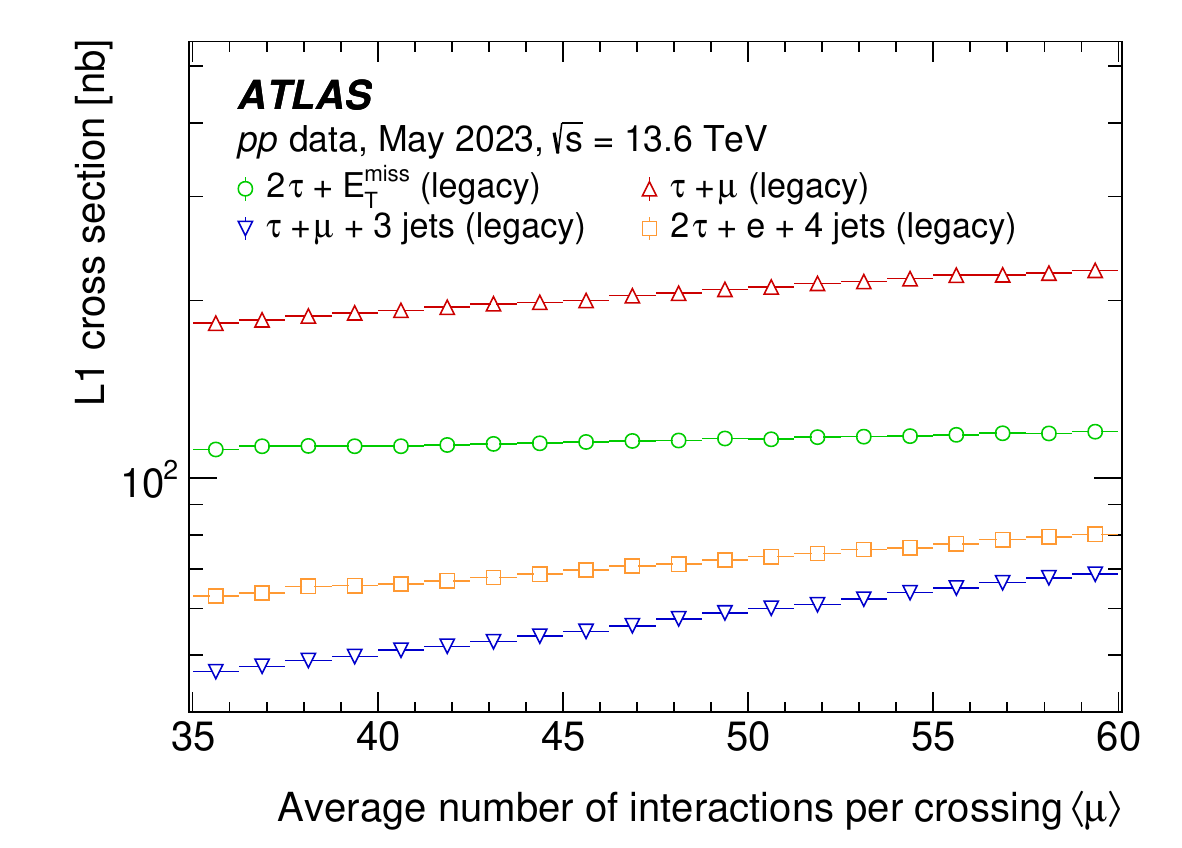}}
\caption{L1 rates and cross-sections for the unprescaled $\tau(+X)$ triggers in \(pp\) collisions recorded at $\sqrt{s} = 13.6$ \tev in 2023. Figures~\protect\subref{fig:menu:2023:tau:rates} and \protect\subref{fig:menu:2023:taupx:rates} show L1 rates as a function of instantaneous luminosity. The L1 rates at a reference luminosity of $\SI{2.0e34}{cm^{-2}s^{-1}}$ are indicated by the numbers to the right in kHz. Figures~\protect\subref{fig:menu:2023:tau:xsection} and \protect\subref{fig:menu:2023:taupx:xsection} show L1 trigger cross-sections as a function of the average number of interactions per bunch-crossing. Statistical uncertainties are smaller than the marker size. Details of these trigger chains can be found in~\cref{tab:menu}.}
\end{figure}

\begin{figure}[ht!]
\centering
\subfloat[]{\label{fig:menu:2025:tau:rates}\includegraphics[width=0.45\textwidth]{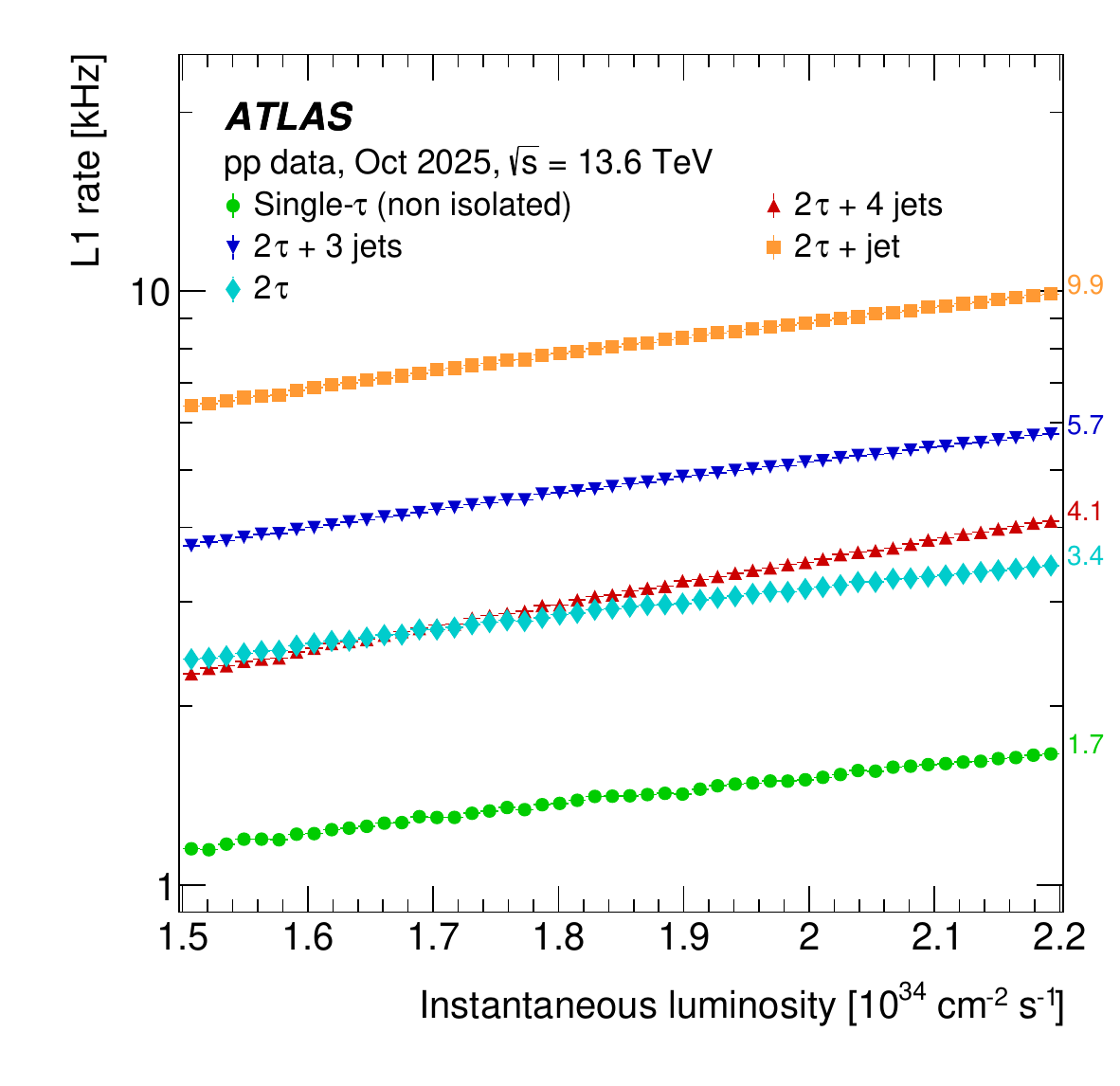}}
\subfloat[]{\label{fig:menu:2025:tau:xsection}\includegraphics[width=0.45\textwidth]{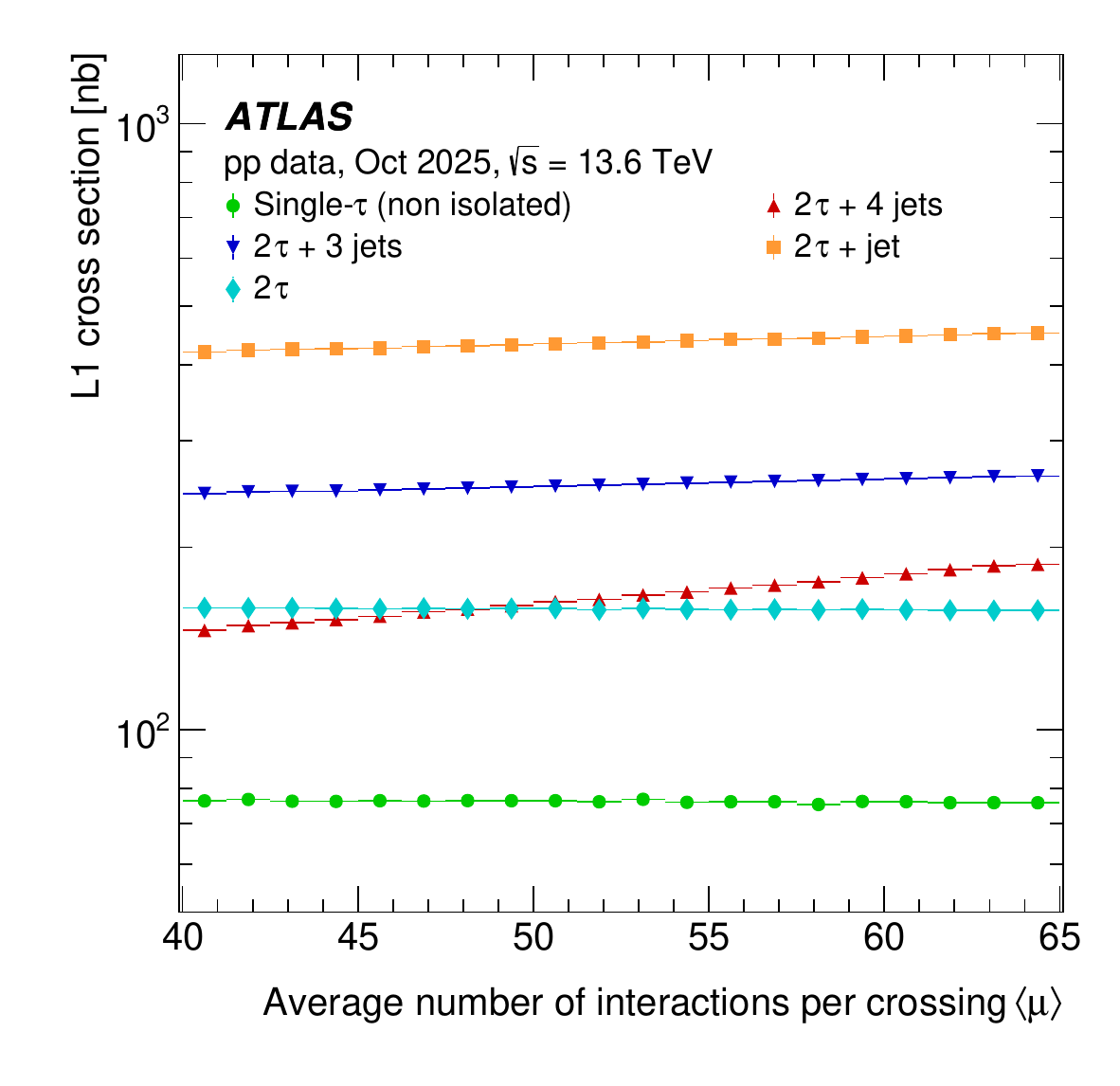}}

\subfloat[]{\label{fig:menu:2025:taupx:rates}\includegraphics[width=0.45\textwidth]{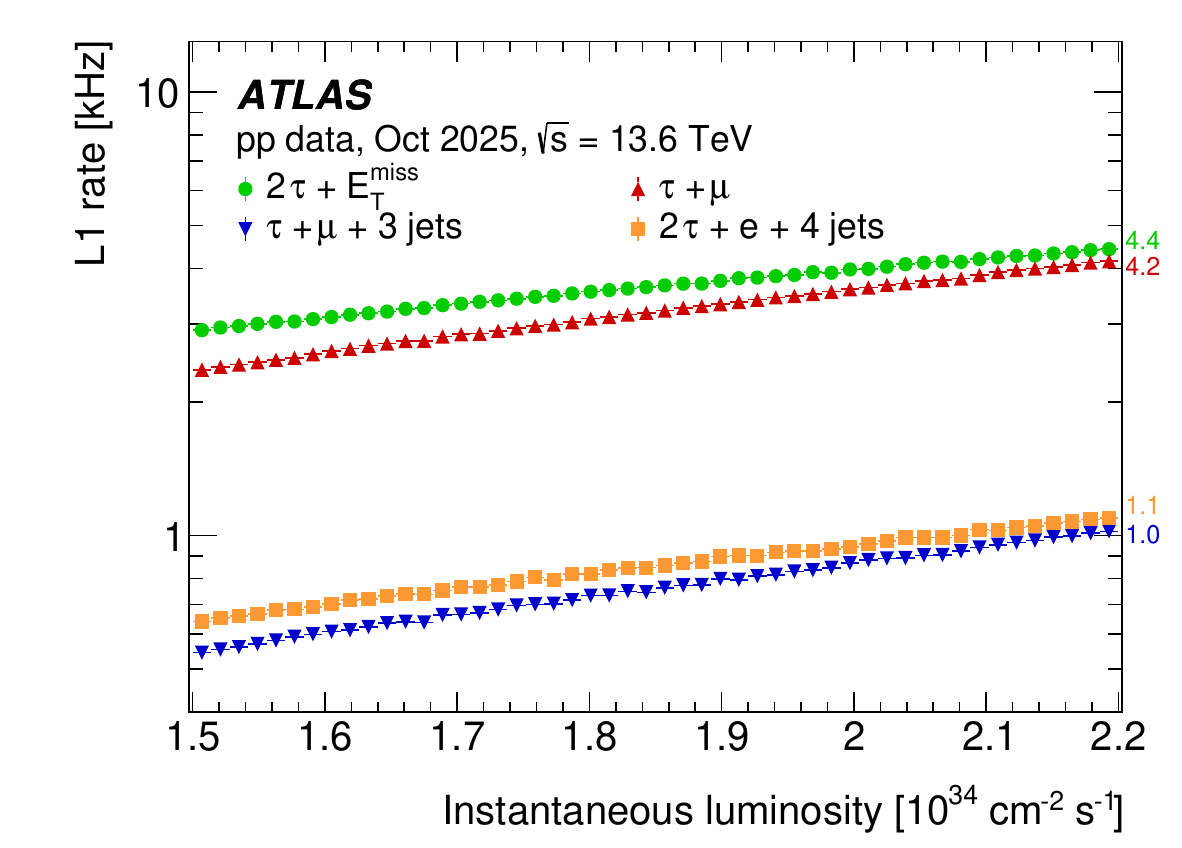}}
\subfloat[]{\label{fig:menu:2025:taupx:xsection}\includegraphics[width=0.45\textwidth]{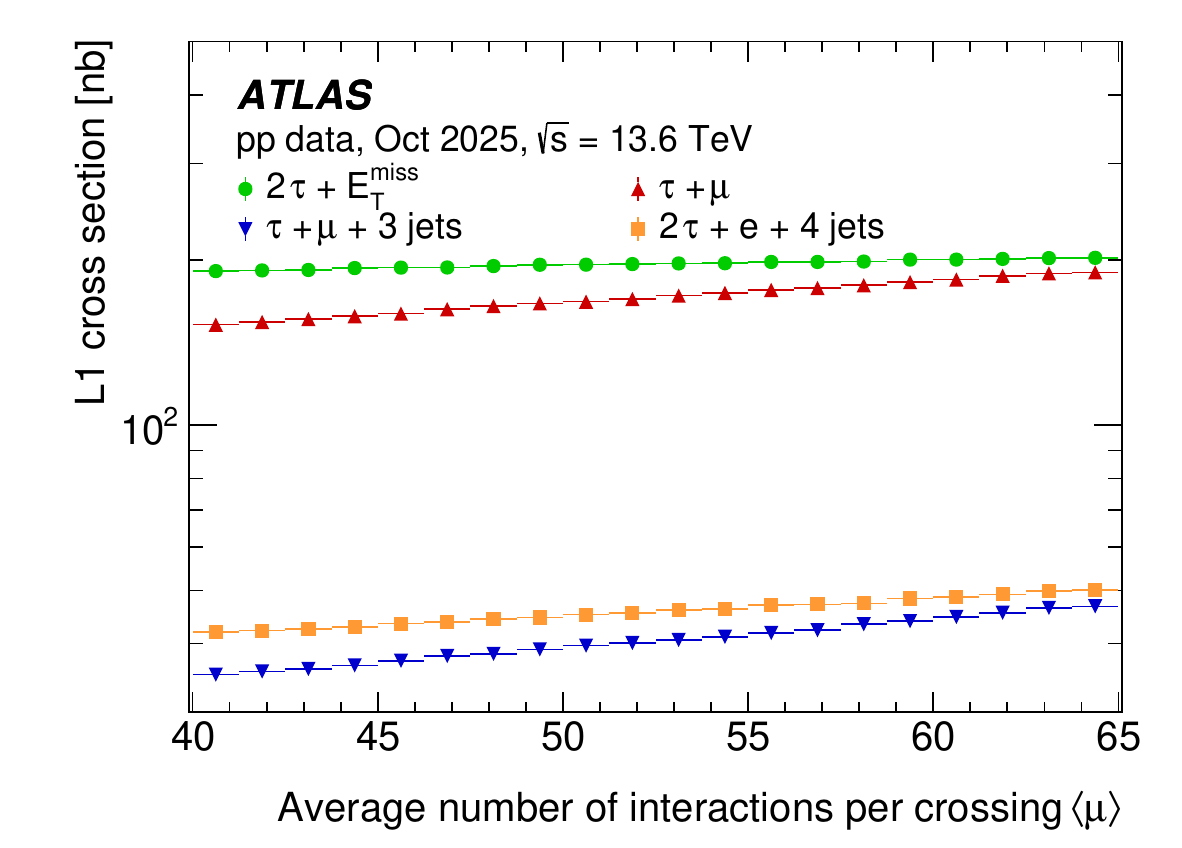}}
\caption{L1 rates and cross-sections for the unprescaled $\tau(+X)$ triggers in \(pp\) collisions recorded at $\sqrt{s} = 13.6$ \tev in 2025. Figures~\protect\subref{fig:menu:2025:tau:rates} and \protect\subref{fig:menu:2025:taupx:rates} show L1 rates as a function of instantaneous luminosity. The L1 rates at a reference luminosity of $\SI{2.2e34}{cm^{-2}s^{-1}}$ are indicated by the numbers to the right in kHz. Figures~\protect\subref{fig:menu:2025:tau:xsection} and \protect\subref{fig:menu:2025:taupx:xsection} show L1 trigger cross-sections as a function of the number of interactions per bunch-crossing. Statistical uncertainties are smaller than the marker size. Details of these trigger chains can be found in~\cref{tab:menu}.}
\end{figure}


\section{Performance}\label{sec:perf}

This section describes the procedure to evaluate tau trigger scale factors (SFs) that are required to do physics analyses. The SFs are defined as the ratio of the efficiency of the trigger observed in the background-subtracted data to that expected from signal MC simulations. The SFs are applied to MC simulations to correct for any differences seen between the trigger efficiency measured in data and MC simulations.

A well-defined approach, known as tag-and-probe, is used to estimate the tau trigger efficiency. It consists of selecting events triggered by the presence of a lepton (tag) and containing a \tauhad candidate (probe) in the final state fulfilling a set of identification requirements collectively defining a given working point. The tag lepton can either be an electron ($e$) or a muon ($\mu$). The method leads to an event selection with the least biased probe object, while maintaining good selection purity due to the tight selections applied on the tag object. The data samples, MC samples, and object selection described in \cref{sec:datamc} are used in the studies in this section.

\subsection{Tag-and-probe analysis}\label{sec:perf:tnp}

The tag-and-probe analysis targets $\Ztautau \rightarrow \tau_{\mathrm{lep}} + \tauhad + 3\nu$ and $\ttbar \rightarrow [b + e/\mu + \nu][b + \tauhad + 2\nu]$ events.
For these signal processes, the analysis tag consists of the presence of an identified $e/\mu$ in the final state, whereas the probe consists of a \tauhadvis candidate.
Performing the measurement in two orthogonal phase spaces guarantees complementarity and a wide coverage of the \pt range of \tauhadvis, with the low and high \pt regions populated by \Ztautau and \ttbar processes, respectively.

To measure the efficiency of \tauhad triggers, dedicated \emph{tag-and-probe} triggers are deployed online to collect data. These triggers are designed as a combination of the unprescaled single $e/\mu$ trigger and a dedicated \tauhad trigger. The \pt thresholds for the single electron and muon are 26 and \SI{24}{\GeV} at the HLT, respectively, while thresholds on the \pt of the \tauhadvis candidate are varied from 25 to \SI{160}{\GeV} at the HLT. The efficiency of the \tauhad trigger, for a given \pt threshold, is estimated from the ratio of events selected by the tag-and-probe trigger to the events selected by the single $e/\mu$ trigger with same lepton selection as in the tag-and-probe trigger.
In this section, the efficiency of the \tauhad trigger for a \pt threshold of \SI{25}{\GeV} at the HLT and \SI{12}{\GeV} at L1 is featured, where both HLT and L1 criteria must be fulfilled.

\begin{table}[ht]
\caption{Selection requirements applied to select offline reconstructed $e/\mu$ and the $\tauhadvis$ candidates.}
\label{tab:perf:tnp:obj_sel}
\centering
\resizebox{0.8\textwidth}{!}{
\begin{tabular}{c|c|c}
\toprule
$e$-tag                             & $\mu$-tag             & $\tauhadvis$-probe                   \\
\midrule
$\pT > \SI{27}{\GeV}$               & $\pT > \SI{25}{\GeV}$ & $\pt > \SI{20}{\GeV}$             \\
$|\eta| < 2.47$                      & $|\eta| < 2.5$        & $|\eta| < 2.5$                    \\
excluding $1.37 < |\eta| < 1.52$    &                       & excluding $1.37 < |\eta| < 1.52$  \\
&                       & $1$ or $3$ tracks                 \\
&                       & $|q| = 1$                         \\
\bottomrule
\end{tabular}
}
\end{table}

\subsubsection{Event selection}\label{sec:perf:tnp:event_sel}

To evaluate the performance of the tau trigger, requirements on the reconstructed objects, as described in \cref{sec:Objectreco}, are applied to select events with \tauhad in signal regions (SRs) targeting \Ztautau and \ttbar final states. The common object selections for the offline $e/\mu/\tauhadvis$, shown in \cref{tab:perf:tnp:obj_sel}, are applied to both final states. The results presented in this section use \emph{Medium} identification criteria for the offline $\tauhadvis$ candidates; however, the measurement is also carried out for \emph{Loose} and \emph{Tight} identification criteria. The \tauhadvis candidate considered by the HLT is required to be within $\Delta R < 0.2$ of the selected offline candidate. Events with exactly one selected \tauhadvis candidate and exactly one electron or muon candidate are used.

The main backgrounds for the efficiency measurement come from \Wjets, $Z\to ee/\mu\mu$, \ttbar and multijet processes where an electron, muon, or jet are misidentified as \tauhad candidates. Regions enriched in background events, also referred to as control regions (CR), are used to estimate the backgrounds in the SRs. In signal-like events, the electric charge of \tauhad is expected to be opposite to that of $e/\mu$. The CRs are further subdivided into same sign (SS) and opposite sign (OS) categories based on the charge of \tauhad and $e/\mu$ candidates to use charge symmetric and asymmetric backgrounds independently.

Tables~\ref{tab:perf:tnp:Ztautau_regions} and ~\ref{tab:perf:tnp:ttbar_regions} summarise the event selection criteria used for the \Ztautau and \ttbar regions, respectively, and \cref{fig:perf:tnp:event_sel:SRs} shows the \pt distributions of \tauhad candidates in the SRs. The requirement on the number of $b$-jets is applied to select events enriched in either \Ztautau or \ttbar. In the SRs, the multijet background is reduced by requiring the $e/\mu$ to be isolated, while \Wjets background is reduced by requiring the transverse mass, denoted as $\mT(e/\mu, \MET)$, reconstructed from the \pt of $e/\mu$ and the \MET to be smaller than \SI{50}{\GeV}. Furthermore, $\Sigma\cos\Delta\phi$ is required to be above $-0.15$, and is defined as:
\[
\Sigma\cos\Delta\phi = \cos\Delta\phi(e/\mu, \MET) + \cos\Delta\phi(\tauhadvis, \MET).
\]
A \Wjets CR, \WCR, is defined by requiring \MET in the event to be above \SI{30}{\GeV}, $\Sigma\cos\Delta\phi$ smaller than zero and $\mT(e/\mu, \MET)$ greater than \SI{50}{\GeV}. A CR enriched in multijet background, \MJCR, is selected by inverting the $e/\mu$ isolation criteria. These CRs are chosen to ensure that the relevant background process dominates the event composition in the respective region, while retaining enough events for reliable background estimates described in Section~\ref{sec:perf:tnp:bkg}.

\begin{table}[ht]
\caption{Event selection criteria in the \Ztautau region, the \Wjets and multijet control regions, for the tag-and-probe analysis.}
\label{tab:perf:tnp:Ztautau_regions}
\centering
\resizebox{\linewidth}{!}{
\begin{tabular}{c|c|c|c}
\toprule
& \Ztautau SR & \WCR & \Ztautau \MJCR \\
\midrule
$e$ isolation & \emph{Tight} & \emph{Tight} & fails \emph{Tight} \\
$\mu$ isolation & \emph{Loose} & \emph{Loose} & fails \emph{Loose} \\
$e/\mu$ and \tauhadvis & opposite electric charge & opposite electric charge & opposite/same electric charge  \\
$m_{\text{T}} (e/\mu, \met )$ & $ < \SI{50}{\GeV}$ & $> \SI{50}{\GeV}$  & $ < \SI{50}{\GeV}$ \\
$\Sigma \cos \Delta \phi$ & $\geq -0.15$ & $\leq 0.$ & $\geq -0.15$\\
$\Delta \phi(\tauhadvis, e/\mu)$ & $ > 2.4$ & $ > 2.4$ & $ > 2.4$\\
Number of $b$-jets of 85\% efficiency working point & 0 & 0 & 0 \\
$ m_\mathrm{vis}(\tauhadvis, e/\mu)$ & $[\SI{45}{\GeV}, \SI{80}{\GeV}]$ & $[\SI{45}{\GeV}, \SI{80}{\GeV}]$ & $[\SI{45}{\GeV}, \SI{80}{\GeV}]$ \\
\met & -- & $> \SI{30}{\GeV}$ & -- \\
\bottomrule
\end{tabular}
}
\end{table}

\begin{table}[ht]
\caption{Event selection criteria in the \ttbar signal region and multijet control region, for the tag-and-probe analysis.}
\label{tab:perf:tnp:ttbar_regions}
\centering
\resizebox{\textwidth}{!}{
\begin{tabular}{c|c|c}
\toprule
& \ttbar SR & \ttbar \MJCR \\
\midrule
$e$ isolation & \emph{Tight} & fails \emph{Tight} \\
$\mu$ isolation & \emph{Loose} & fails \emph{Loose} \\
$e/\mu$ and \tauhadvis & opposite electric charge & opposite/same electric charge  \\
Number of jets & $\geq 2$ & $\geq 2$ \\
Number of $b$-jets of 70\% efficiency working point & $\geq 1$ & $\geq 1$ \\
\bottomrule
\end{tabular}
}
\end{table}

\begin{figure}[ht!]
\centering
\subfloat[]{\label{fig:perf:tnp:event_sel:SRs:Ztautau:1p}\includegraphics[width=0.45\linewidth]{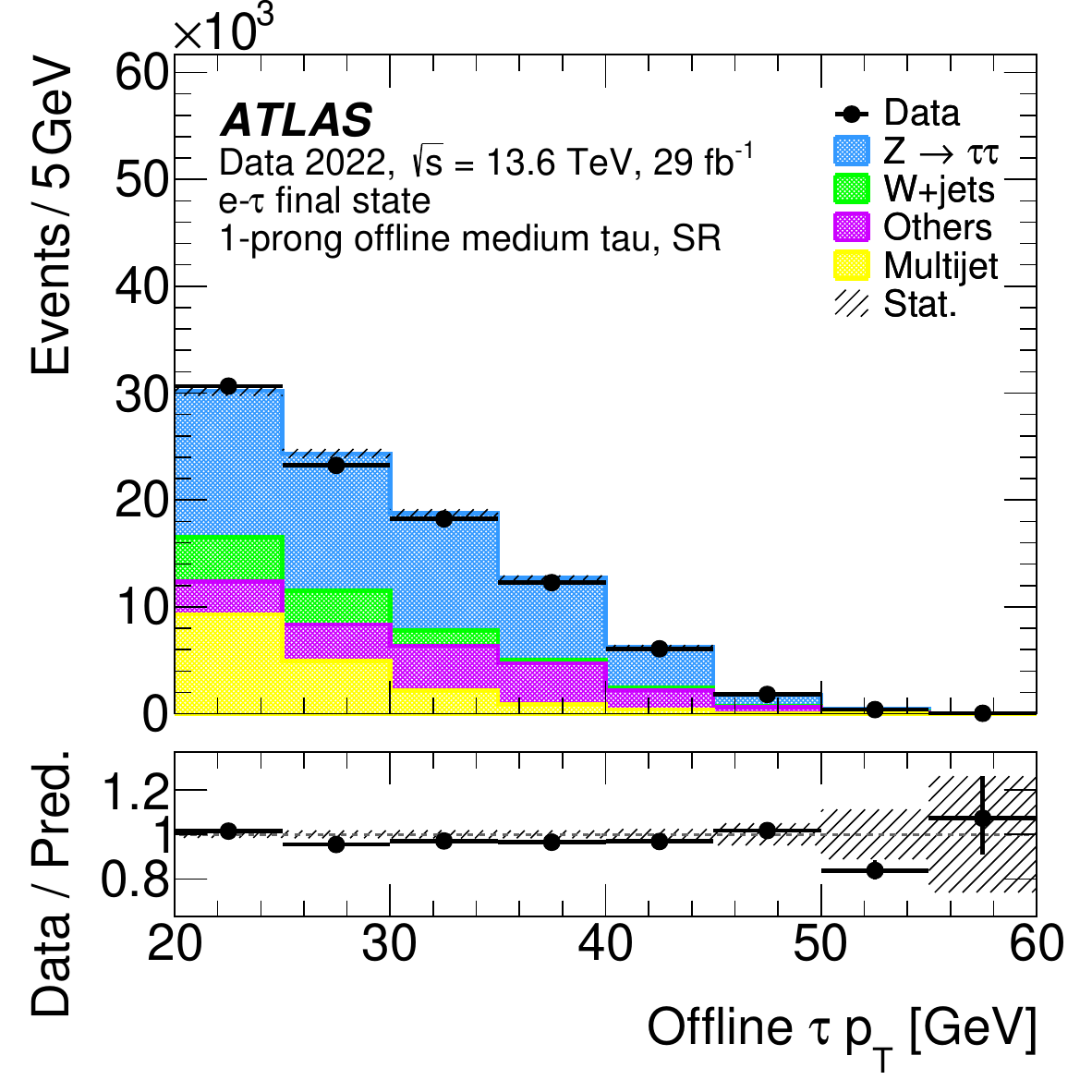}}
\subfloat[]{\label{fig:perf:tnp:event_sel:SRs:Ztautau:3p}\includegraphics[width=0.45\linewidth]{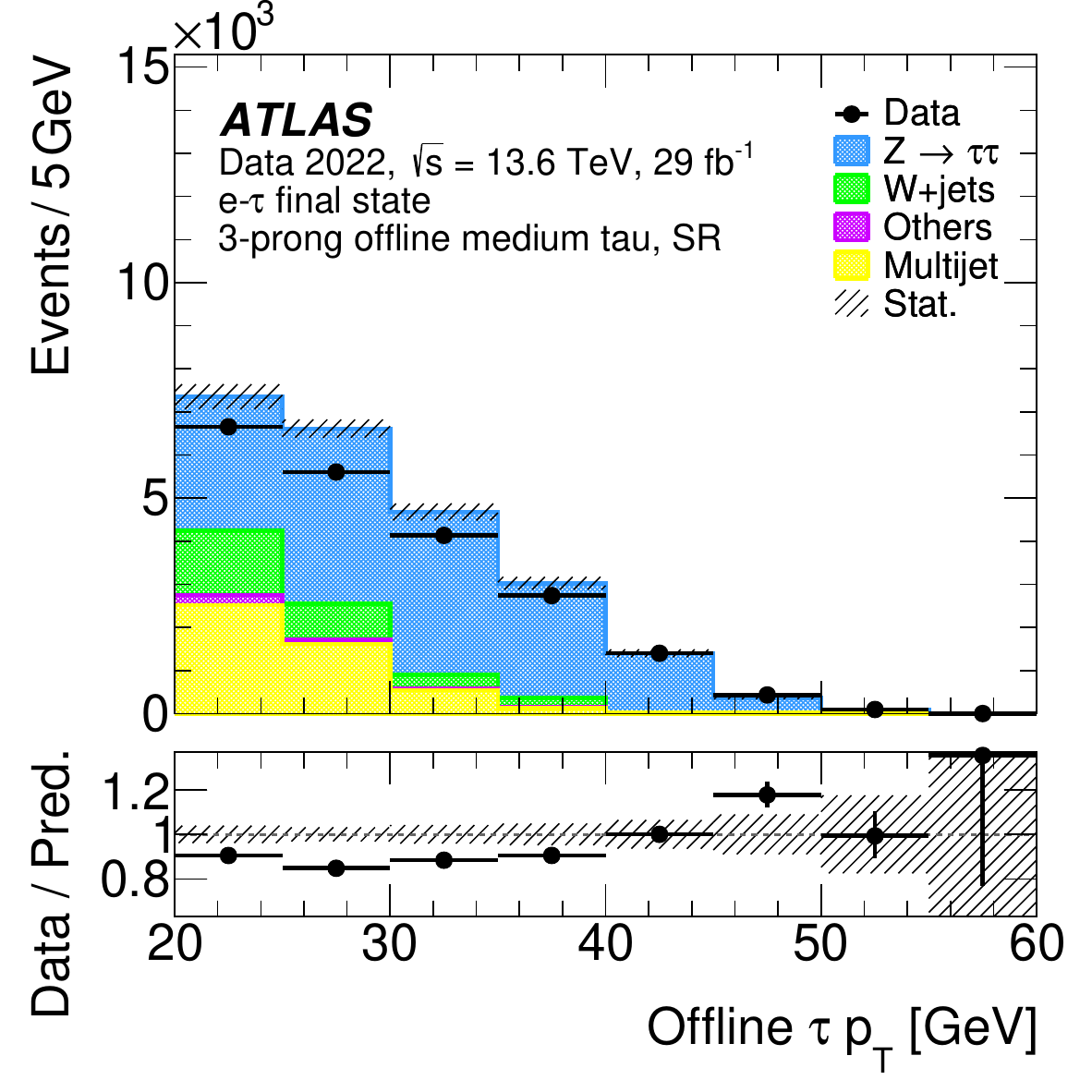}}

\subfloat[]{\label{fig:perf:tnp:event_sel:SRs:ttbar:1p}\includegraphics[width=0.45\linewidth]{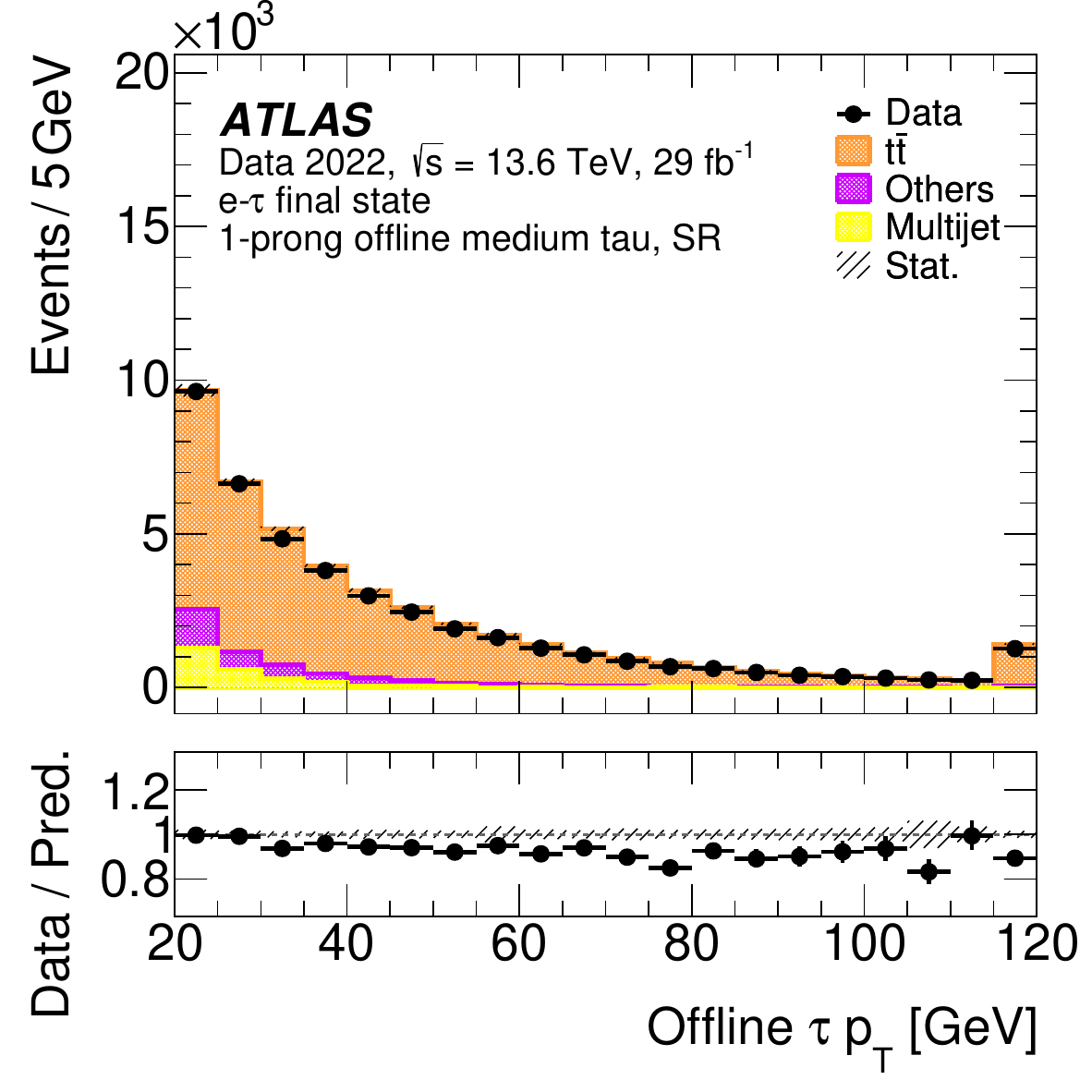}}
\subfloat[]{\label{fig:perf:tnp:event_sel:SRs:ttbar:3p}\includegraphics[width=0.45\linewidth]{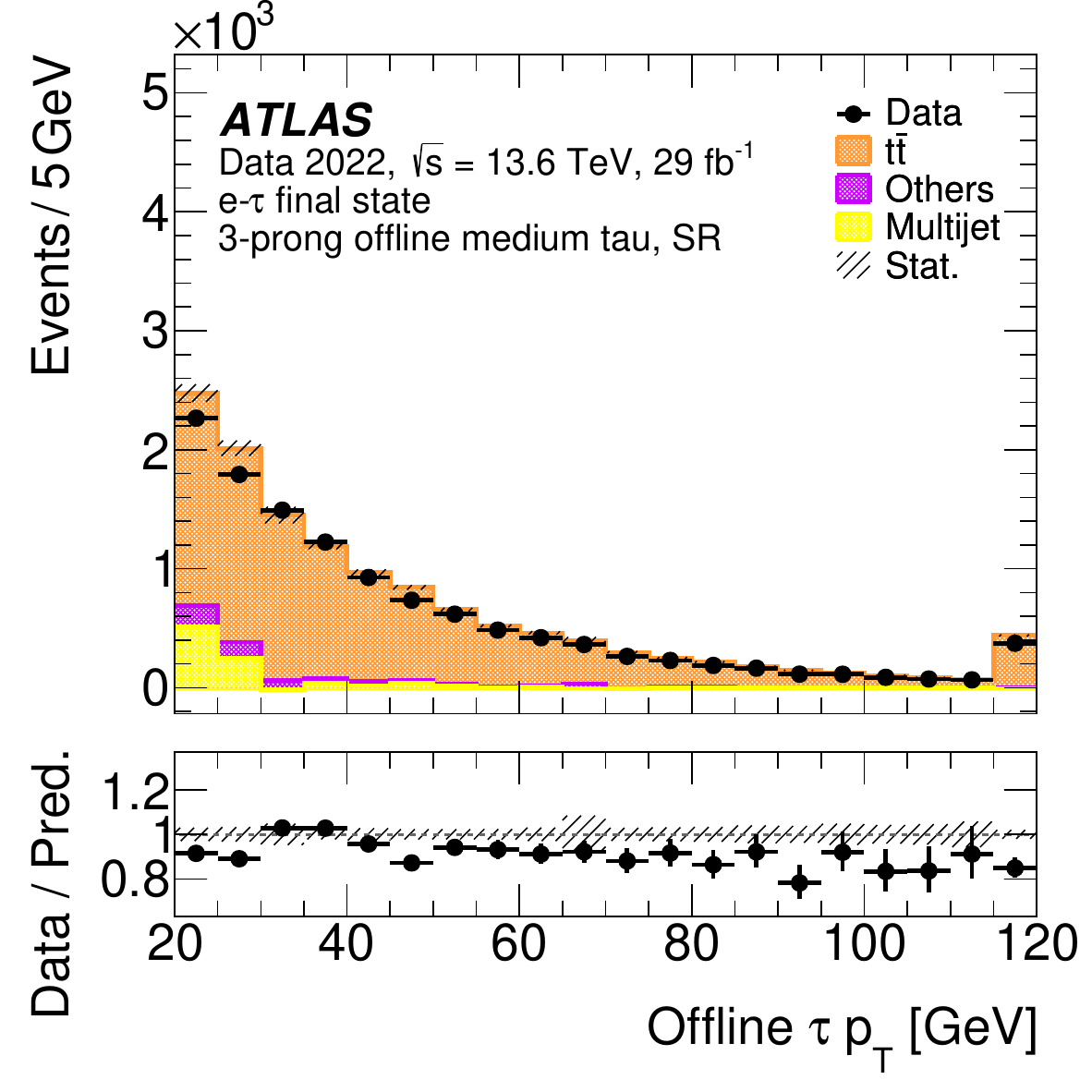}}
\caption{Distributions of the offline \pt of \tauhad candidates in the SRs for the \Ztautau enriched selection after applying the \kW and \rMJ corrections described in \cref{sec:perf:tnp:bkg} for \protect\subref{fig:perf:tnp:event_sel:SRs:Ztautau:1p} $e+\tauhad$ with offline one-prong \tauhad selection, \protect\subref{fig:perf:tnp:event_sel:SRs:Ztautau:3p} $e+\tauhad$ with offline three-prong \tauhad selection; \protect\subref{fig:perf:tnp:event_sel:SRs:ttbar:1p} and \protect\subref{fig:perf:tnp:event_sel:SRs:ttbar:3p} show the distributions in \ttbar enriched regions for $e+\tauhad$ with offline one-prong and three-prong \tauhad selection, respectively, after applying the \rMJ correction. \emph{Others} include all MC processes used in the analysis (as summarised in \cref{tab:samples} for \cref{sec:perf}) apart from the corresponding signal contribution. All the distributions are plotted using the 2022 data sample. Uncertainties shown are statistical only. The modelling differences are covered by the systematic uncertainties in the efficiency scale factors.}
\label{fig:perf:tnp:event_sel:SRs}
\end{figure}

\subsubsection{Background estimation}\label{sec:perf:tnp:bkg}

A combination of data-driven and MC-based approaches are used to model the various backgrounds. The \Wjets background is estimated by using simulated MC samples, with a correction factor, \kW, used to correct its normalisation. This correction factor is calculated in the \WCR, as described in \cref{sec:perf:tnp:event_sel}, and is defined as
\[
\kW = \frac{N^{\mathrm{Data}}-N^{\mathrm{MC}}_{\mathrm{Non~}\Wjets}}{N^{\mathrm{MC}}_{\Wjets}} \Bigg|_{\WCR},
\]
where $N^{\mathrm{Data}}$ is the number of selected data events, $N^{\mathrm{MC}}_{\mathrm{Non~}\Wjets}$, the number of simulated MC events estimated from processes other than \Wjets process, and $N^{\mathrm{MC}}_{\Wjets}$ is the number of \Wjets events estimated from MC simulation. The \kW correction factor is calculated as a function of \pt of the \tauhad separately for one-prong and three-prong \tauhad candidates, and is typically well within 5\% of unity. The modelling of the \pt of the \tauhadvis candidate in the OS \WCR, after applying the \kW correction, is shown in \cref{fig:perf:tnp:bkg:WCR}. Distributions are shown for one-prong and three-prong \tauhad candidates in $e+\tauhad$ final state.

\begin{figure}[ht!]
\centering
\subfloat[]{\label{fig:perf:tnp:bkg:WCR:1p}\includegraphics[width=0.45\linewidth]{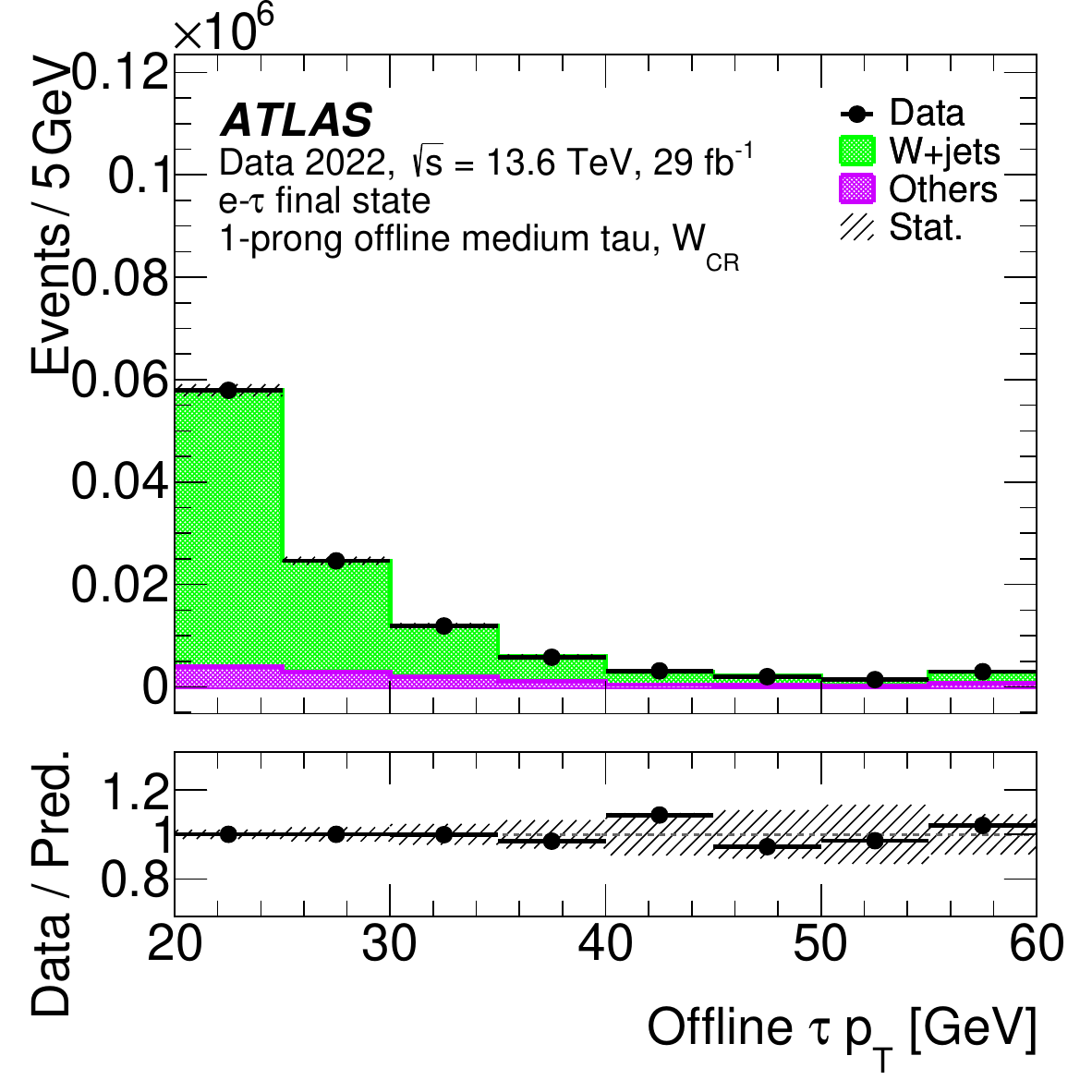}}
\subfloat[]{\label{fig:perf:tnp:bkg:WCR:3p}\includegraphics[width=0.45\linewidth]{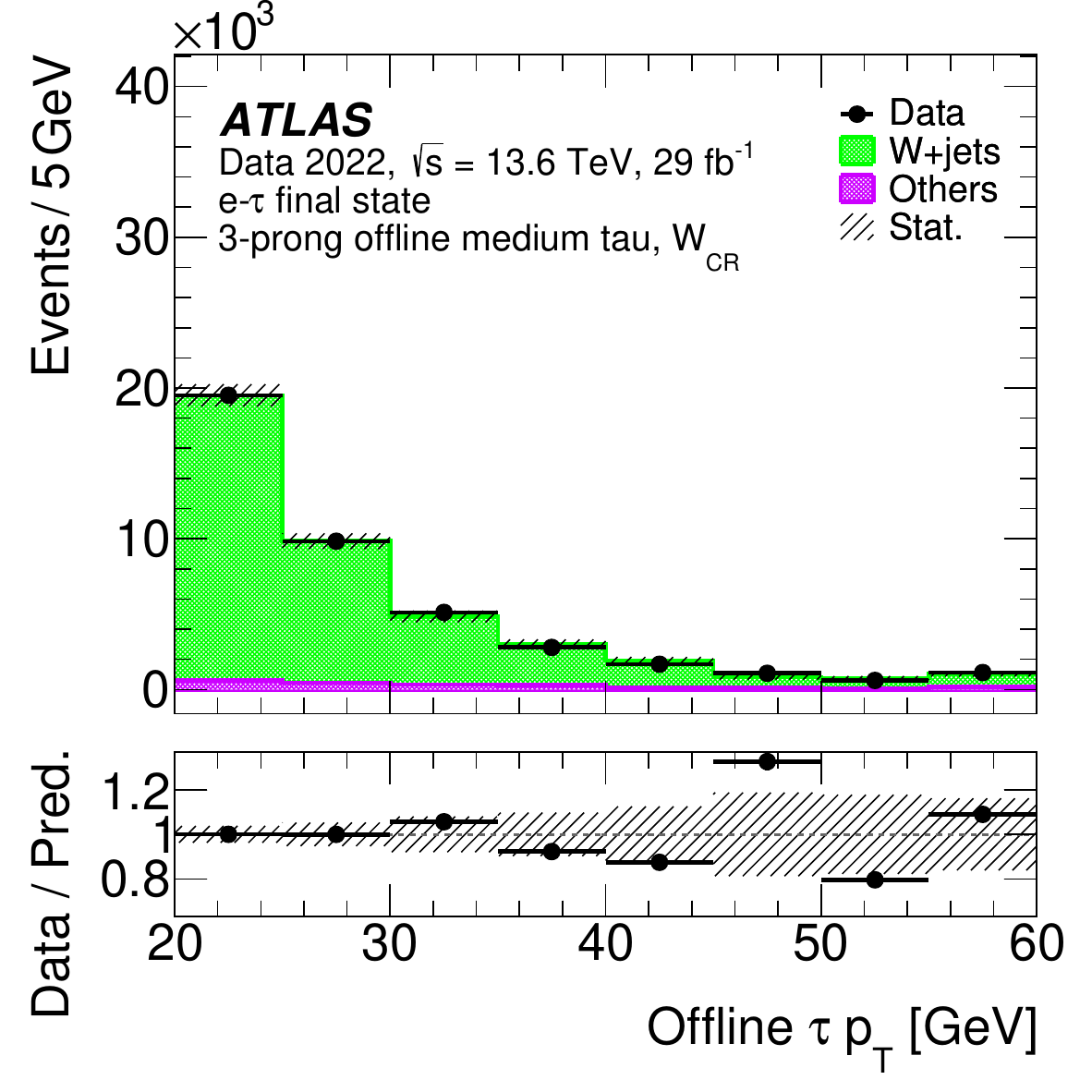}}
\caption{Distributions of the offline \pt of \tauhad candidates in the OS \WCR of the \Ztautau enriched selection after applying the \kW correction for $e+\tauhad$ with offline \protect\subref{fig:perf:tnp:bkg:WCR:1p} one-prong and \protect\subref{fig:perf:tnp:bkg:WCR:3p} three-prong \tauhad selection. \emph{Others} include all MC processes used in the analysis (and summarised in \cref{tab:samples}) apart from W+jets. All the distributions are plotted using the 2022 data sample. Uncertainties shown are statistical only.}
\label{fig:perf:tnp:bkg:WCR}
\end{figure}

The contribution from the multijet background is estimated by using a data-driven technique. It relies on the assumption of no charge correlation between the lepton ($e/\mu$) and the misidentified \tauhad candidate in multijet enriched events. The contribution from multijet background in the OS region is thus proportional to the contribution in the SS region,
\[
N^{\mathrm{OS}}_{\mathrm{MJ}} = \rMJ \cdot N^{\mathrm{SS}}_{\mathrm{MJ}}.
\]
The factor, \rMJ, is defined as the ratio of the number of multijet events in the \MJCR with OS selection, $N^{\mathrm{OS}}_{\mathrm{MJ}}$ and SS selection, $N^{\mathrm{SS}}_{\mathrm{MJ}}$,
\[
\rMJ = \frac{\left(N^{\mathrm{Data}}-N^{\mathrm{MC}}\right)\bigg|_{\MJCR}^{\mathrm{OS}}}{\left(N^{\mathrm{Data}}-N^{\mathrm{MC}}\right)\bigg|_{\MJCR}^{\mathrm{SS}}},
\]
where $N^{\mathrm{Data}}$ and $N^{\mathrm{MC}}$ are the number of observed data and expected MC events in the \MJCR, respectively. Furthermore, \rMJ is calculated for each trigger threshold as a function of the \pt, prongness, and identification working point of the \tauhad, and is typically well within 5\% of unity.

The modelling of the \tauhadvis \pt distribution in the \MJCR OS region after applying the \rMJ correction is shown in \cref{fig:perf:tnp:bkg:MJCR} for the \Ztautau and \ttbar enriched regions. Distributions are shown for one-prong and three-prong \tauhad candidates in $e+\tauhad$ and $\mu+\tauhad$ final states.

\begin{figure}[ht!]
\centering
\subfloat[]{\label{fig:perf:tnp:bkg:MJCR:Ztautau:1p}\includegraphics[width=0.45\linewidth]{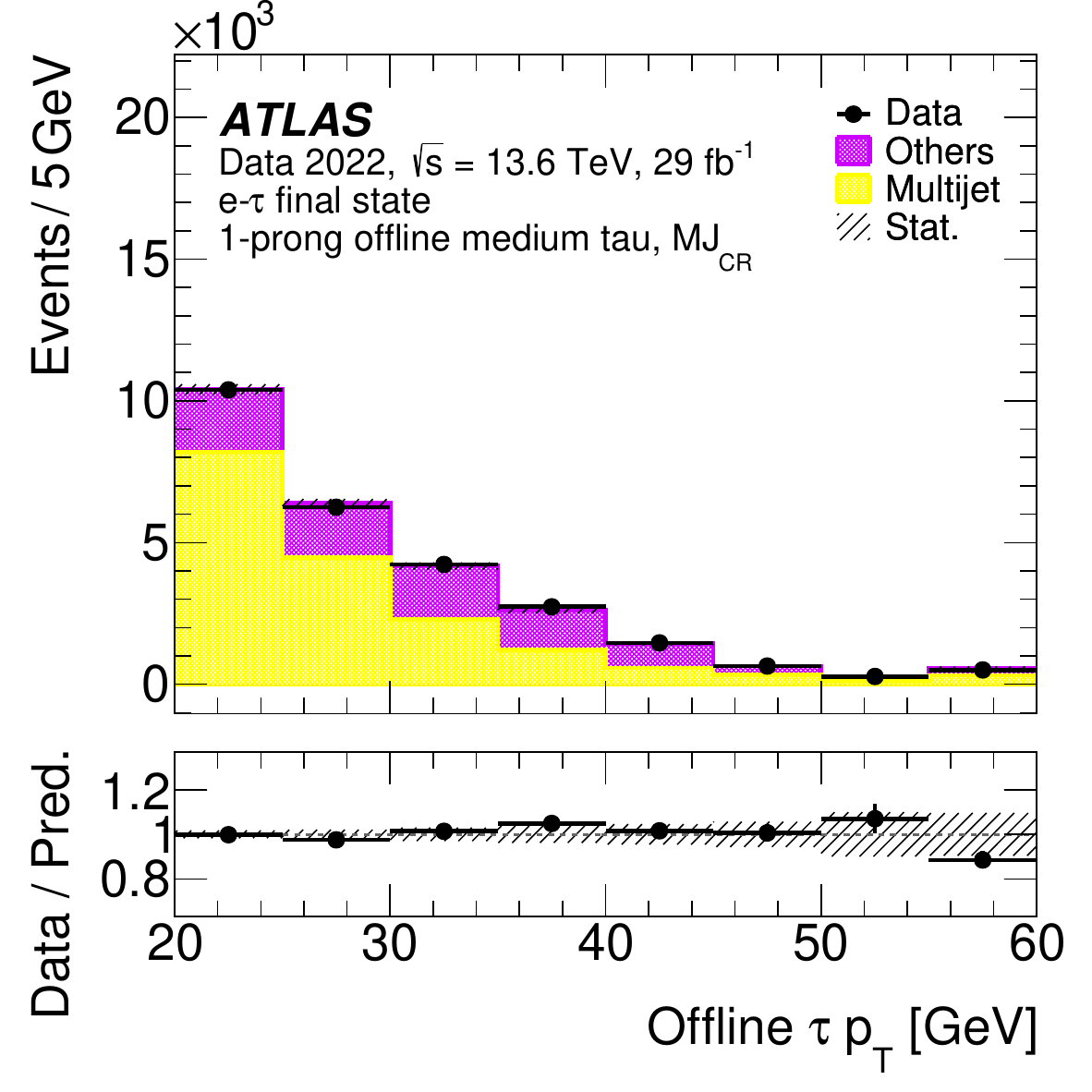}}
\subfloat[]{\label{fig:perf:tnp:bkg:MJCR:Ztautau:3p}\includegraphics[width=0.45\linewidth]{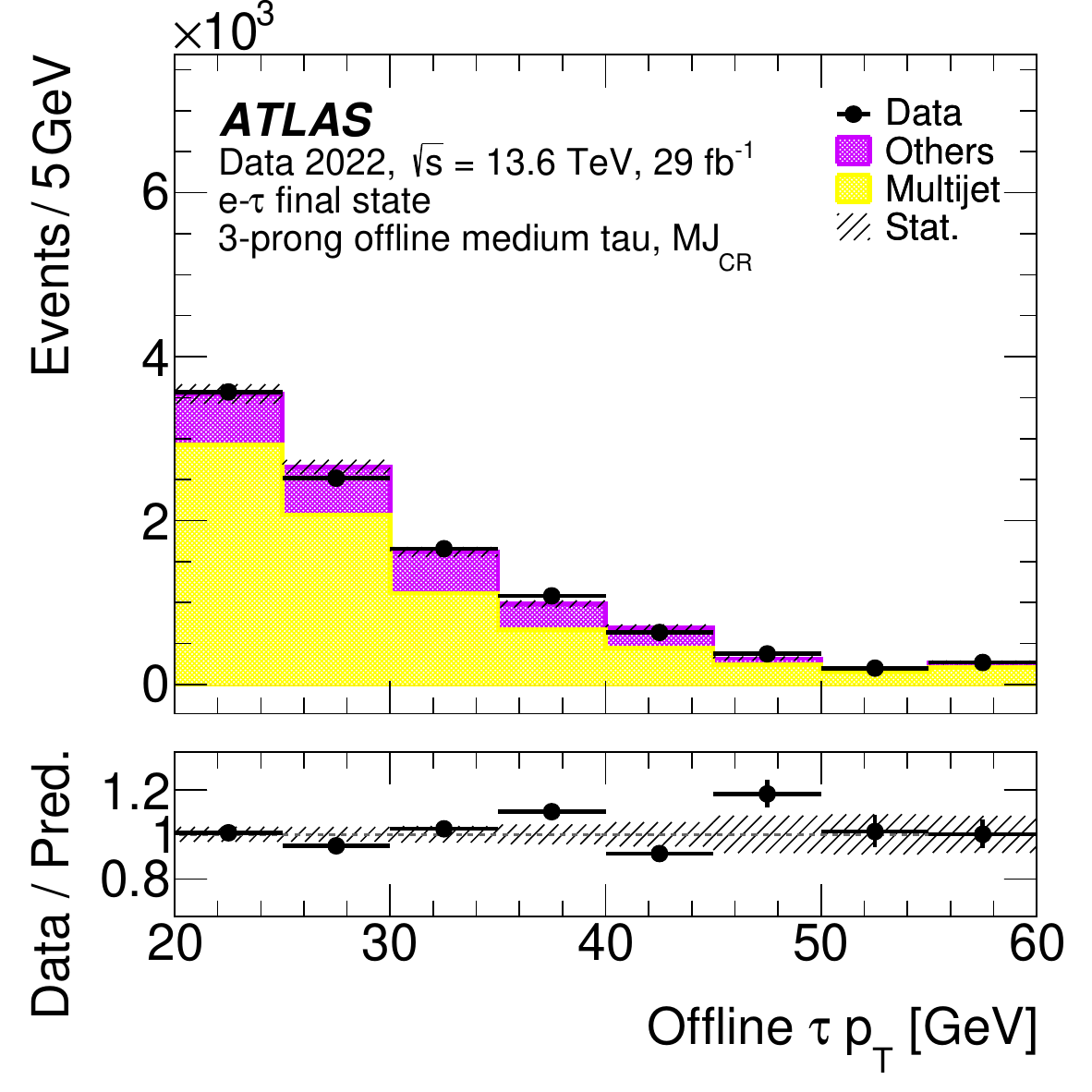}}

\subfloat[]{\label{fig:perf:tnp:bkg:MJCR:ttbar:1p}\includegraphics[width=0.45\linewidth]{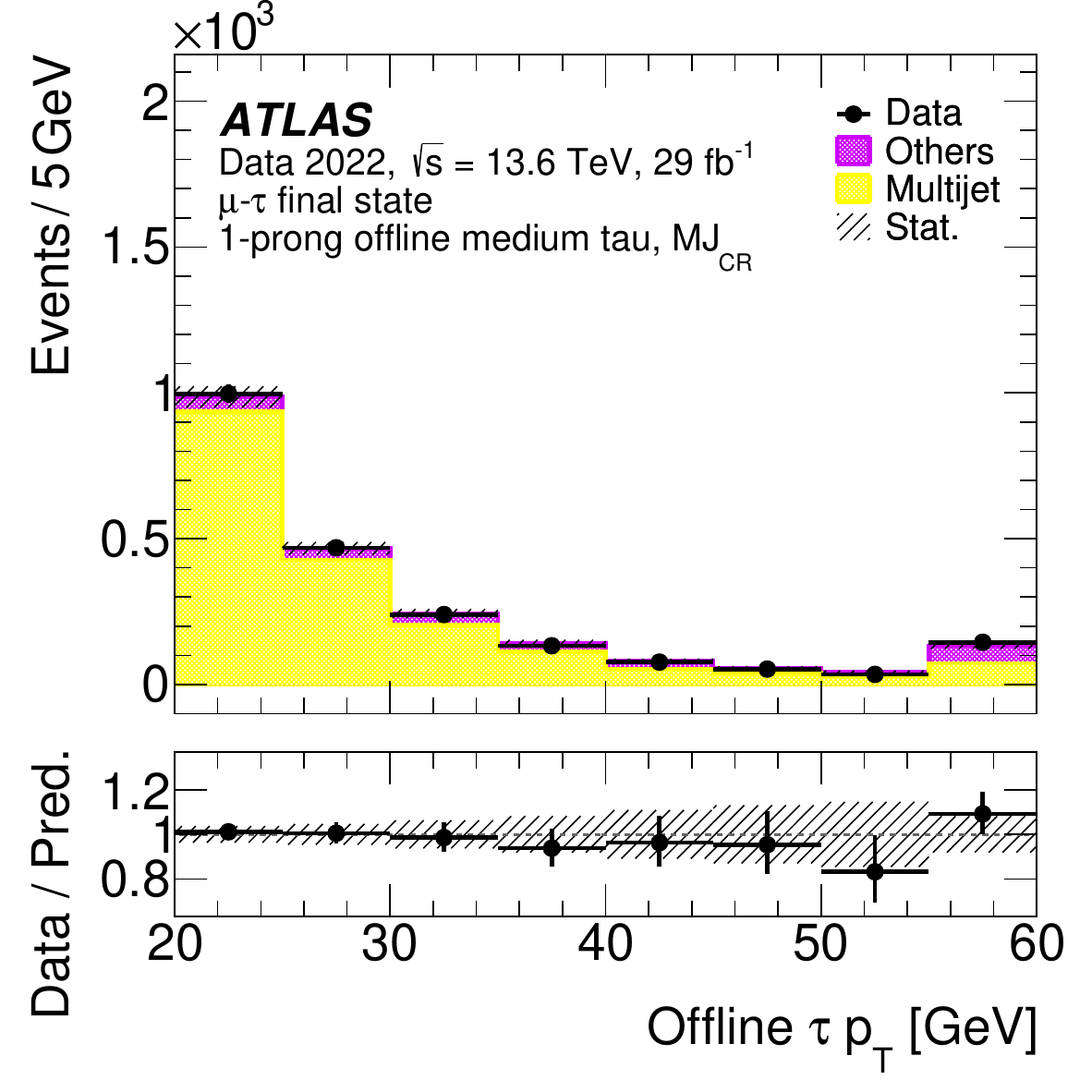}}
\subfloat[]{\label{fig:perf:tnp:bkg:MJCR:ttbar:3p}\includegraphics[width=0.45\linewidth]{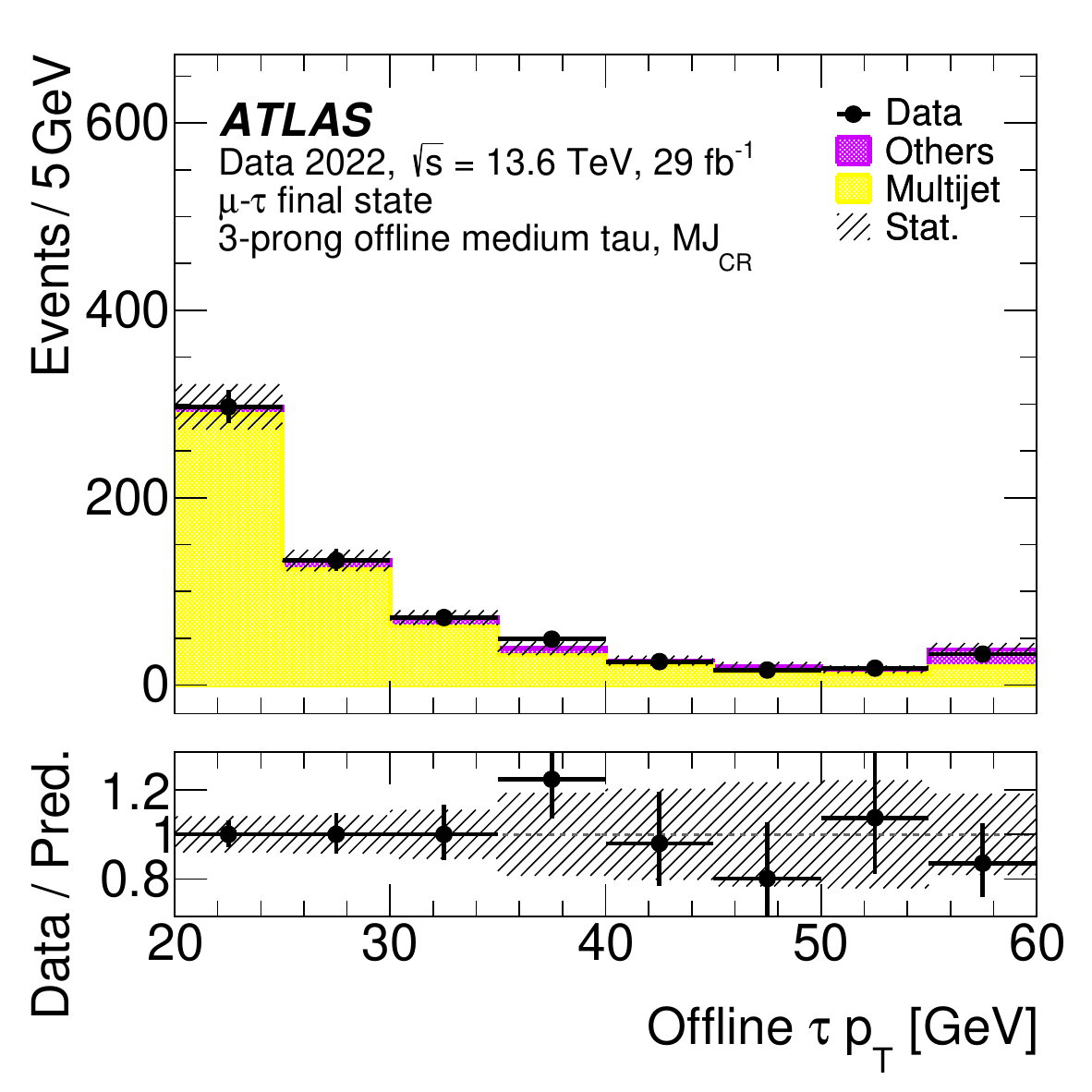}}
\caption{Distributions of the offline \pt of \tauhad candidates in the OS \MJCR after applying the \rMJ correction for the \Ztautau enriched selection in $e+\tauhad$ final state with \protect\subref{fig:perf:tnp:bkg:MJCR:Ztautau:1p} offline one-prong \tauhad or \protect\subref{fig:perf:tnp:bkg:MJCR:Ztautau:3p} offline three-prong \tauhad; figures \protect\subref{fig:perf:tnp:bkg:MJCR:ttbar:1p} and \protect\subref{fig:perf:tnp:bkg:MJCR:ttbar:3p} show the distributions for \ttbar enriched selection in $\mu+\tauhad$ final state with \protect\subref{fig:perf:tnp:bkg:MJCR:ttbar:1p} offline one-prong \tauhad or \protect\subref{fig:perf:tnp:bkg:MJCR:ttbar:3p} offline three-prong \tauhad. \emph{Others} include all MC processes used in the analysis (and summarised in \cref{tab:samples}), while \emph{Multijet} are multijet events estimated as described in \cref{sec:perf:tnp:bkg}. All distributions are plotted using the 2022 data sample. Uncertainties shown are statistical only.}
\label{fig:perf:tnp:bkg:MJCR}
\end{figure}

\subsection{Tau trigger efficiency and scale factor measurement}%

The efficiency of a tau trigger with a given \pt threshold is defined as the ratio of the number of events selected by the corresponding tag-and-probe trigger and satisfying the signal selection as defined in \cref{tab:perf:tnp:Ztautau_regions,tab:perf:tnp:ttbar_regions} to the events selected by the single $e/\mu$ trigger satisfying the same signal selection:

\[
\epsilon_{\mathrm{trig}} = \frac{N_\mathrm{events}^\mathrm{SR, pass \: tag-and-probe} - N_\mathrm{background}^\mathrm{SR, pass \: tag-and-probe} }{N_\mathrm{events}^\mathrm{SR} - N_\mathrm{background}^\mathrm{SR}}.
\]

The efficiency is calculated using data and signal MC events for \Ztautau and \ttbar processes. The signal purity for both \Ztautau and \ttbar signal regions is above 50\% (60\%) for 1-prong (3-prong) \tauhadvis candidates. The expected background estimated from various sources is subtracted from the data before comparing the observed efficiency in data with the one expected from the MC simulation to derive SFs, which combine both processes.

\subsubsection{Systematic uncertainties}

Systematic uncertainties play an important role in the evaluation of SFs, as they help to quantify the robustness of the measurements and their dependence on various modelling assumptions. These uncertainties originate from different aspects of the data and MC simulation processes, such as object reconstruction, identification, isolation, and triggering. For the tau trigger efficiency and SF measurements, specific attention is given to uncertainties associated with \tauhad identification. In addition, a modelling systematic uncertainty in the background estimate is applied, based on alternative values of the \kW and \rMJ factors when using a slightly modified definition of the \WCR.
Additionally, uncertainties from other objects (e.g. electrons, muons, jets or \btagged jets) are considered. The complete list of systematic uncertainties considered is presented in \cref{tab:perf:sfs:systematics}.

\begin{table}[htbp]
\centering
\caption{Overview of systematic uncertainties used in the scale factors measurement.}
\label{tab:perf:sfs:systematics}
\begin{tabular}{lp{0.7\linewidth}}
\toprule
\textbf{Category}   & \textbf{Description}                                                                                  \\
\midrule
Electron            & Uncertainties associated with the trigger, reconstruction, identification, and isolation of an electron.  \\
Muon                & Uncertainties associated with the trigger, reconstruction, identification, and isolation of a muon.       \\
Tau                 & Uncertainties associated with the reconstruction and identification of offline \tauhad candidates. The uncertainties are evaluated separately for one-prong and three-prong \tauhad candidates decays in distinct \pt regions: 20-\SI{30}{\GeV}, 30-\SI{40}{\GeV}, and $\geq \SI{40}{\GeV}$. \\
\kW                 & Modelling and statistical uncertainties in the \kW factors used to estimate fake contributions in OS and SS regions, for both one-prong and three-prong tau decays.  \\
\rMJ                & Modelling and statistical uncertainties in the \rMJ factors applied to estimate the multijet background for one-prong and three-prong tau decays. \\
Flavour             & Uncertainties in $b$-, $c$-, and light-flavour jet tagging efficiencies, including extrapolation uncertainties from $c$-tagging. \\
Jet                 & Uncertainties in the jet vertex tagging efficiency for central and forward jets.                      \\
\bottomrule
\end{tabular}
\end{table}

The largest systematic uncertainties in the measured SFs originate from the jet, flavour, \kW, and \rMJ categories, contributing between 0.1--1.4\% uncertainty in the efficiency plateau, depending on the channel.
The total uncertainty in the SF measurement is computed as the sum in quadrature of all systematic uncertainties and the statistical uncertainty.

\subsubsection{Combined scale factor}

\cref{fig:perf:sfs:comb:2022} presents the efficiency results as a function of the offline \tauhadvis \pT for one-prong and three-prong \tauhad candidates, based on 2022 data and associated MC simulation using a combination of electron and muon tag-and-probe triggers.
\cref{fig:perf:sfs:comb:2023} shows the efficiency results based on data collected in 2023 and associated MC simulation for \tauhad triggers with \pt thresholds of 40 and 60 {\GeV} at HLT. These have been seeded with legacy L1 tau trigger candidates with \pt thresholds of 25 and 40 {\GeV}, respectively. All these results are shown for the $\mu+\tauhad$ final state. Similar results are obtained for the $e+\tauhad$ final state. The same offline tau selection is applied in both \cref{fig:perf:sfs:comb:2022,fig:perf:sfs:comb:2023}, with the offline \pT requirement set to a value \SI{5}{\GeV} below the trigger \pT threshold of the respective tau trigger chain. The ratio of the efficiencies between the observed values in the background-subtracted data and those in MC simulation yield the SFs for the particular channel in question, which can be seen in the bottom panels.

\begin{figure}[!htbp]
\centering
\subfloat[]{\label{fig:perf:sfs:comb:2022:etau:1p}\includegraphics[width=0.45\linewidth]{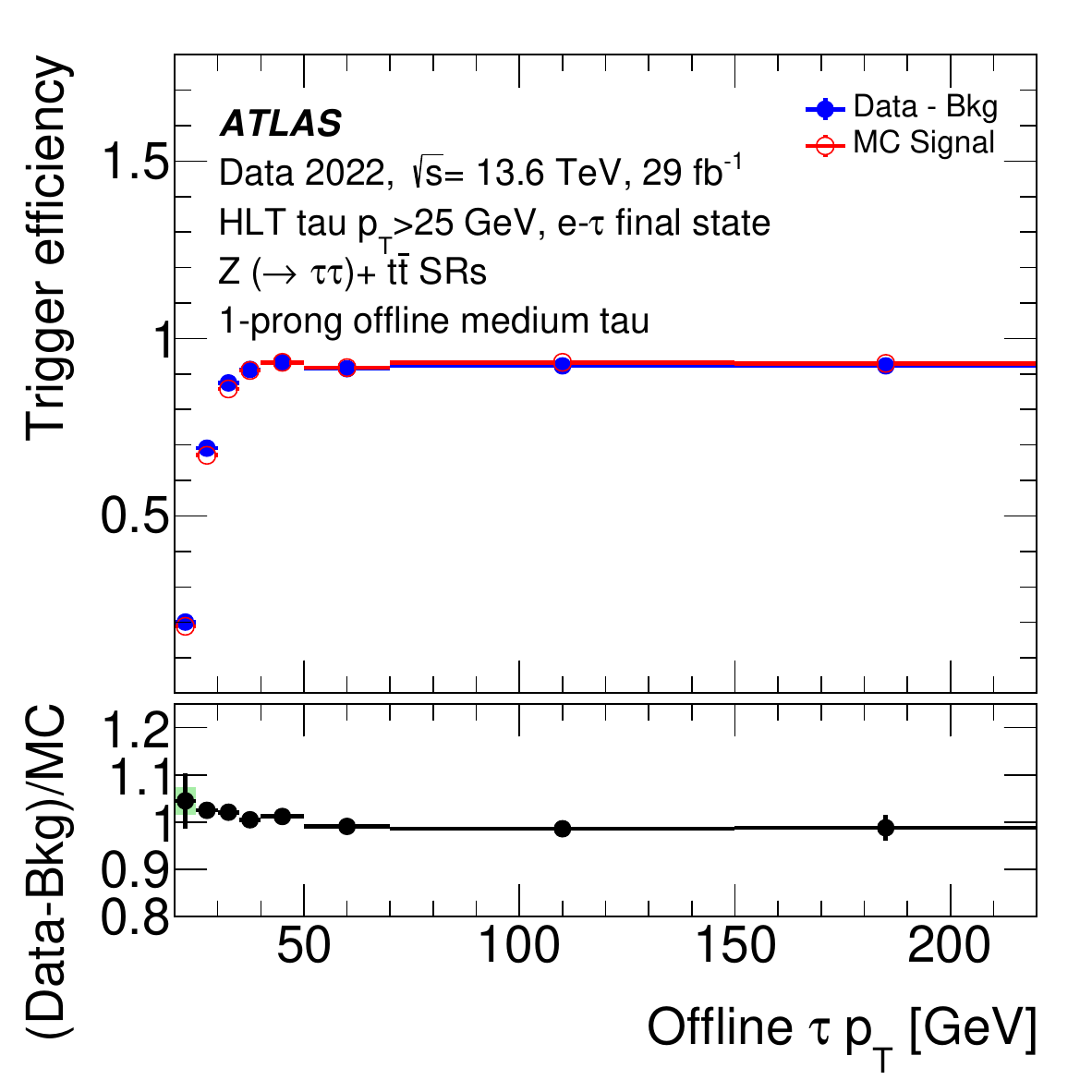}}
\subfloat[]{\label{fig:perf:sfs:comb:2022:etau:3p}\includegraphics[width=0.45\linewidth]{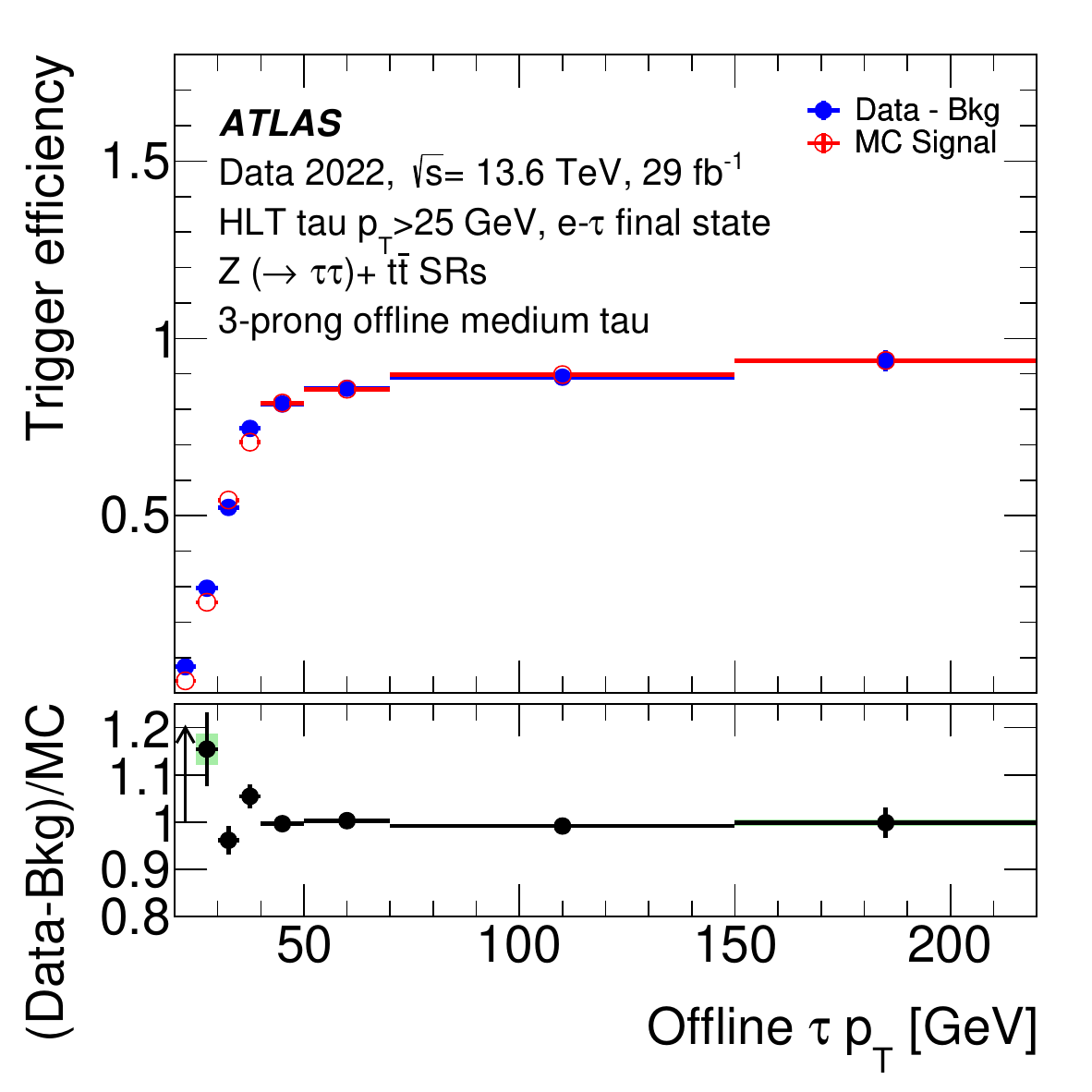}}

\subfloat[]{\label{fig:perf:sfs:comb:2022:mutau:1p}\includegraphics[width=0.45\linewidth]{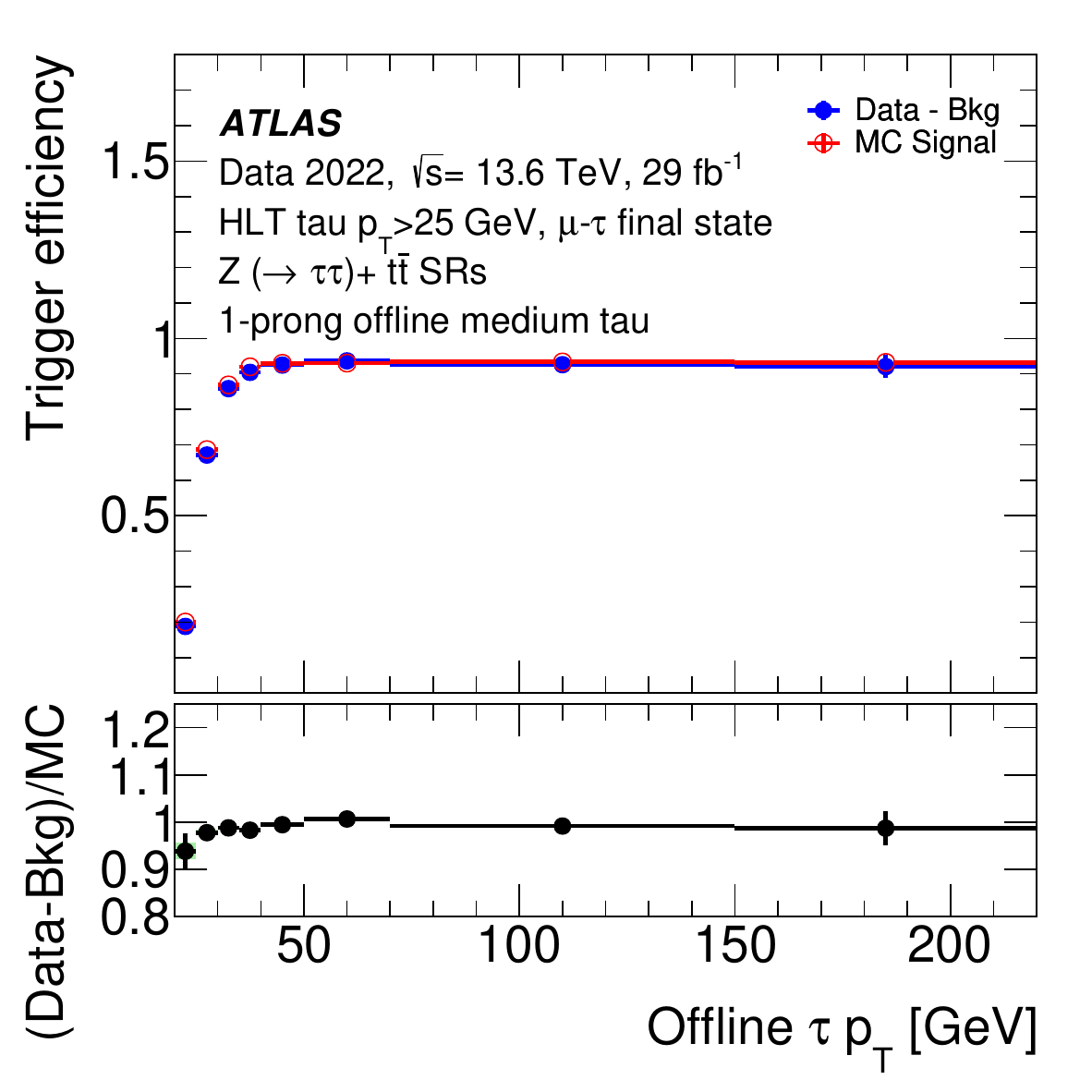}}
\subfloat[]{\label{fig:perf:sfs:comb:2022:mutau:3p}\includegraphics[width=0.45\linewidth]{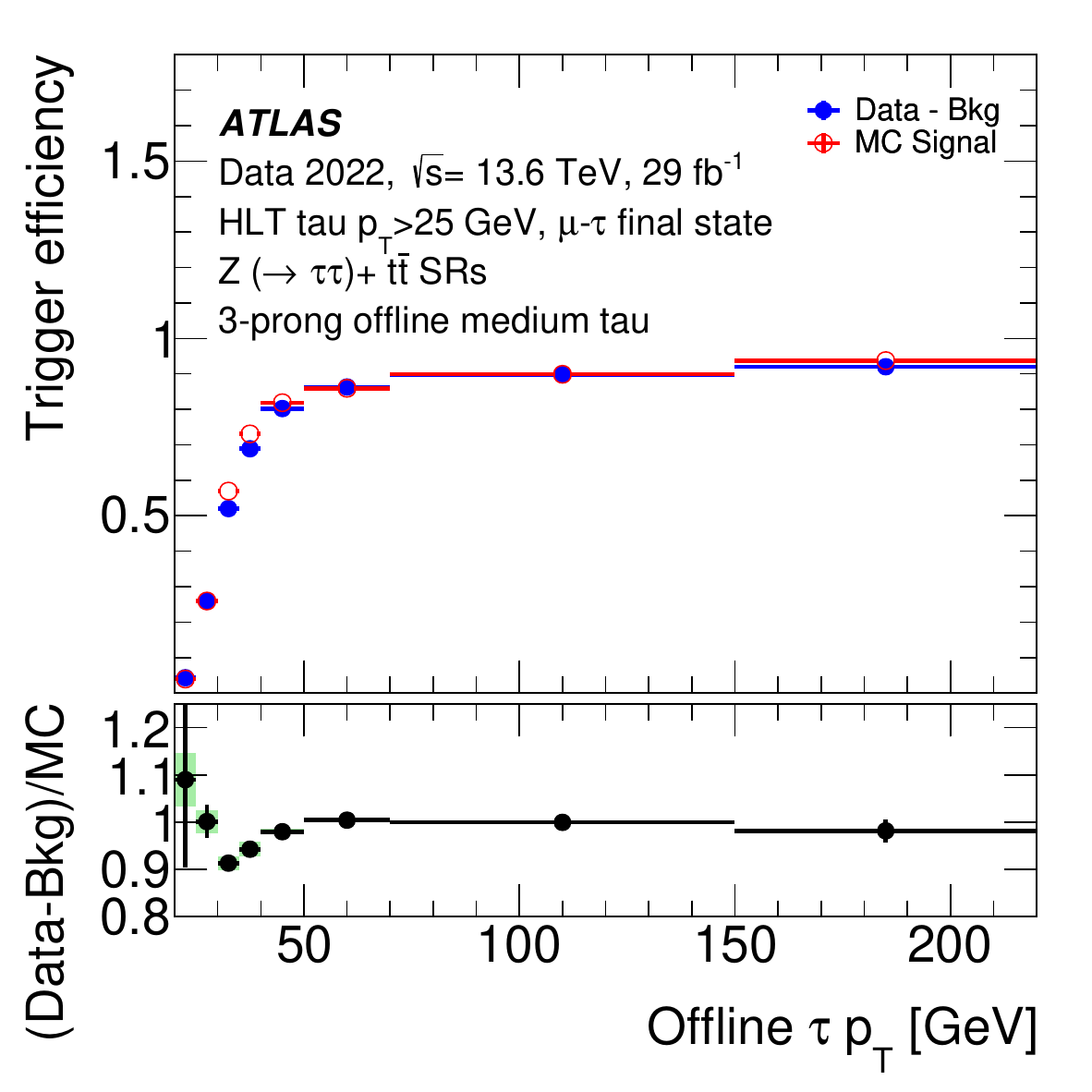}}
\caption{Efficiency of the tau trigger as a function of the offline \tauhadvis \pt for offline one-prong (\protect\subref{fig:perf:sfs:comb:2022:etau:1p}, \protect\subref{fig:perf:sfs:comb:2022:mutau:1p}) and offline three-prong (\protect\subref{fig:perf:sfs:comb:2022:etau:3p}, \protect\subref{fig:perf:sfs:comb:2022:mutau:3p}) candidates. These measurements are performed using $e+\tauhad$ (\protect\subref{fig:perf:sfs:comb:2022:etau:1p}, \protect\subref{fig:perf:sfs:comb:2022:etau:3p}) and $\mu+\tauhad$ (\protect\subref{fig:perf:sfs:comb:2022:mutau:1p}, \protect\subref{fig:perf:sfs:comb:2022:mutau:3p}) tag and probe triggers, with a trigger level \tauhadvis having $\pT > \SI{25}{\GeV}$. Efficiencies are computed from the sum of \Ztautau and \ttbar signal region events estimated both for data with background subtraction and MC signal. The offline \tauhadvis candidates are required to have $\pT > \SI{20}{\GeV}$. Bottom panels show the ratio between signal and data-background efficiencies with statistical uncertainties only and the bands represent the systematic uncertainty. All the distributions are shown using the 2022 data sample.}
\label{fig:perf:sfs:comb:2022}
\end{figure}

\begin{figure}[!htbp]
\centering
\subfloat[]{\label{fig:perf:sfs:comb:2023:mutau40:1p}\includegraphics[width=0.45\linewidth]{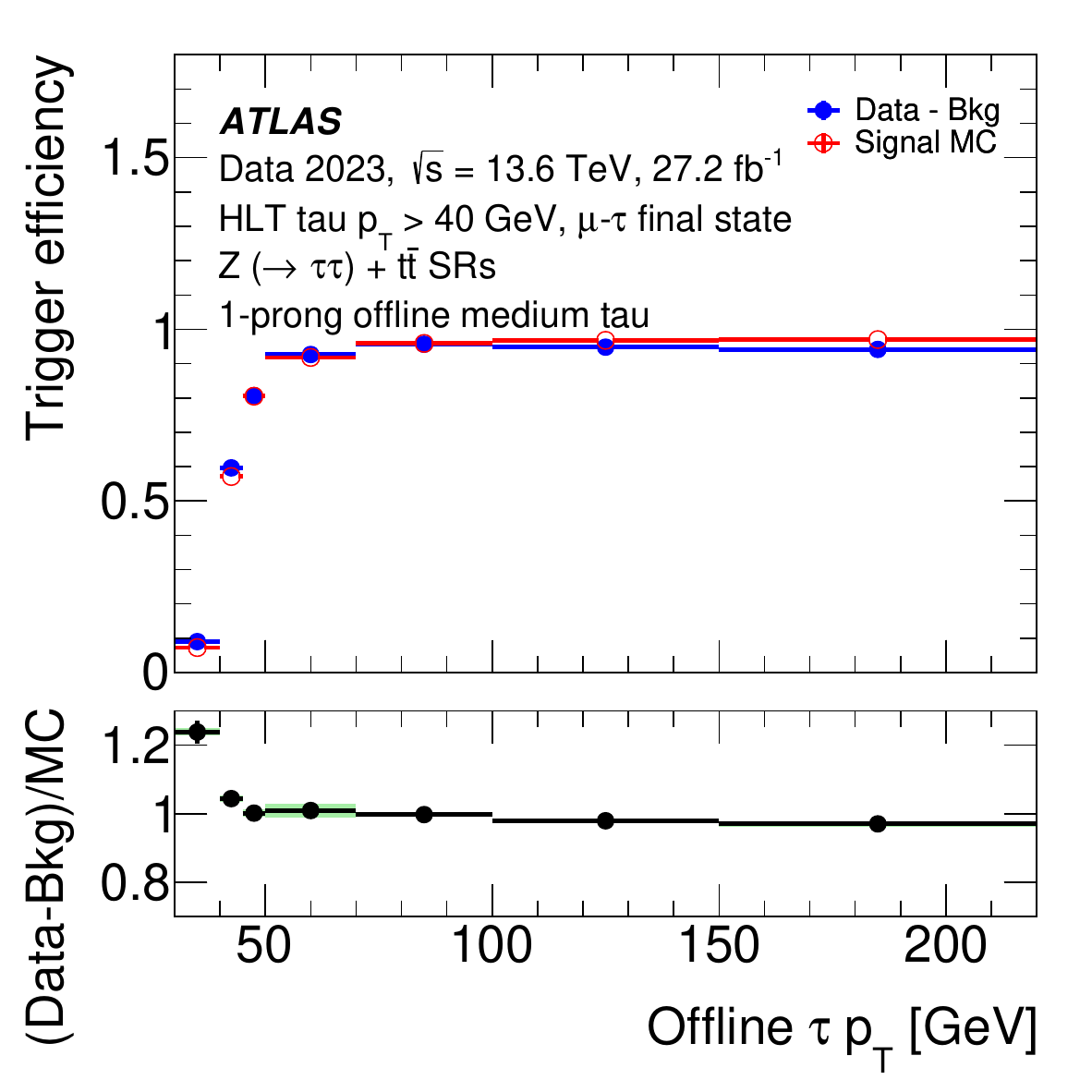}}
\subfloat[]{\label{fig:perf:sfs:comb:2023:mutau40:3p}\includegraphics[width=0.45\linewidth]{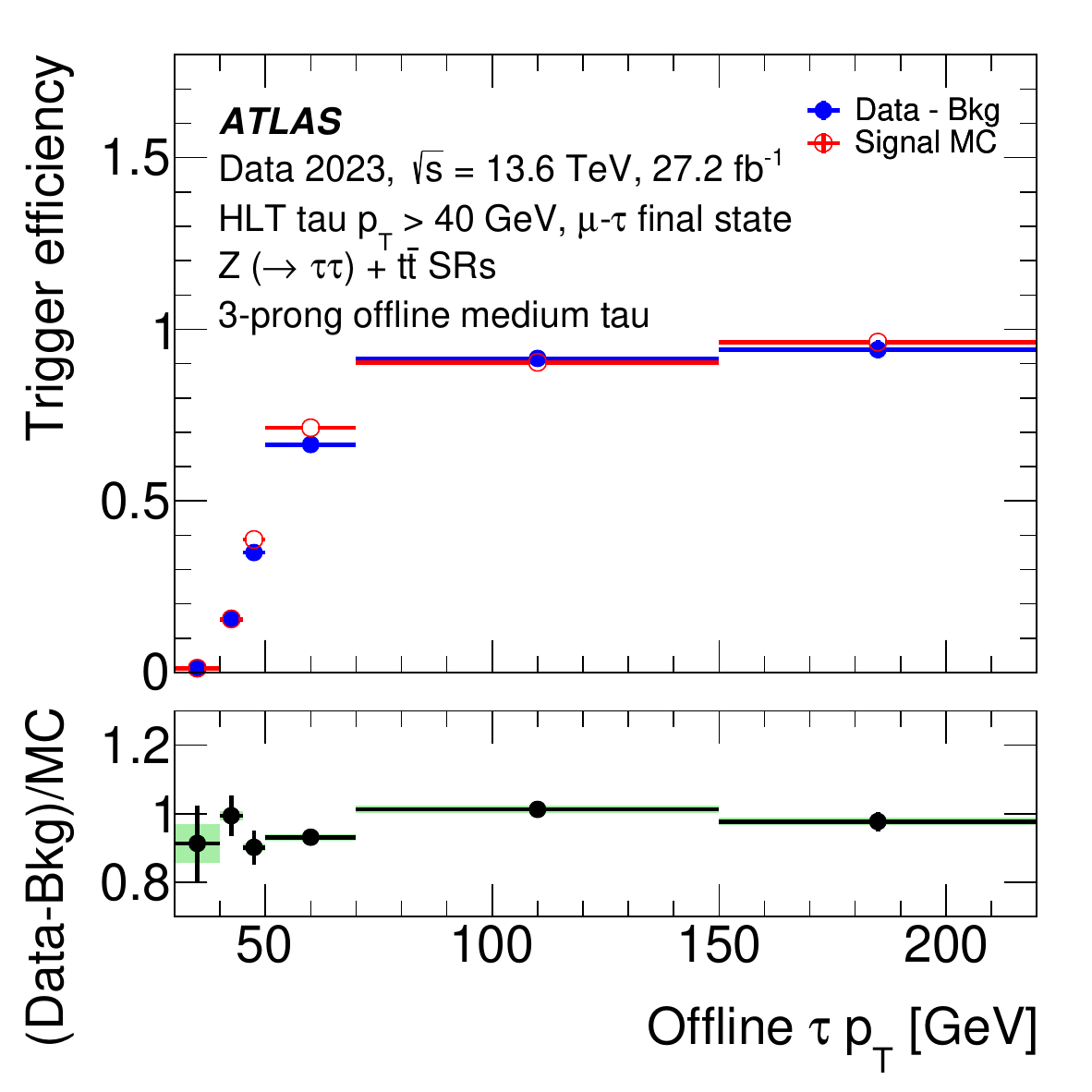}}

\subfloat[]{\label{fig:perf:sfs:comb:2023:mutau60:1p}\includegraphics[width=0.45\linewidth]{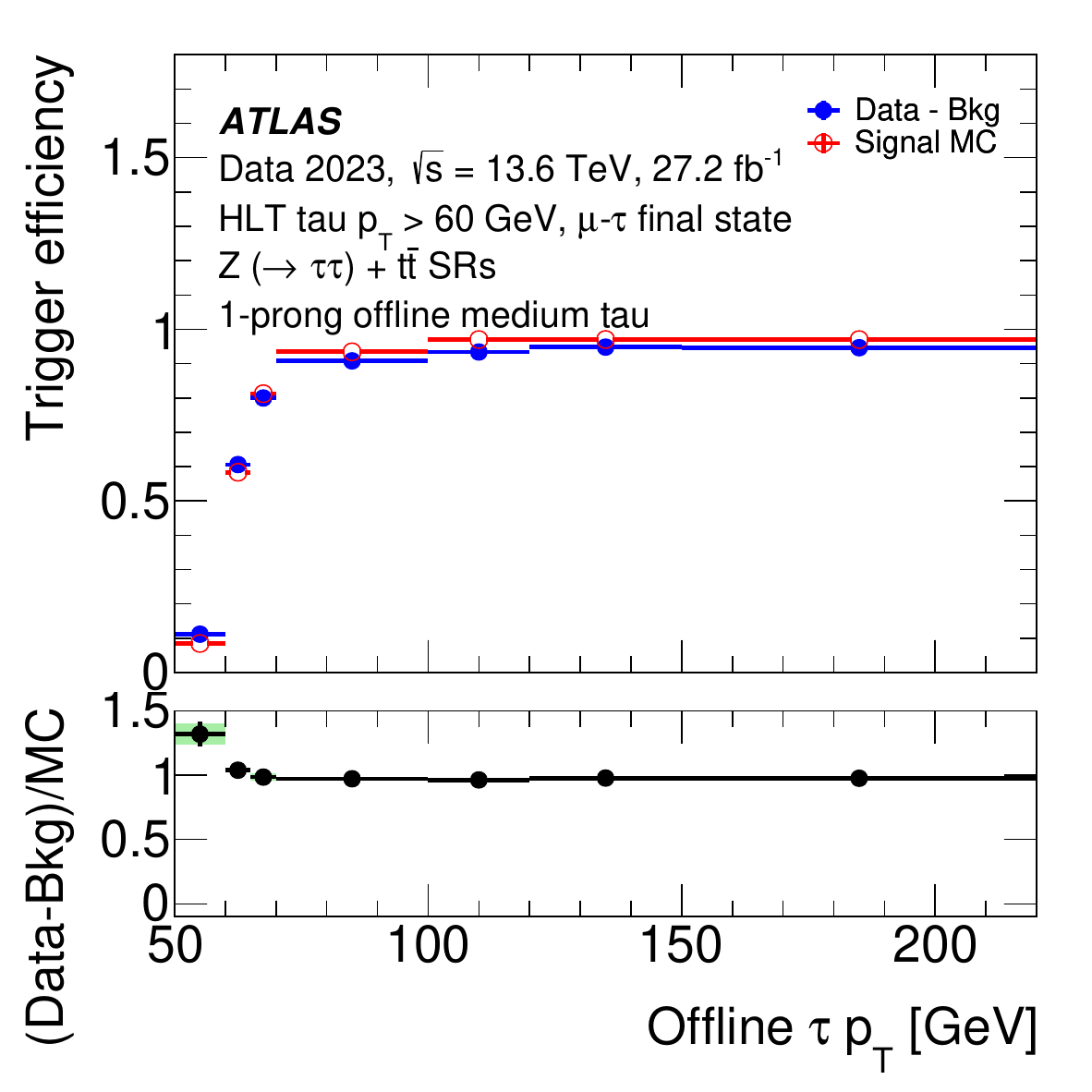}}
\subfloat[]{\label{fig:perf:sfs:comb:2023:mutau60:3p}\includegraphics[width=0.45\linewidth]{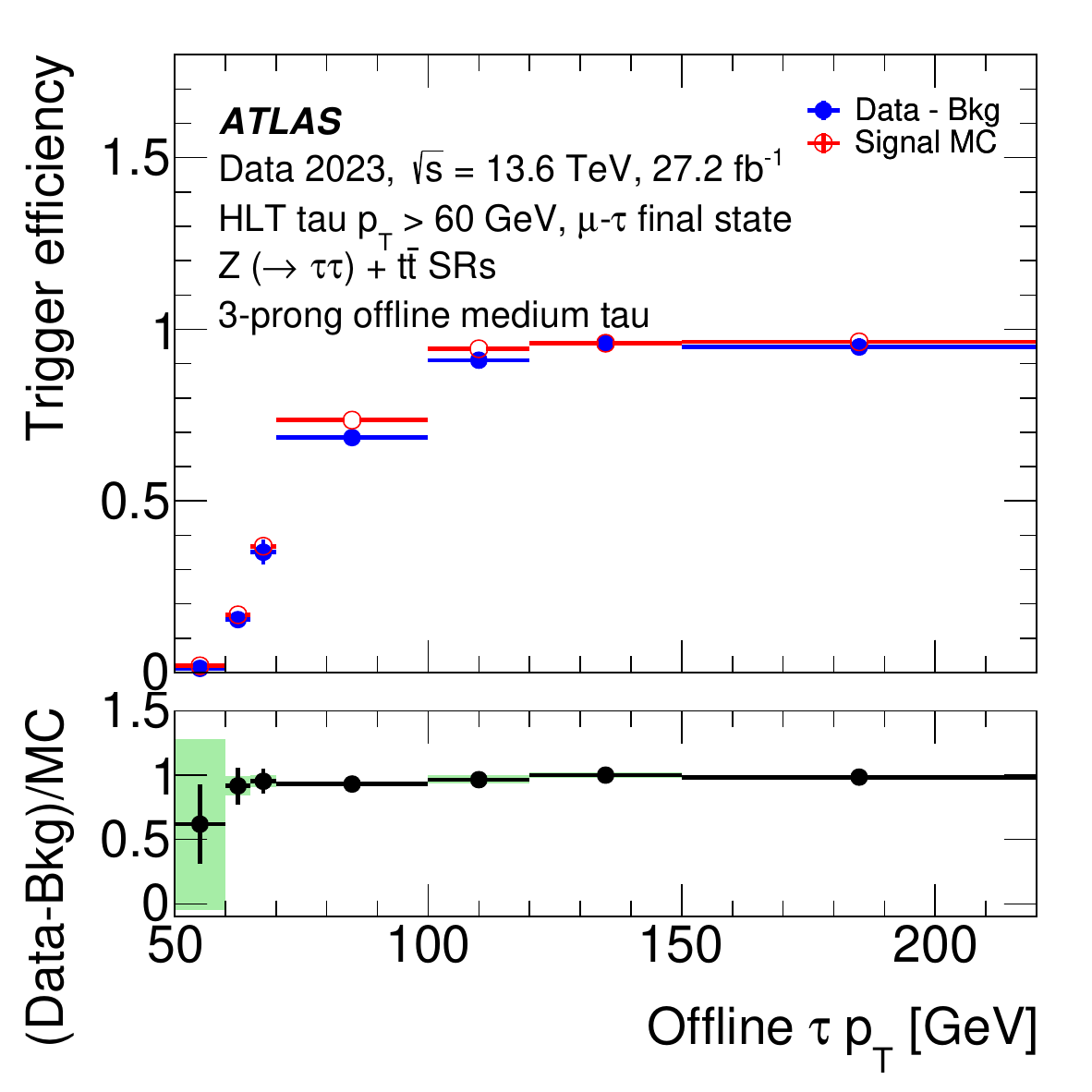}}
\caption{Efficiency of the tau trigger as a function of the offline \tauhadvis \pt for offline one-prong (\protect\subref{fig:perf:sfs:comb:2023:mutau40:1p}, \protect\subref{fig:perf:sfs:comb:2023:mutau60:1p}) and offline three-prong (\protect\subref{fig:perf:sfs:comb:2023:mutau40:3p}, \protect\subref{fig:perf:sfs:comb:2023:mutau60:3p}) candidates. These measurements are performed using $\mu+\tauhad$ tag and probe triggers, with a trigger level \tauhadvis having $\pT > \SI{40}{\GeV}$ in (\protect\subref{fig:perf:sfs:comb:2023:mutau40:1p}, \protect\subref{fig:perf:sfs:comb:2023:mutau40:3p}) and $\pT > \SI{60}{\GeV}$ in (\protect\subref{fig:perf:sfs:comb:2023:mutau60:1p}, \protect\subref{fig:perf:sfs:comb:2023:mutau60:3p}). Efficiencies are computed from the sum of \Ztautau and \ttbar signal region events estimated both for data with background subtraction and MC signal. The offline \tauhadvis candidates are required to have $\pT > \SI{30}{\GeV}$ (\protect\subref{fig:perf:sfs:comb:2023:mutau40:1p}, \protect\subref{fig:perf:sfs:comb:2023:mutau40:3p}) and \SI{50}{\GeV} (\protect\subref{fig:perf:sfs:comb:2023:mutau60:1p}, \protect\subref{fig:perf:sfs:comb:2023:mutau60:3p}). Bottom panels show the ratio between signal and data-background efficiencies with statistical uncertainties only and the bands represent the systematic uncertainty. All the distributions are shown using the 2023 data sample.}
\label{fig:perf:sfs:comb:2023}
\end{figure}
\begin{figure}[!htbp]
\centering
\subfloat[]{\label{fig:perf:sfs:comb:sf:1p}\includegraphics[width=0.45\linewidth]{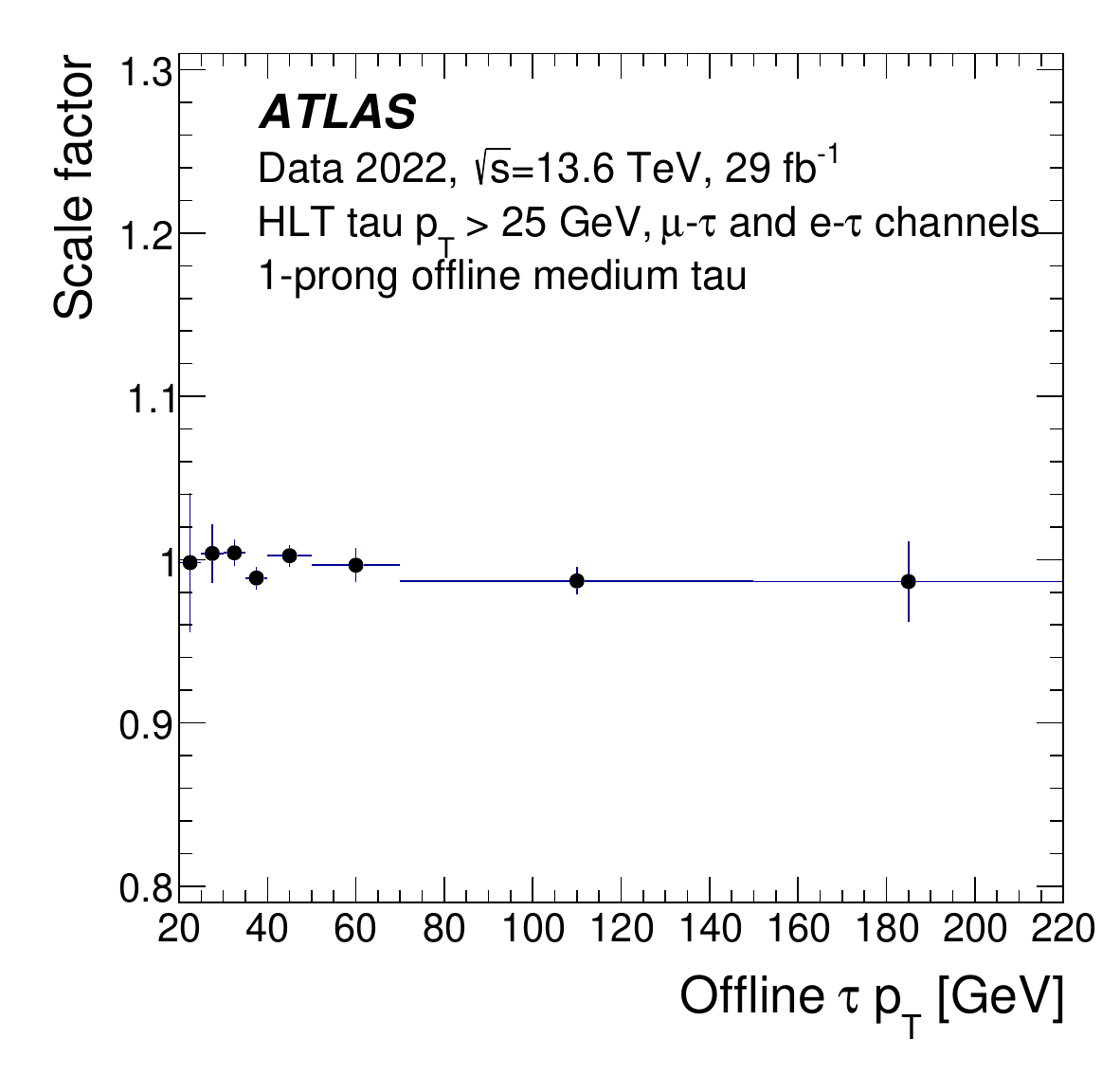}}
\subfloat[]{\label{fig:perf:sfs:comb:sf:3p}\includegraphics[width=0.45\linewidth]{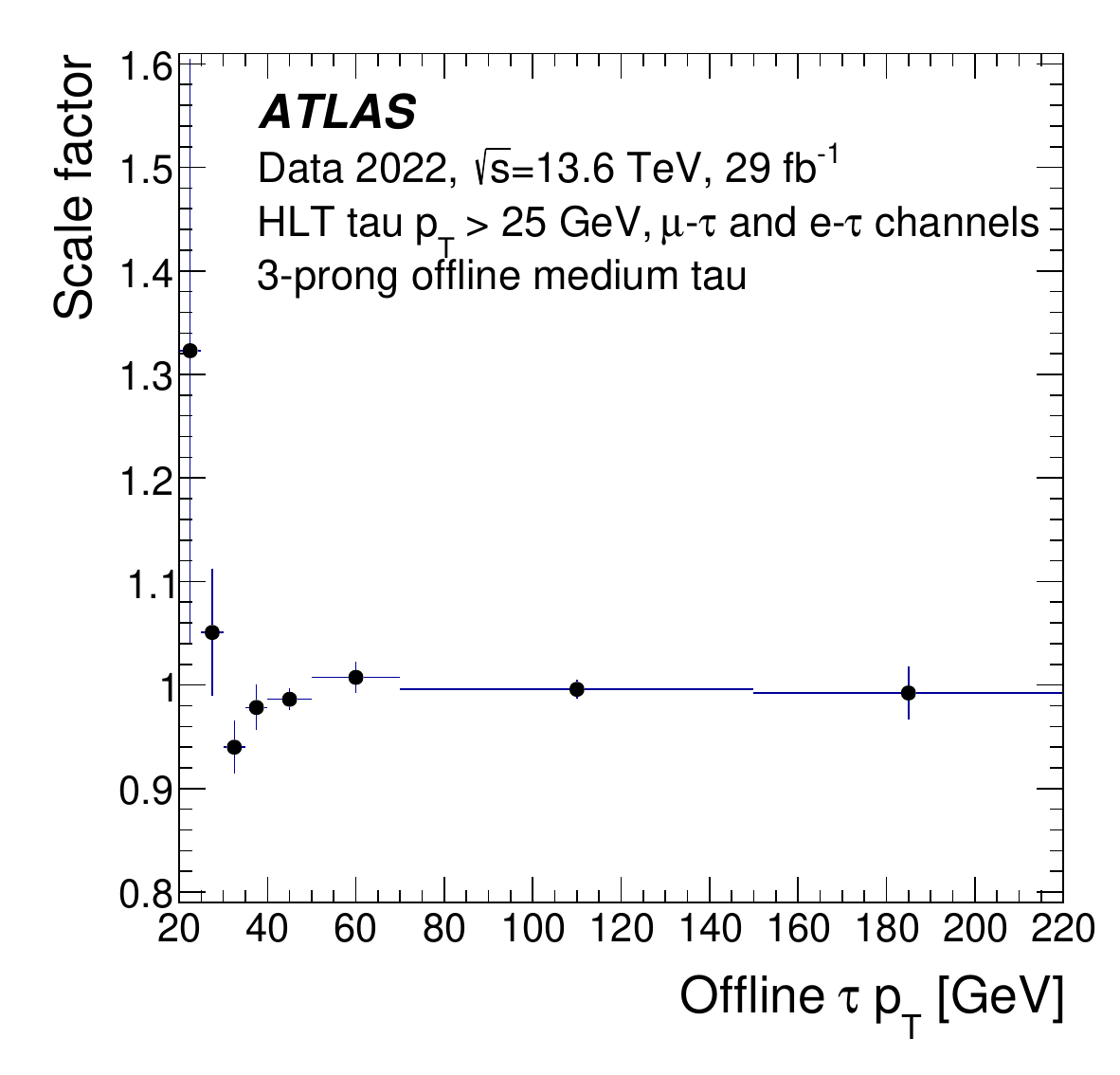}}
\caption{$\tauhad$ efficiency scale factors for \emph{Medium} \protect\subref{fig:perf:sfs:comb:sf:1p} offline one-prong and \protect\subref{fig:perf:sfs:comb:sf:3p} offline three-prong \tauhad, obtained combining the scale factors of the $e + \tauhad$ and $\mu + \tauhad$ channels weighted by their total uncertainties in the 2022 data sample. Total uncertainties are shown.}
\label{fig:perf:sfs:comb:sf}
\end{figure}
The final SF estimation involves evaluating the consistency of the SFs obtained from the electron and muon final states. Given the observed agreement between the two measurements, the final SF is derived as the weighted sum of the SFs from the electron and muon channels:
\[
\mathrm{SF}_{\mathrm{combined}} = \frac{w_e \cdot \mathrm{SF}_e + w_\mu \cdot \mathrm{SF}_\mu}{w_e + w_\mu},
\]
and its associated uncertainty as:
\[
\sigma^2_{\mathrm{SF}} = \frac{1}{w_e + w_\mu},
\]
where $w_e = 1/\sigma_e^2$ and $w_\mu = 1/\sigma_\mu^2$ are the weights determined by the total uncertainties in the SFs for the electron and muon final states, respectively.

A final set of SFs, to be used in physics analyses, is provided as a function of the \pt of the offline \tauhadvis candidate for one-prong and three-prong \tauhad candidates, and for different offline \tauhad identification working points. An example of the SFs is given in \cref{fig:perf:sfs:comb:sf}. In the plateau region of the efficiency curve for a given single tau trigger, the data to simulation correction factors range from $0.95$--$1.05$ with a total uncertainty of $2$--$3\%$ over a wide range of the transverse momentum of hadronic tau candidate.


\section{Possible future improvements}\label{sec:future}

The HLT \tauhad trigger strategy was designed and optimised for the \RunThr data-taking period, which ended in June 2026. An approximate 4-year shutdown will lead to the High-Luminosity LHC (HL-LHC)~\cite{Aberle:2749422} phase, with the accelerator delivering an instantaneous luminosity of \SI{5.0e34}{cm^{-2}s^{-1}}, corresponding to an average number of about 140 collisions per bunch crossing. Experiments will also be equipped with upgraded detectors, including the new all-silicon ATLAS Inner Tracker (ITk)~\cite{ATLAS-TDR-25,ATLAS-TDR-30,IDTR-2023-01}.

This section describes the work being carried out to improve the algorithms used for \tauhad triggers, increasing their performance in denser final states and their robustness against increased \pileup levels. The expected performance of the explored strategies is assessed on the simulated MC samples for \RunThr, and the observed performance is measured in test trigger selections that operated during the 2026 data-taking period.

\subsection{Optimising computing resources}%

As described in \cref{sec:trigger}, the requirements to select \tauhad candidates are applied in sequential steps, so the usage of computing resources can be optimised by applying further selections rejecting fake candidates at an early stage. Selections based on fast tracking are implemented and validated in \bjet trigger selections and are also used for chains involving multiple \bjet and \tauhad candidates~\cite{ATL-DAQ-PUB-2026-002,TRIG-2022-03} in \RunThr.

The computing footprint of reconstruction algorithms can also be reduced by introducing new tools for the fast extraction of event features.
HLT track reconstruction is one of the most time-consuming tasks~\cite{ATL-DAQ-PUB-2023-002}, largely due to the limited information about the primary vertex (PV) position along the beam axis.

A transformer-based model, adapted from GN2, is designed to provide an estimate of the positions of the PV along the beam line before track reconstruction, using information from pixels and strips, which is processed by a clustering algorithm. The specific inputs to the model are the \pt and $\eta$ of \tauhadvis candidates reconstructed at L1 along with the spatial coordinates of the Si detector clusters within the RoI defined to run the first-stage FTF instance. The output is an estimate of the $z$ position of the PV along with the associated uncertainty. Both quantities are obtained by minimising a regression loss derived from a Gaussian likelihood function~\cite{Nix1994},
\[
\mathrm{Loss} \left( z_{\mathrm{truth}}, z_{\mathrm{predicted}}, \sigma \right) = -\ln\left( \frac{1}{\sqrt{2\pi\sigma^2}} \cdot \exp\left( -\frac{1}{2} \left(\frac{z_{\mathrm{predicted}} - z_{\mathrm{truth}}}{\sigma} \right)^2 \right) \right),
\]
where $z_{\mathrm{truth}}$ and $z_{\mathrm{predicted}}$ are the true and predicted values of the $z$ coordinate of PV, respectively, while $\sigma$ is the uncertainty in the predicted value.

The model is trained using a simulated MC sample of \ttbar+jets and tested on \tauhad candidates in a simulated MC sample of \gtautau events, where the reconstructed candidates are matched to true $\tau$ leptons. The \pt of the candidate is required to be in the range of 20--\SI{250}{\GeV} while $|\eta| < 2.5$ is also required. More detailed information about the MC samples is indicated in \cref{tab:samples}. \cref{fig:future:HitZ:residuals} shows the distribution of residuals for the predicted $z$-coordinate with respect to its true value for all \tauhadvis candidates in the test sample. In addition, these residuals are also shown for candidates in which the uncertainty on the predicted value is below \SI{10}{mm}, which comprise 66.5\% of the sample.

\begin{figure}[ht!]
\centering
\includegraphics[width=0.5\linewidth]{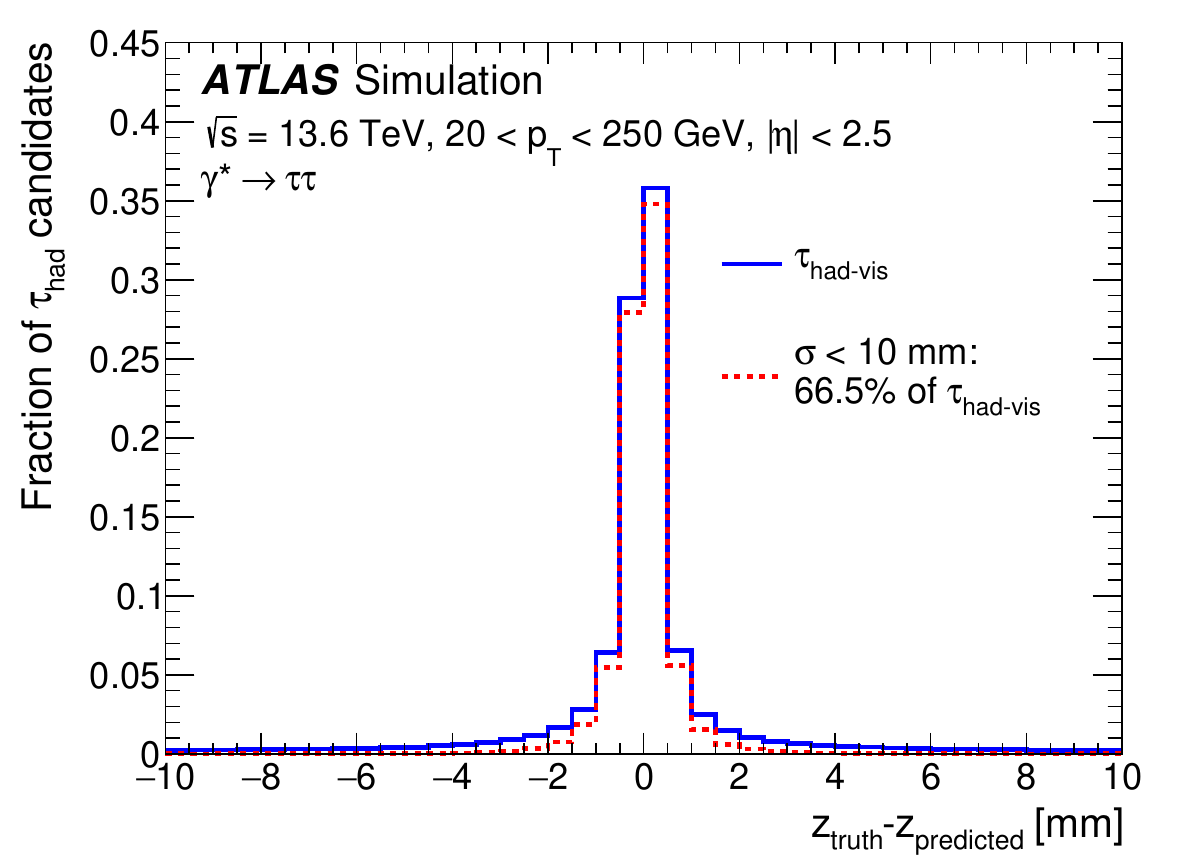}
\caption{Distribution of the residuals obtained by subtracting the predicted $z_{\mathrm{predicted}}$ value and the true $z_{\mathrm{truth}}$ for all \tauhadvis candidates in the \gtautau test sample
and only for those satisfying the condition $\sigma < \SI{10}{mm}$.}
\label{fig:future:HitZ:residuals}
\end{figure}

\begin{figure}[ht!]
\centering
\subfloat[]{\label{fig:future:HitZ:data_perf:eff}\includegraphics[width=0.43\linewidth, trim=0 -5mm 0 0]{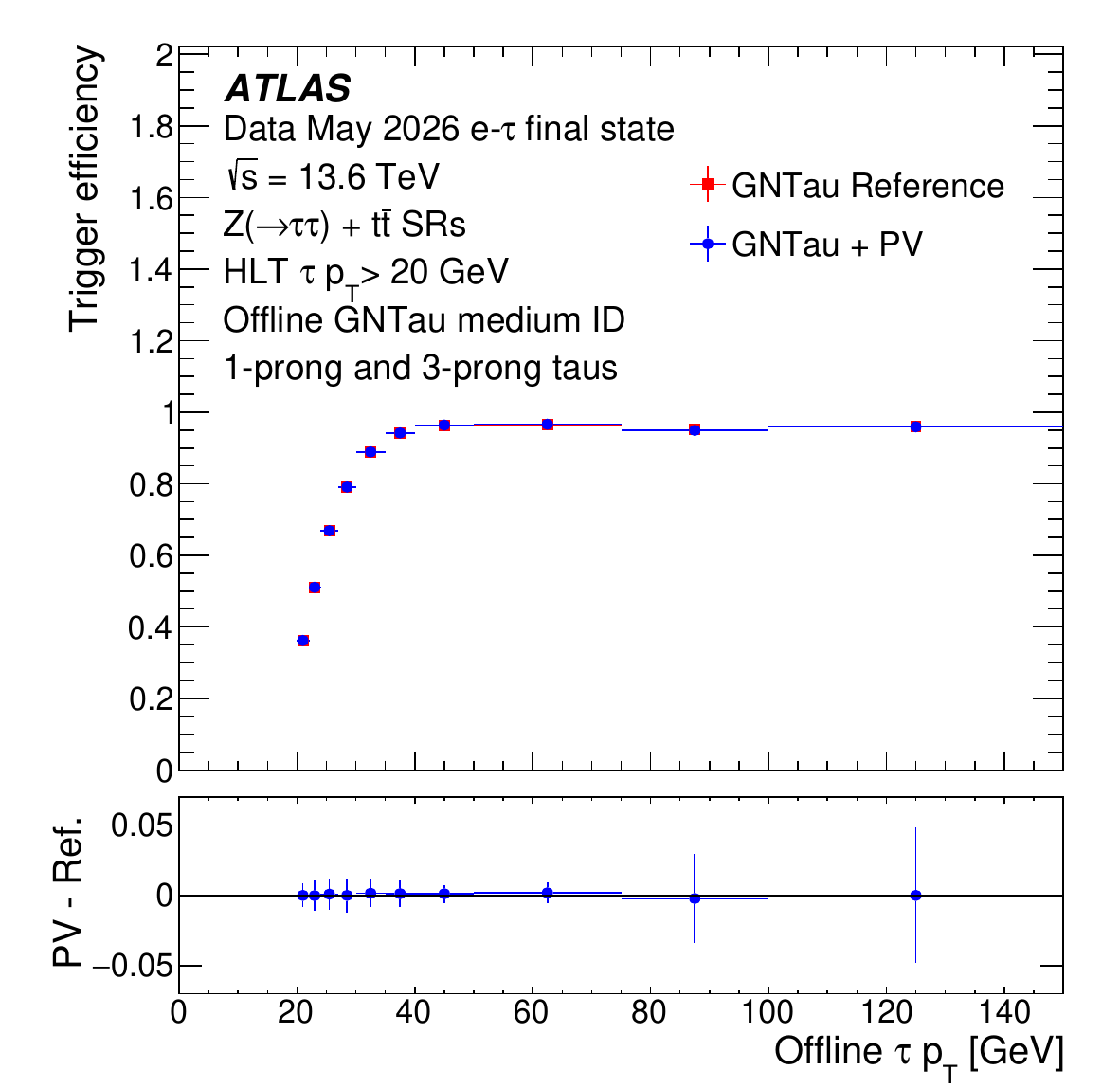}}
\subfloat[]{\label{fig:future:HitZ:data_perf:rate}\includegraphics[width=0.47\linewidth, trim= 0 10mm 0 0, clip]{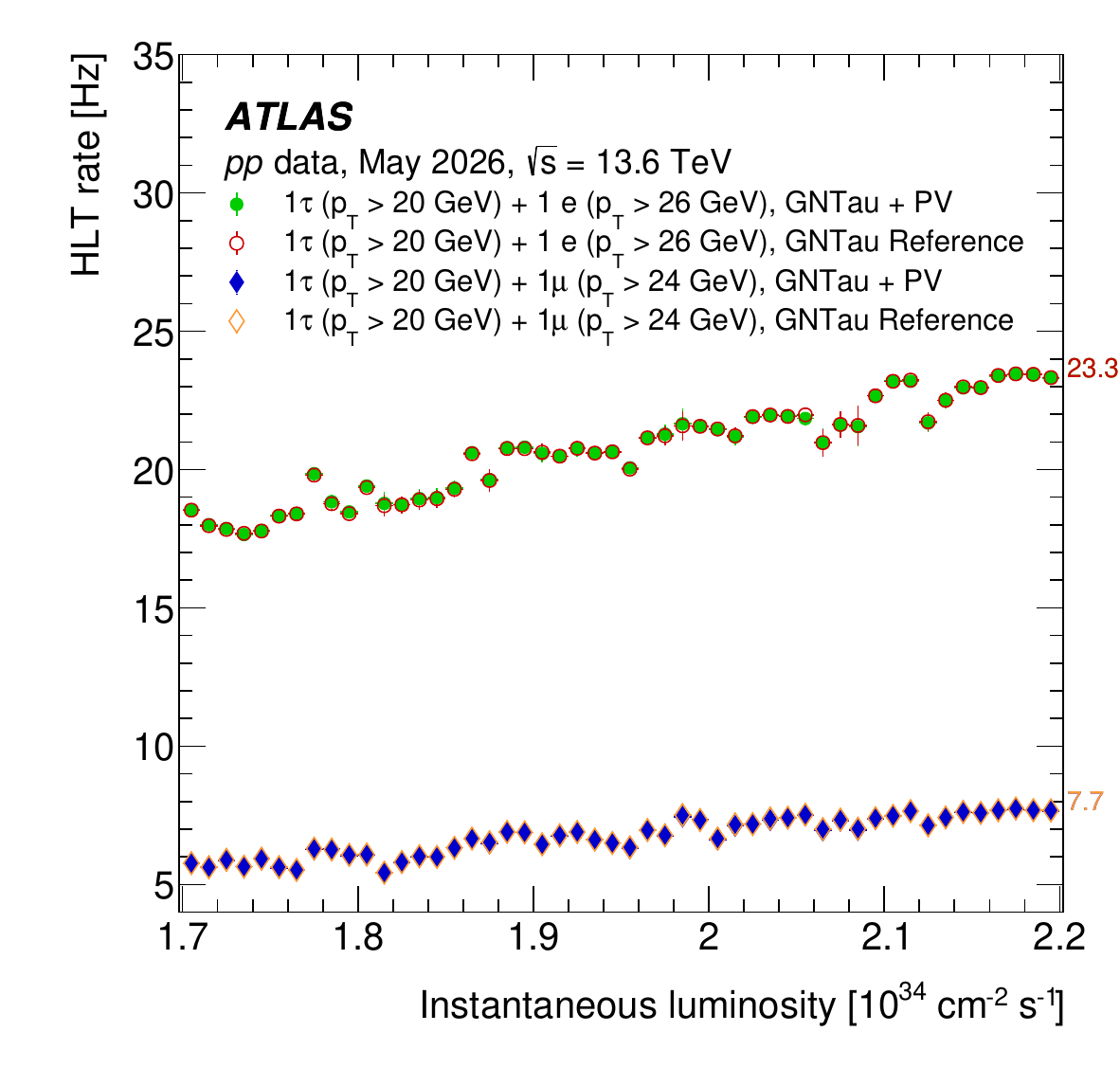}}
\caption{Physics performance, as measured on a fraction of the 2026 data sample, for the reference tau trigger requiring a candidate with $\pT > \SI{20}{\GeV}$ and the corresponding one where the estimated PV position is used to reduce the $z$-width of the first tracking stage RoI. \protect\subref{fig:future:HitZ:data_perf:eff} Efficiencies as a function of the offline \tauhadvis \pt for one-prong and three-prong candidates measured using $e+\tauhad$ tag-and-probe triggers. They are computed from the sum of \Ztautau and \ttbar signal region events estimated in data without background subtraction. The offline \tauhadvis candidates are required to have $\pT > \SI{20}{\GeV}$. The bottom panel shows the efficiency difference between the reference and the test chains. Only statistical uncertainties are shown. \protect\subref{fig:future:HitZ:data_perf:rate}  Trigger acceptance rates for the $e+\tauhad$ and $\mu+\tauhad$ triggers.}
\end{figure}

Based on these results, new test chains were developed for MC-based studies and deployed online as support trigger chains during 2026 data taking. In the test chains the tau reconstruction workflow is modified, so the new algorithm is always run on the first-stage tracking RoI. If the obtained $\sigma$ value is below a chosen threshold $\sigma_{\text{thr}}$, the $z_{\text{predicted}}$ value is used to update the first-stage tracking RoI, whose $z$-width is reduced from the standard value of \SI{180}{mm} to three times  $\sigma_{\text{thr}}$. If instead the uncertainty is above the threshold, the reconstruction proceeds with the standard fast tracking RoI definitions. Similar hit-based algorithms used to identify the interaction vertex position along the z-axis are already used in the ATLAS trigger~\cite{Konstantinidis:2002pxa}.

The physics performance of the test \tauhad trigger with a momentum threshold of \SI{20}{\GeV} and the value of $\sigma_{\text{thr}}$ set to \SI{10}{mm} is evaluated in a fraction of 2026 data sample using the tag-and-probe method in $e+\tauhad$ final state, as described in \cref{sec:perf}. The offline \tauhad candidates are required to pass identification requirements based on the newest \texttt{GNTau} algorithm.

The efficiency and rates for the test triggers are shown in \cref{fig:future:HitZ:data_perf:eff} and \cref{fig:future:HitZ:data_perf:rate}, respectively, and found to be almost identical to the ones measured for the analogous reference triggers. The number of pixel detector clusters processed for track seeding in the first tracking stage is reduced by 60\% as shown in \cref{fig:future:HitZ:tracking:nhits}. This leads to a 68\% decrease in the execution time for the corresponding track seeding task, as shown in \cref{fig:future:HitZ:tracking:timing}.

\begin{figure}[ht!]
\centering
\subfloat[]{\label{fig:future:HitZ:tracking:nhits}\includegraphics[width=0.48\linewidth]{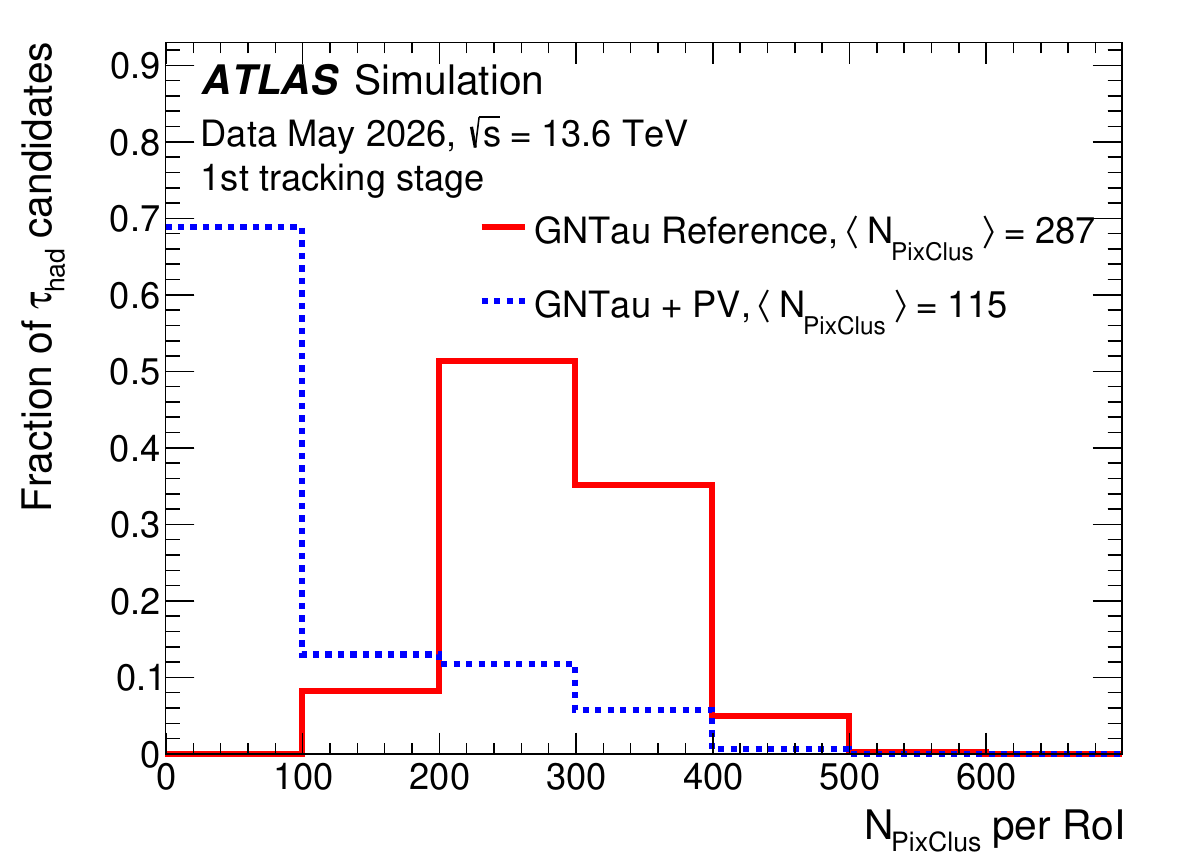}}
\subfloat[]{\label{fig:future:HitZ:tracking:timing}\includegraphics[width=0.48\linewidth]{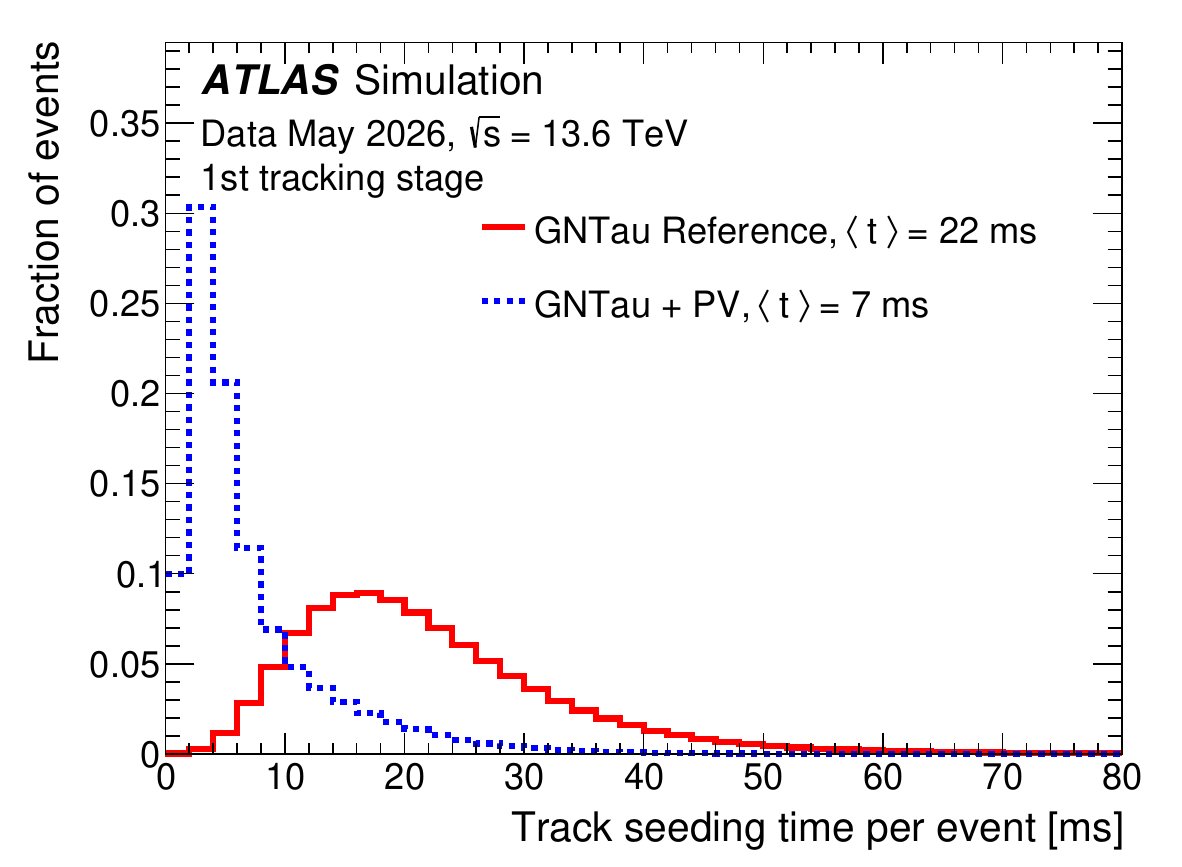}}
\caption{Distributions related to the first tracking stage processing, as measured on a fraction of the 2026 data sample, for the reference tau trigger requiring a trigger level candidate having $\pT > \SI{20}{\GeV}$ and the corresponding one where the estimated PV position is used to reduce the $z$-width of the first tracking stage RoI.  \protect\subref{fig:future:HitZ:tracking:nhits} Number of pixel detector clusters processed by the first tracking stage for the two triggers. \protect\subref{fig:future:HitZ:tracking:timing} Track pattern recognition execution time for the two triggers. The averages of the distributions are included in the legends.}
\end{figure}

\subsection{Improving robustness against track density}%

The selection for the \tauhad trigger, as described in \cref{sec:HLT:Tracks}, is based on track counting and thus its performance can deteriorate in the high \pileup environment at the HL-LHC. For example, an increase in track density can reduce the correspondence between the prongness at true level and the number of tracks reconstructed in the core region of \tauhadvis candidates. This is verified by calculating the efficiency to appropriately count the number of tracks of candidates with $\pT > \SI{20}{\GeV}$ and $|\eta| < 2.5$ from simulated MC samples of different physics processes. The efficiency of the track counting-based candidate selection is 98\% for \gtautau and \Ztautau, while it drops to 93\% for \ttbar events.
As a solution to these undesired features, and to move towards a modern end-to-end approach to tau identification, the track-based candidate selection step can be removed, with the number of tracks being incorporated in the training of the \texttt{GNTau} algorithm. %

\begin{figure}[ht!]
\centering
\subfloat[]{\label{fig:future:ID:ROC_IDPerf:gtautau}\includegraphics[width=0.45\linewidth]{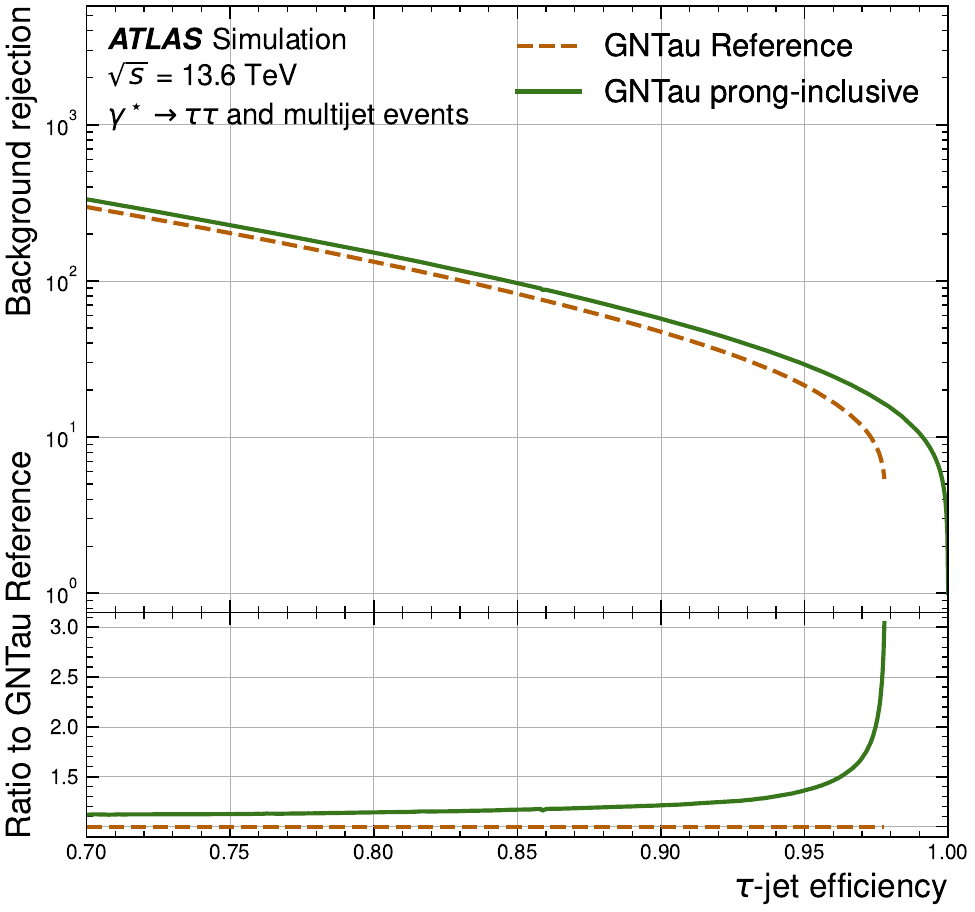}}
\subfloat[]{\label{fig:future:ID:ROC_IDPerf:ttbar}\includegraphics[width=0.45\linewidth]{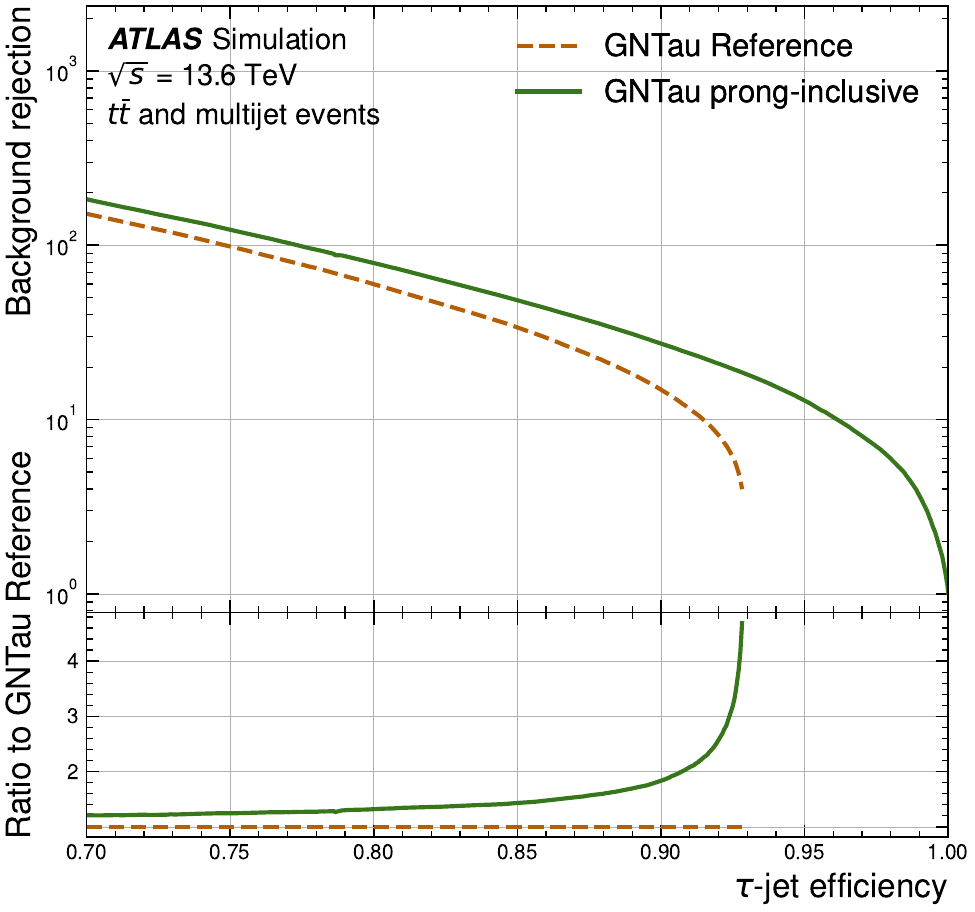}}
\caption{ROC curves illustrating the performance of the reference \texttt{GNTau} training, based on track counting preselection and sample splitting based on prongness, versus the new proposed \texttt{GNTau} prong-inclusive approach, with no preselection and sample splitting. At 92\% efficiency WP, the background rejection improves a factor 1.2 for \tauhadvis candidates in \gtautau events \protect\subref{fig:future:ID:ROC_IDPerf:gtautau} and a factor 2.6 for those in \ttbar events \protect\subref{fig:future:ID:ROC_IDPerf:ttbar}. Binomial uncertainties in the ROCs are considered in the figure, though they are not visible due to their small size.}
\label{fig:future:ID:ROC_IDPerf}
\end{figure}

The reference \texttt{GNTau} strategy, described in \cref{sec:HLT:Tracks}, foresees a training in which the sample is split into three categories based on the track-based prongness classification: zero-prong, one-prong and multi-prong corresponding to candidates with $N_{\mathrm{core}} = 0$, $N_{\mathrm{core}} = 1$, $N_{\mathrm{core}} \geq 2$, respectively. Tau identification is then trained on each of the three prongness classes with the respective exclusive model. In the proposed prong-inclusive approach, training is instead performed inclusively on the overall sample without any preselection, reducing the number of models to be trained and maintained. In both cases, training is performed using true \tauhadvis candidates in simulated samples of \tauhad decays from \gtautau, \Ztautau and \ttbar events as the signal target and light-quark and gluon jets in multijet events as background. The MC samples used in the training, summarised in \cref{tab:samples}, are the same ones used to train the \texttt{GNTau} model explained in \cref{sec:HLT:Tracks}. The performance of the two models is compared in terms of the inclusive \tauhadvis efficiency, i.e. the efficiency on the whole sample before applying the track-counting selection. For the reference model, the predictions for the three prongness categories are merged together to evaluate the overall performance.

\cref{fig:future:ID:ROC_IDPerf} shows the comparison of ROC curves for the reference \texttt{GNTau} model with the proposed prong-inclusive one, obtained evaluating the selection efficiency for true \tauhadvis candidates in \gtautau and \ttbar events versus the rejection against light-jets in multijet events.  It can be seen that the reference \texttt{GNTau} model does not reach an efficiency above 98\%. The endpoint in its ROC curve corresponds to the efficiency and rejection of the selection criteria for the track-based selection, applied before evaluating and using the \texttt{GNTau} discriminant. The results confirm the expected dependence on the event topology and show how a single training without any selection based on track counting allows the definition of better-performing working points, especially at high efficiency, which is particularly important for trigger operation.

\begin{table}[htbp]
\centering
\caption{Comparison between the reference and extended \texttt{GNTau} models.}
\label{tab:future:ID:GNTauExt}
\renewcommand{\arraystretch}{1.1} %
\begin{tabular}{c|cc|ccc|c}
\toprule
& \multicolumn{2}{c|}{Number of inputs} & \multicolumn{3}{c|}{Model hyperparameters} & \multirow{3}{*}{\begin{tabular}[c]{@{}c@{}}Trainable\\ parameters\end{tabular}} \\ \cmidrule{2-6}
Model & Tracks & Clusters & \begin{tabular}[c]{@{}c@{}}Attention\\ heads\end{tabular} & \begin{tabular}[c]{@{}c@{}}Hidden\\ dimension\end{tabular} & \begin{tabular}[c]{@{}c@{}}Transformer\\ layers\end{tabular} & \\
\midrule
Reference & 10 & 8 & 2 & 128 & 2 & 343K \\
Extended & 20 & 16 & 8 & 256 & 4 & 2.3M \\
\bottomrule
\end{tabular}
\end{table}

\begin{figure}[ht!]
\centering
\subfloat[]{\label{fig:future:ID:ROC_GNTauExt:gtautau}\includegraphics[width=0.4\linewidth]{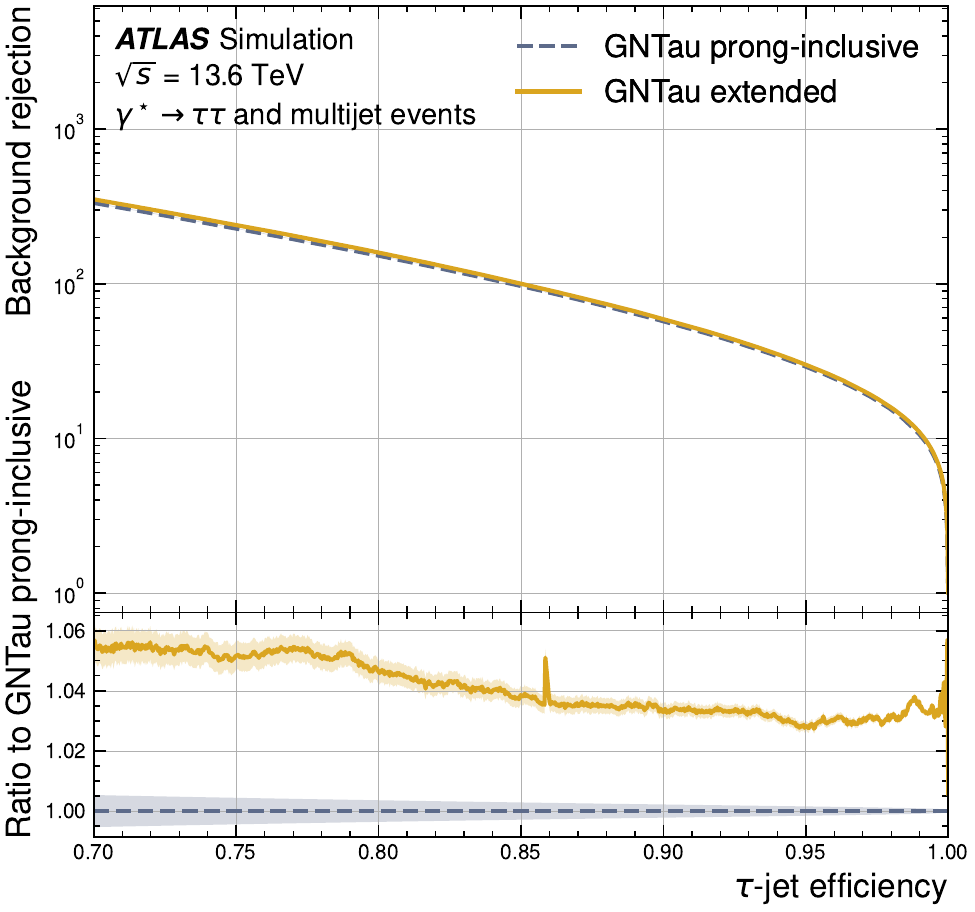}}
\subfloat[]{\label{fig:future:ID:ROC_GNTauExt:ttbar}\includegraphics[width=0.4\linewidth]{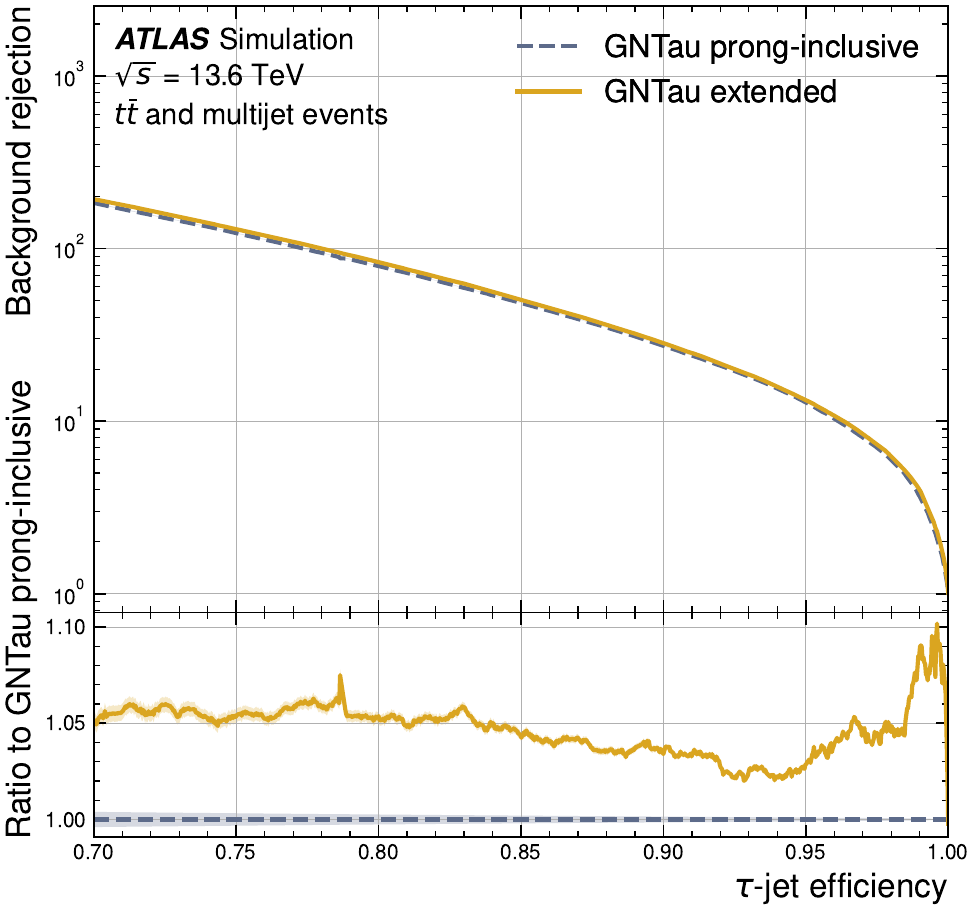}}
\caption{ROC curves illustrating the performance of the prong-inclusive \texttt{GNTau} model (dashed) and of the extended model in \gtautau events \protect\subref{fig:future:ID:ROC_GNTauExt:gtautau} and in \ttbar events \protect\subref{fig:future:ID:ROC_GNTauExt:ttbar}. The shaded bands represent the binomial uncertainties in the ROC curves.}
\label{fig:future:ID:ROC_GNTauExt}
\end{figure}

Another aspect of the reference \texttt{GNTau} strategy is the choice of the hyperparameters for the transformer model described in \cref{sec:HLT:ID:GNTau}. The parameters are optimised for \RunThr pile-up conditions. These include the maximum number of charged-particle tracks and energy clusters in the calorimeter accepted as inputs, respectively set to 10 and 8.

Increasing the track density or the number of \pileup interactions can significantly impact the number of tracks and energy clusters associated with each \tauhadvis candidate for identification purposes. The limits adopted on the number of tracks and energy clusters are already exceeded in the \RunThr training samples, particularly for energy clusters, whose multiplicity peaks at around 20 in the training data sample. To retain enough margin for future increases in object multiplicity, the maximum number of associated tracks and energy clusters accepted as inputs is therefore doubled. To help the model to accommodate such larger number of input information, its internal structure is also modified, as described in~\cref{tab:future:ID:GNTauExt} leading to a deeper and more complex network. The extended model is trained and tested on the same samples using the same prong-inclusive model with larger number of inputs. The performance comparison is reported in \cref{fig:future:ID:ROC_GNTauExt}, where the extended model shows up to a 5\% increase in rejection, hinting at a deeper understanding of the task. \cref{fig:future:ID:GNTauExt_eff} shows the expected and observed improvements in efficiency for the proposed trigger based on the extended \texttt{GNTau} model.
\begin{figure}[ht!]
\centering
\subfloat[]{\label{fig:future:ID:GNTauExt_eff:mc}\includegraphics[width=0.4\linewidth]{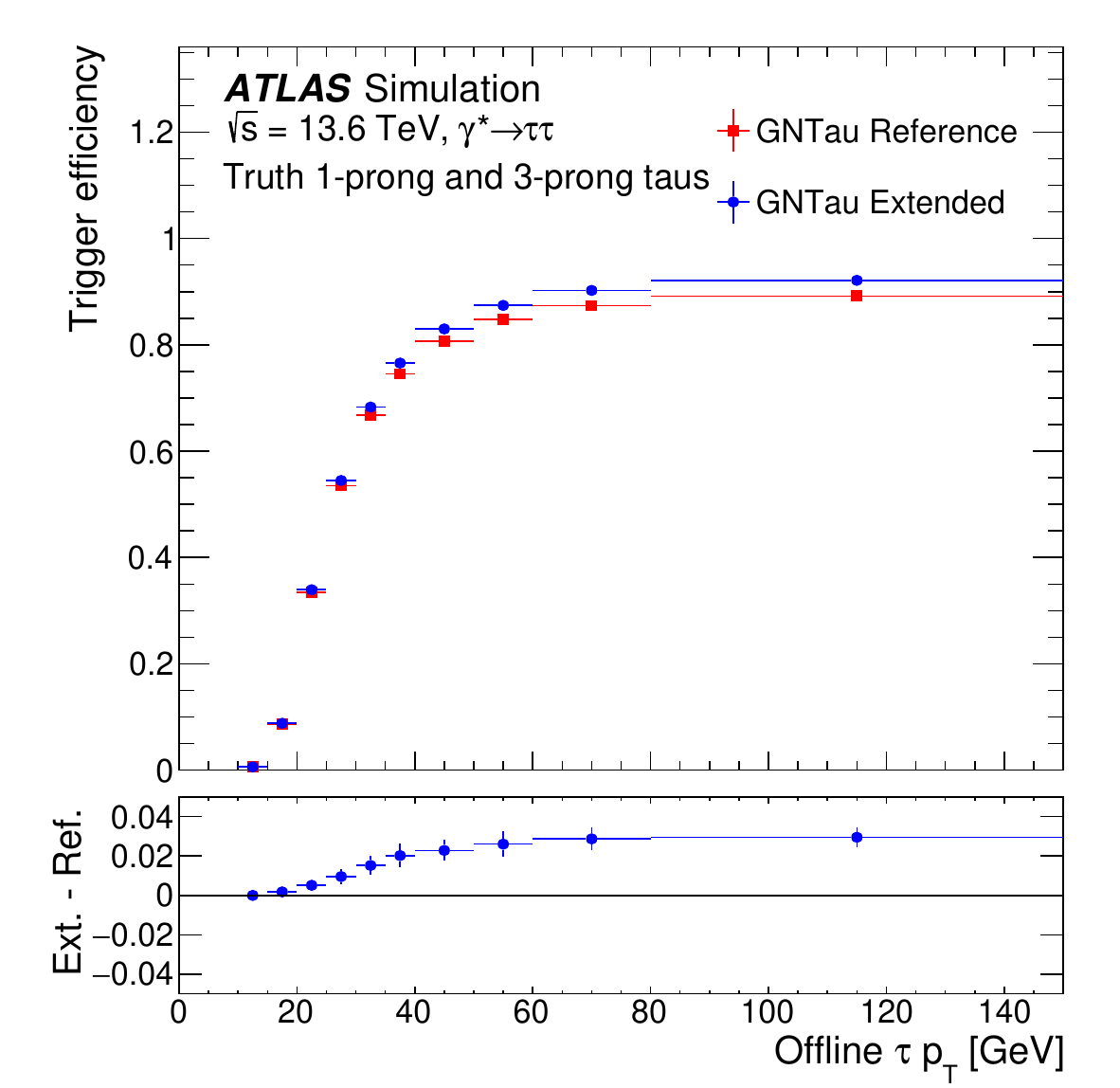}}
\subfloat[]{\label{fig:future:ID:GNTauExt_eff:data}\includegraphics[width=0.4\linewidth]{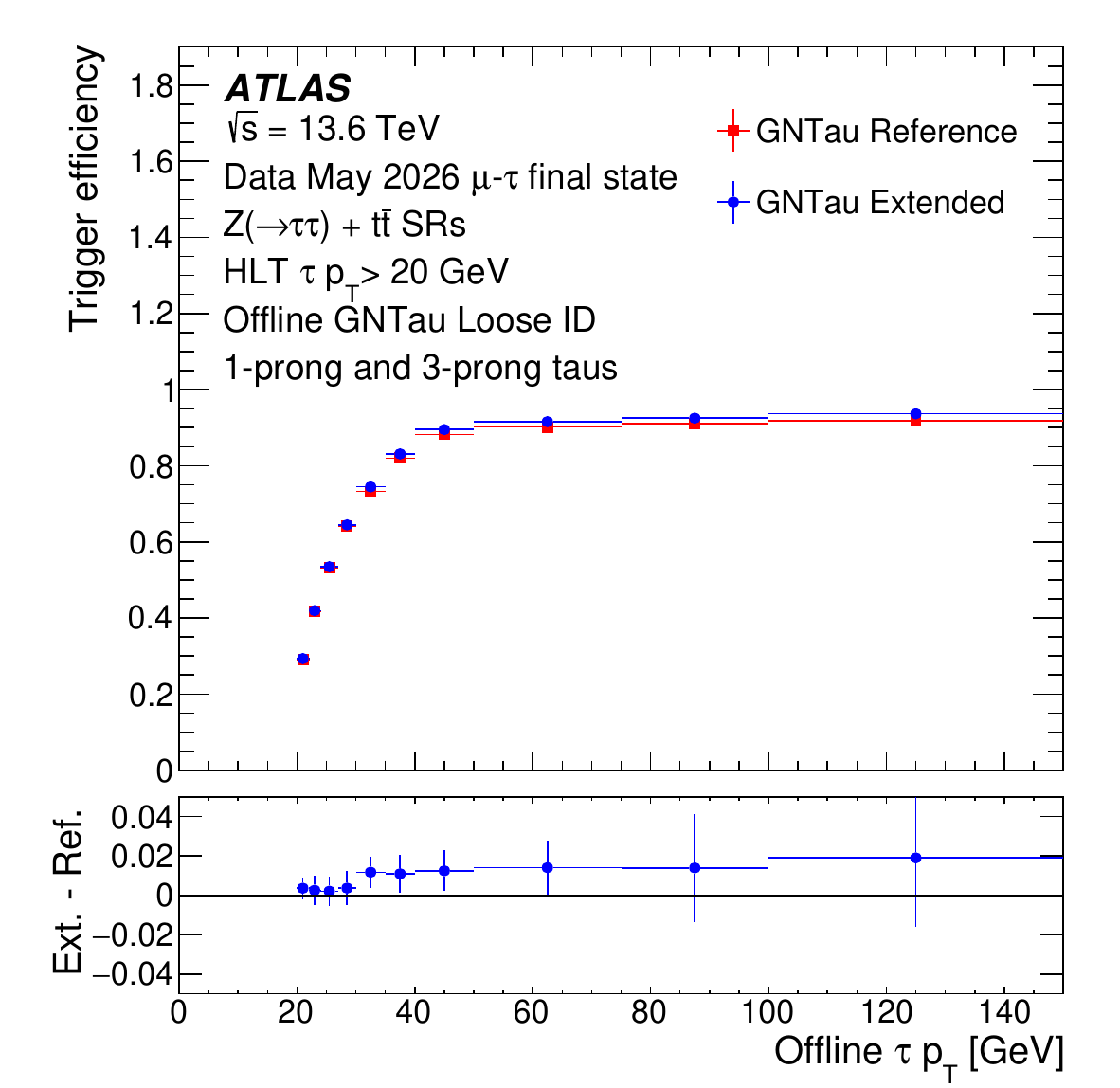}}
\caption{Efficiencies of one-prong and three-prong candidates for selection criteria based on reference and extended \texttt{GNTau} models, measured as a function of the offline \tauhadvis \pt on \protect\subref{fig:future:ID:GNTauExt_eff:mc} \gtautau simulated samples and \protect\subref{fig:future:ID:GNTauExt_eff:data} on a fraction of the 2026 data sample. For data, efficiencies are computed using $\mu+\tauhad$ tag-and-probe triggers, from the sum of \Ztautau and \ttbar signal region events estimated in data without background subtraction. The offline \tauhadvis candidates are required to have $\pT > \SI{20}{\GeV}$. The bottom panels show the efficiency difference between the reference and the extended selections. Only statistical uncertainties are shown.}
\label{fig:future:ID:GNTauExt_eff}
\end{figure}

\cref{fig:future:ID:GNTauExt_eff:mc} shows the expected performance computed using a \gtautau simulated MC sample  while \cref{fig:future:ID:GNTauExt_eff:data} shows the observed performance in a fraction of 2026 data using the tag-and-probe method in $\mu+\tauhad$ final state, as was done in \cref{fig:perf:sfs:comb:2023} of \cref{sec:perf:tnp}. The offline \tauhad candidates are required to pass a \emph{Loose} identification requirement based on the offline \texttt{GNTau} algorithm.

\begin{figure}[ht!]
\centering
\includegraphics[width=0.5\linewidth]{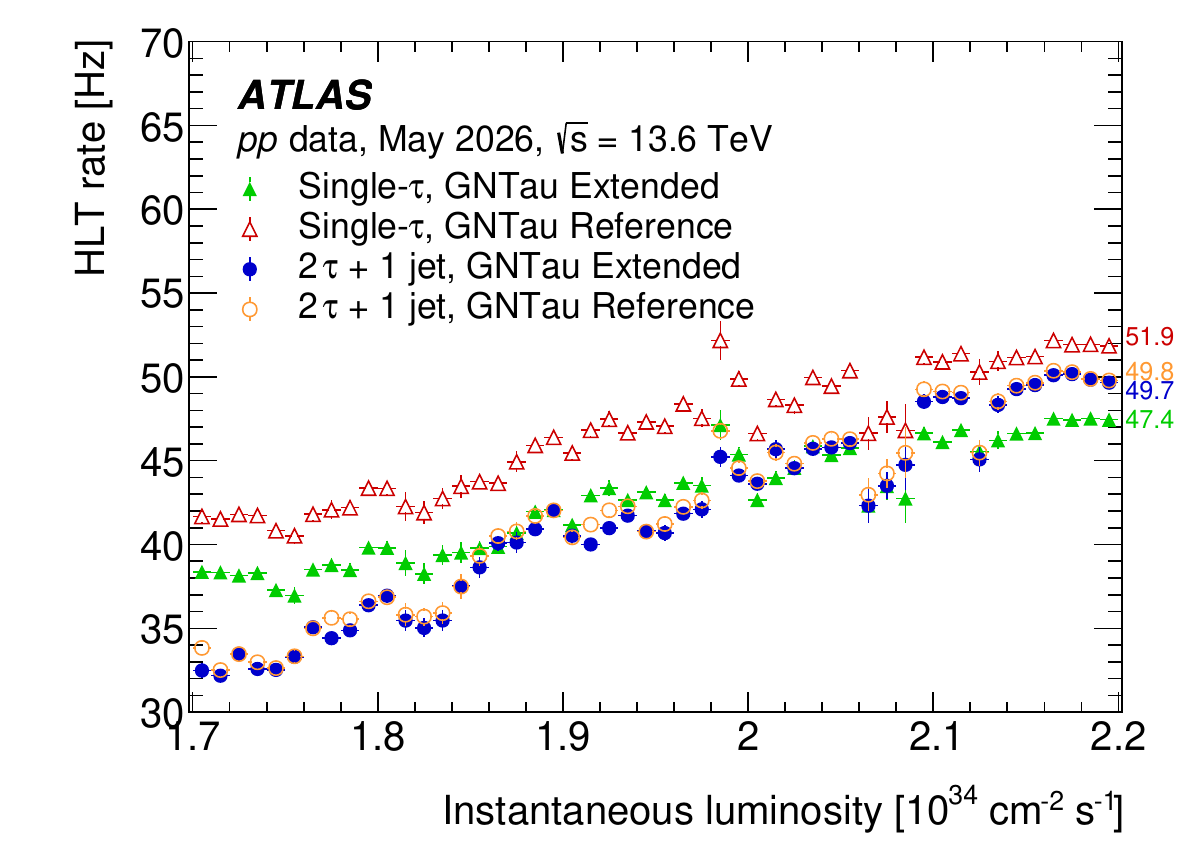}
\caption{HLT rates for single and di-$\tau$ triggers based on the reference and extended \texttt{GNTau} models, as measured in \(pp\) collisions recorded at $\sqrt{s} = 13.6$ \tev in May 2023. The rates at a reference luminosity of $\SI{2.2e34}{cm^{-2}s^{-1}}$ are indicated by the numbers to the right in Hz. The error bars represent statistical uncertainties.}
\label{fig:future:ID:GNTauExt_rates}
\end{figure}

To measure the tagging performance on real data, the extended \texttt{GNTau} model was deployed in test chains for the 2026 data-taking campaign, with the lowest \pt requirement among all single-\tauhad triggers and among all di-\tauhad triggers available in the trigger system. The identification requirements were optimised to achieve rates within the allocated budget while targeting higher signal efficiencies as indicated by MC studies.
The observed rates for triggers based on both the reference and extended \texttt{GNTau} models are shown in ~\cref{fig:future:ID:GNTauExt_rates}
as a function of the instantaneous luminosity. The observed rates of the new triggers are either consistent or lower than the reference ones.


\section{Summary and conclusions}%

Many updates to the architecture for the selection algorithms of hadronic tau triggers at the ATLAS experiment were implemented for Run 3. A summary of the various updates, including the commissioning of the Phase-I L1Calo system, and detailed \RunThr performance studies of the hadronic tau triggers are presented. With the increased granularity of information used by the tau trigger at L1, more sophisticated algorithms, such as a BDT, are used to identify physics objects. The isolation criterion applied at L1 is redefined and optimised to maximise the selection efficiency of the signal like events. New and improved selection algorithms and strategies such as GNN at the HLT further improve the identification of \tauhadvis candidates at trigger level, despite the harsher environment in \RunThr.

The performance studies of different tau triggers and trigger efficiencies relative to the offline quantities are presented using \SI{13.6}{\TeV} \pp collisions data collected during 2022 and 2023. In the plateau region of the efficiency curve for a given single tau trigger, efficiency varies between $90$--$98\%$, while the data to simulation correction factors range from $0.95$--$1.05$ with an uncertainty of $2$--$3\%$ over a wide range of the transverse momentum of the \tauhadvis candidate.

Finally, concrete ideas to improve the tau trigger algorithms for the high pile-up conditions of the HL-LHC, with preliminary studies performed during Run 3 are presented.


\let\clearpage\clearpagebak
\section*{Acknowledgements}

%
%

%
%

We thank CERN for the very successful operation of the LHC and its injectors, as well as the support staff at
CERN and at our institutions worldwide without whom ATLAS could not be operated efficiently.

The crucial computing support from all WLCG partners is acknowledged gratefully, in particular from CERN, the ATLAS Tier-1 facilities at TRIUMF/SFU (Canada), NDGF (Denmark, Norway, Sweden), CC-IN2P3 (France), KIT/GridKA (Germany), INFN-CNAF (Italy), NL-T1 (Netherlands), PIC (Spain), RAL (UK) and BNL (USA), the Tier-2 facilities worldwide and large non-WLCG resource providers. Major contributors of computing resources are listed in Ref.~\cite{ATL-SOFT-PUB-2026-001}.

We gratefully acknowledge the support of ANPCyT, Argentina; YerPhI, Armenia; ARC, Australia; BMWFW and FWF, Austria; ANAS, Azerbaijan; CNPq and FAPESP, Brazil; NSERC, NRC and CFI, Canada; CERN; ANID, Chile; CAS, MOST, NSFC and FRFCU, China; Minciencias, Colombia; MEYS CR, Czech Republic; DNRF and DNSRC, Denmark; IN2P3-CNRS and CEA-DRF/IRFU, France; SRNSFG, Georgia; BMFTR, HGF and MPG, Germany; GSRI, Greece; RGC and Hong Kong SAR, China; ICHEP and Academy of Sciences and Humanities, Israel; INFN, Italy; MEXT and JSPS, Japan; CNRST, Morocco; NWO, Netherlands; RCN, Norway; MNiSW, Poland; FCT, Portugal; MNE/IFA, Romania; MSTDI, Serbia; MSSR, Slovakia; ARIS and MVZI, Slovenia; DSI/NRF, South Africa; MICIU/AEI, Spain; SRC and Wallenberg Foundation, Sweden; SERI, SNSF and Cantons of Bern and Geneva, Switzerland; NSTC, Taipei; TENMAK, T\"urkiye; STFC/UKRI, United Kingdom; DOE and NSF, United States of America.

Individual groups and members have received support from BCKDF, CANARIE, CRC and DRAC, Canada; CERN-CZ and PRIMUS, Czech Republic; COST, ERC, ERDF, Horizon 2020 and Marie Sk{\l}odowska-Curie Actions, European Union; Investissements d'Avenir Labex, Investissements d'Avenir Idex and ANR, France; DFG and AvH Foundation, Germany; Herakleitos, Thales and Aristeia programmes co-financed by EU-ESF and the Greek NSRF, Greece; BSF-NSF and MINERVA, Israel; NCN and NAWA, Poland; La Caixa Banking Foundation, CERCA and AGAUR programs from Generalitat de Catalunya and PROMETEO and GenT Programmes Generalitat Valenciana, Spain; G\"{o}ran Gustafssons Stiftelse, Sweden; The Royal Society and Leverhulme Trust, United Kingdom; Eric and Wendy Schmidt Fund for Strategic Innovation, United States of America.

In addition, individual members wish to acknowledge support from Chile: Agencia Nacional de Investigaci\'on y Desarrollo (FONDECYT 1230812, FONDECYT 1230987, FONDECYT 1240721, FONDECYT 1240864, FONDECYT 3240661); China: Fundamental Research Funds for the Central Universities (010-63263105), Chinese Ministry of Science and Technology (MOST-2023YFA1605700, MOST-2023YFA1609300), National Natural Science Foundation of China (NSFC 12275265, NSFC-W2543005); Czech Republic: Czech Science Foundation (GACR - 24-11373S), Ministry of Education Youth and Sports (ERC-CZ-LL2327, FORTE CZ.02.01.01/00/22\_008/0004632); EU: H2020 European Research Council (ERC - 101002463); European Union: European Research Council (BARD No. 101116429, ERC - 101219398, ERC 101089007), European Regional Development Fund (HE COFUND GA No.101081355, ERDF), Marie Sklodowska-Curie Actions (GAP-101168829); France: Agence Nationale de la Recherche (ANR-22-EDIR-0002, ANR-24-CE31-0504-01); China: Research Grants Council (GRF); Italy: Istituto Nazionale di Fisica Nucleare (LHC-MIUR - 28003/2025), Ministero dell'Università e della Ricerca (NextGenEU I53D23000820006 M4C2.1.1, SOE2024\_0000023); Japan: Japan Society for the Promotion of Science (JSPS KAKENHI  JP25H0063, JSPS KAKENHI JP22H04944, JSPS KAKENHI JP24K23939, JSPS KAKENHI JP24KK0251, JSPS KAKENHI JP25H00650, JSPS KAKENHI JP25H01291, JSPS KAKENHI JP25K01011, JSPS KAKENHI JP25K01023, JSPS KAKENHI JP25KK0047); Poland: Polish National Science Centre (NCN OPUS 2023/51/B/ST2/02507, NCN OPUS nr 2022/47/B/ST2/03059, UMO-2019/34/E/ST2/00393, UMO-2021/42/E/ST2/00350, UMO-2022/47/O/ST2/00148, UMO-2023/49/B/ST2/04085, UMO-2023/51/B/ST2/00920, UMO-2024/53/N/ST2/00869); Spain: Agència de Gestió d'Ajuts Universitaris i de Recerca. (AGAUR - 2023 BP 00141), Ministry of Science and Innovation (RYC2019-028510-I, RYC2020-030254-I, RYC2021-031273-I, RYC2022-038164-I), Ministerio de Ciencia, Innovación y Universidades/Agencia Estatal de Investigaci\'on (PID2022-142604OB-C22); Sweden: Carl Trygger Foundation (Carl Trygger Foundation CTS 22:2312), Swedish Research Council (Swedish Research Council 2023-04654, VR 2021-03651, VR 2022-03845, VR 2022-04683, VR 2023-03403, VR 2024-05451, VR 2025-05940), Knut and Alice Wallenberg Foundation (KAW 2023.0366); United Kingdom: The Binks Trust, Royal Society (NIF-R1-231091); United States of America: U.S. Department of Energy (ECA DE-AC02-76SF00515), John Templeton Foundation (John Templeton Foundation 63206), Neubauer Family Foundation.

%
%


%
%
%
%
%
%
%
%
%

%
%
%
%
%
%
%
%
%
%

%
%

%
%
\printbibliography
\clearpage
\input{atlas_authlist}
%

%
%

%

%
%
%
%

\end{document}

%% file: atlas_authlist.tex
 
\begin{flushleft}
\hypersetup{urlcolor=black}
{\Large The ATLAS Collaboration}

\bigskip

\AtlasOrcid[0000-0002-6665-4934]{G.~Aad}$^\textrm{\scriptsize 102}$,
\AtlasOrcid[0000-0001-7616-1554]{E.~Aakvaag}$^\textrm{\scriptsize 17}$,
\AtlasOrcid[0000-0002-5888-2734]{B.~Abbott}$^\textrm{\scriptsize 121}$,
\AtlasOrcid[0000-0002-0287-5869]{S.~Abdelhameed}$^\textrm{\scriptsize 83b}$,
\AtlasOrcid[0000-0002-1002-1652]{K.~Abeling}$^\textrm{\scriptsize 54}$,
\AtlasOrcid[0000-0001-5763-2760]{N.J.~Abicht}$^\textrm{\scriptsize 48}$,
\AtlasOrcid[0000-0002-8496-9294]{S.H.~Abidi}$^\textrm{\scriptsize 30}$,
\AtlasOrcid[0009-0003-6578-220X]{M.~Aboelela}$^\textrm{\scriptsize 44}$,
\AtlasOrcid[0000-0002-9987-2292]{A.~Aboulhorma}$^\textrm{\scriptsize 36e}$,
\AtlasOrcid[0000-0001-5329-6640]{H.~Abramowicz}$^\textrm{\scriptsize 154}$
% ...
\bigskip
\\

% ...
$^{17}$Department for Physics and Technology, University of Bergen, Bergen; Norway.\\
% ...
$^{30}$Physics Department, Brookhaven National Laboratory, Upton NY; United States of America.\\
% ...
$^{36}$$^{(a)}$Facult\'e des Sciences Ain Chock, Universit\'e Hassan II de Casablanca;$^{(b)}$Facult\'{e} des Sciences, Universit\'{e} Ibn-Tofail, K\'{e}nitra;$^{(c)}$Facult\'e des Sciences Semlalia, Universit\'e Cadi Ayyad, LPHEA-Marrakech;$^{(d)}$LPMR, Facult\'e des Sciences, Universit\'e Mohamed Premier, Oujda;$^{(e)}$Facult\'e des sciences, Universit\'e Mohammed V, Rabat;$^{(f)}$Institute of Applied Physics, Mohammed VI Polytechnic University, Ben Guerir; Morocco.\\
% ...
$^{44}$Physics Department, Southern Methodist University, Dallas TX; United States of America.\\
% ...
$^{48}$Fakult\"{a}t Physik, Technische Universit{\"a}t Dortmund, Dortmund; Germany.\\
% ...
$^{54}$II. Physikalisches Institut, Georg-August-Universit\"{a}t G\"ottingen, G\"ottingen; Germany.\\
% ...
$^{83}$$^{(a)}$Khalifa University of Science and Technology, Abu Dhabi;$^{(b)}$New York University Abu Dhabi, Abu Dhabi;$^{(c)}$United Arab Emirates University, Al Ain;$^{(d)}$University of Sharjah, Sharjah; United Arab Emirates.\\
% ...
$^{102}$CPPM, Aix-Marseille Universit\'e, CNRS/IN2P3, Marseille; France.\\
% ...
$^{121}$Homer L. Dodge Department of Physics and Astronomy, University of Oklahoma, Norman OK; United States of America.\\
% ...
$^{154}$Raymond and Beverly Sackler School of Physics and Astronomy, Tel Aviv University, Tel Aviv; Israel.\\
% ...
$^{a}$ Also at Affiliated with an institute formerly covered by a cooperation agreement with CERN.\\
$^{b}$ Also at An-Najah National University, Nablus; Palestine.\\
$^{c}$ Also at Borough of Manhattan Community College, City University of New York, New York NY; United States of America.\\
$^{d}$ Also at Center for Interdisciplinary Research and Innovation (CIRI-AUTH), Thessaloniki; Greece.\\
$^{e}$ Also at Centre of Physics of the Universities of Minho and Porto (CF-UM-UP); Portugal.\\
$^{f}$ Also at CERN, Geneva; Switzerland.\\
% ...

\end{flushleft}
